\documentclass[twocolumn]{aa} 
\usepackage{lscape}
\usepackage{graphicx}
\usepackage{amssymb}
\usepackage{array}
\usepackage{threeparttable}
\usepackage{hyperref}
\usepackage{textcomp}
\usepackage{float}
\usepackage{natbib}		
\usepackage{longtable}
\bibpunct{(}{)}{;}{a}{}{,}  % A&A style
\usepackage{color}

\def\XMM{XMM-\textit{Newton}\ }
 
\begin{document}
    \title{Kinematic analysis of a sample of X-ray luminous distant galaxy clusters}
   \subtitle{The $L_X$ - $\sigma_v$ relation in the $z>0.6$ universe}

   \author{A. Nastasi   \inst{\ref{MPE}}
	\and
	H. B\"ohringer  \inst{\ref{MPE}}
	\and
	R. Fassbender	\inst{\ref{MPE}}
	\and
	A. de Hoon	\inst{\ref{AIP}}
	\and
	G. Lamer	\inst{\ref{AIP}}
	\and
	J. J. Mohr 	\inst{\ref{MPE}, \ref{LMU}, \ref{EXC}}
	\and
	N. Padilla      \inst{\ref{PUC}}
	\and
	G. W. Pratt	\inst{\ref{CEA}}
	\and
	H. Quintana     \inst{\ref{PUC}}
	\and
	P. Rosati	\inst{\ref{FER}}
	\and
	J. S. Santos	\inst{\ref{INAF}}
	\and
	A. D. Schwope    \inst{\ref{AIP}}
	\and
	R. \v Suhada 	\inst{\ref{LMU}}
	\and
	M. Verdugo	\inst{\ref{UW}}
          }

  \institute{Max-Planck-Institut f\"ur extraterrestrische Physik (MPE),
              Giessenbachstrasse~1, 85748 Garching, Germany \\
              \email{alessandro.nastasi@mpe.mpg.de} \label{MPE}
	\and
	Astrophysikalisches Institut Potsdam (AIP), An der Sternwarte~16, 14482 Potsdam, Germany \label{AIP}
        \and
	Department of Physics and Earth Science, University of Ferrara, Italy \label{FER}
        \and
        Institute of Astrophysics, Pontificia Universidad Cat\'olica de Chile, Casilla 306, Santiago 22, Chile\label{PUC}
	\and
	CEA Saclay, Service d'Astrophysique, L'Orme des Merisiers, B\^at. 709, 91191 Gif-sur-Yvette Cedex, France \label{CEA}
	\and
	INAF - Osservatorio Astronomico di Arcetri, Firenze, Italy \label{INAF}
	\and
	Department of Physics, Ludwig-Maximilians Universit\"at M\"unchen, Scheinerstr. 1, 81679 Munich, Germany \label{LMU}
	\and
	Excellence Cluster Universe, Boltzmannstr. 2, 85748 Garching, Germany \label{EXC}
	\and
	Institut f\"ur Astronomie, Universit\"at Wien, T\"urkenschanzstra\ss e 17, 1180 Wien, Austria \label{UW}
             }

   \date{Received 18 July 2013 / Accepted 14 November 2013}

  \abstract
    {}
   {Observations and cosmological simulations show galaxy clusters as a family of nearly \textit{self-similar} objects with properties that can be described by scaling relations as a function of mass and time. Here we study the scaling relations between the galaxy velocity dispersion ($\sigma _v$) and X-ray quantities, such as X-ray bolometric luminosity ($L^{Bol} _{X,500}$) and temperature ($T_X$) in galaxy clusters at high redshifts (0.64 $\leq$ z $\leq$ 1.46). We also compare our results with the analogous study of the local HIFLUGCS sample.}
   {For the analysis, we use a set of 15 distant galaxy clusters extracted from the literature and selected via different methods. We also use a sample of 10 newly discovered clusters selected via their X-ray emission by the \XMM Distant Cluster Project (XDCP), with more than 10 confirmed spectroscopic members per cluster. For both samples, the same method was used to determine $\sigma_v$. We also study the evolution of this scaling relation by comparing the high redshift results with the data from the HIFLUGCS sample, which is taken as a representative of the conditions in the local Universe. For such an analysis, we restrict the study to the clusters in the common $L^{Bol} _{X,500}$ range. We also investigate the $L_X - T_X$ and the $\sigma_v - T_X$ relations for the 15 clusters from the literature sample.}
   {We report the results of the X-ray and kinematic analysis of ten newly detected high redshift clusters and provide their spectroscopic and kinematic details in Appendix A. For the entire distant sample, we find a slope fully consistent with the one typical of local clusters, albeit with a large associated uncertainty ($\sim$\,26\%). We repeat the fit by freezing the slope to the value found for the HIFLUGCS systems restricted to the same luminosity range as our sample to investigate the evolution of the amplitude alone. We find a positive offset of $\Delta A/A$\,=\,0.44\,$\pm$\,0.22 if the self-similar evolution is neglected, hence indicating the possible need for including evolutionary effects. However, the $L_X - T_X$ relation is found to be in good agreement with the local relation without any significant redshift evolution. Finally, the $\sigma_v - T_X$ relation appears to slightly deviate from the theoretical expectation that galaxies and gas particles have a similar specific kinetic energy. However,
 the associated uncertainty is currently too large for making any conclusive statement in this regard.}
   {}


   \titlerunning{Kinematic analysis of a sample of X-ray luminous distant galaxy clusters}
   \authorrunning{A.Nastasi et al.}

   \maketitle
%________________________________________________________________

\section{Introduction}
\label{par:intro}
Galaxy clusters are the most massive collapsed objects in our Universe. Because of their relatively recent formation process, they are also very sensitive probes of the underlying cosmological framework. In addition, they are excellent laboratories for testing models of galaxy formation and the role of merging, environment, and radiative feedback in this context.
In a scenario where clusters form only via pure gravitational forces, they should appear at the end of their formation process as a family of \textit{self-similar} objects; that is, those that are less massive should be the scaled down versions of those that are more massive, with the mass being to first order the only parameter for scaling all the other quantities\footnote{More elaborate models, such as Navarro-Frenk \& White \citep[NFW, ][]{Navarro1997a}, have formation time as second parameter.}. Self-similarity hence predicts that cluster observables, such as the X-ray temperature ($T_X$) of the intracluster medium (ICM), the X-ray bolometric luminosity computed within R$_{500}$\footnote{R$_{500}$ is defined as the radius within which the average mass density is 500 times the critical density of the universe.} ($L^{Bol} _{X,500}$), the Sunyaev-Zel'dovich (SZ) Compton parameter, and galaxy velocity dispersion ($\sigma _v$) should correlate with the \textit{mass} (and among each other) via tight \textit{
scaling relations} \citep[e.g., ][]{Quintana1982, Kaiser1986a}. The possibility of defining and efficiently calibrating such correlations is extremely important when assessing the total mass of the systems, which is a physical quantity that is not directly observable, but has a key role relating clusters to the cosmological framework and, hence, enables their usage as cosmological probes. 
Several studies on nearby clusters have indeed shown the existence of strong correlations between the X-ray observables of galaxy clusters and the velocity dispersion of their galaxy members \citep[e.g. ][]{Mushotzky1984, Ortiz-Gil2004}, but these relations often exhibit slopes that deviate from the simple self-similar expectations \citep{Wu1999}. Hence, they point toward the influence of additional physical effects and a variety of dynamical states of galaxy clusters. In addition, these observed trends also show a considerable scatter because of effects like ongoing merging processes \citep[][]{Ricker2001}, the presence of cool-cores \citep[][]{Fabian1994, Pratt2009a}, or non-gravitational processes that heat the ICM, e.g. central AGN feedback \citep[][]{Cavaliere2002, Puchwein2008, McCarthy2010}. The study of the scaling relations between cluster observables and their \textit{evolution} with redshift can hence provide important information on the physical processes at work throughout the evolutionary path 
of such complex systems. In this context, it is therefore clear how important it is to push the above studies toward systems at higher redshifts and also to consider observables that are related to different physical components of the clusters.

In this paper, we study the relation between the bolometric X-ray luminosity, a quantity connected to the physical status of the hot ICM, and the 1D velocity dispersion along the line-of-sight ($\sigma _v$) of the member galaxies for a sample of galaxy clusters selected in the redshift range 0.64 $\leq$ z $\leq$ 1.46. We compare our results with the self-similar expectation ($L^{Bol}_X \propto \sigma _v ^4$) and the empirical relations observed for the local HIFLUGCS sample at \textlangle z\textrangle\,=\,0.05 \citep{Zhang2011}. We also investigate how the X-ray temperature of the ICM correlates with the galaxy velocity dispersion to test the assumption of isothermal and hydrostatic equilibrium between gas particles and galaxies. Finally, the $L_X - T_X$ relation is also investigated and compared with the local predictions found by \cite{Pratt2009a}.

The manuscript is structured as follows: in Sec.\,\ref{par:sampleSelec}, we present the distant sample of our study, which composed of 15 clusters extracted from the literature and ten newly discovered clusters drawn from XMM-\textit{Newton} Distant Cluster Project (XDCP) survey and the local HIFLUGCS sample. The X-ray analysis and the spectroscopic reduction of the XDCP sample are then described in Sec.\,\ref{par:dataAnalysis} with a discussion on the method we used in the kinematic analysis of the XDCP sample to estimate the $\sigma _v$ values. The results and the comparison with the HIFLUGCS sample are then presented in Sec.\,\ref{par:results}. Sections\,\ref{par:discussion} and \ref{par:summary} provide a concluding discussion and summary, respectively. Finally, we show the z- or H-band image of the 10 new XDCP systems in Appendix A, with their spectroscopic and kinematic details. In Appendix B, we compare the kinematic quantities of a sample of literature clusters estimated with our method with ones 
provided by the different authors.

Throughout the entire paper, we assume standard $\Lambda$CDM cosmology with $\Omega_m$ = 0.3, $\Omega_{\Lambda}$ = 0.7, and $H_0$ = 70 km s$^{-1}$ Mpc$^{-1}$.
 \begin{table*}[t] 
\centering
\caption{Properties of the clusters selected from the literature for our study. The second references in the \textit{Ref} column contain the values of the X-ray properties ($L^{Bol}_{X,500}$ and \textit{kT}) we adopt in this paper, because these are updated and/or consistent with our cosmology. The parameters $\sigma_v^{lit}$ and N$^{lit}_{gal}$ are the velocity dispersion and the number of members used to compute it, respectively, quoted from the literature. They generally differ from the $\sigma_v^{clip}$ and N$^{clip}_{gal}$ values we obtained with the 3$\sigma$ clipping procedure described in Sec.\,\ref{par:kinAnalysis}. For those clusters without a public redshift dataset the latter values are not computable (``n.a.'') and the ones quoted in the literature are used in our study. Systems defined by the authors as experiencing major merger events are indicated in the last column with a check mark and marked with red squares in Figs.\,\ref{fig:Lx_sigma_distOnly}, \ref{fig:Lx_sigma_HIFLUGCS_samedL}, \ref{fig:Lx_Tx}, and \ref{fig:sigma_Tx}.}
\label{Tab:Cluster_literature}     
\begin{tabular}{l l c c r@{\,$\pm$\,}l c r@{\,$\pm$\,}l r@{\,$\pm$\,}l c l l c}        
\hline\hline \\ 
\vspace{-0.6cm}\\
Cluster ID	& \hspace{-0.3cm} Ref		&   z   & N$^{lit}_{gal}$ &\multicolumn{2}{c}{$\sigma^{lit}_v$}& N$^{clip}_{gal}$ &\multicolumn{2}{c}{$\sigma^{clip}_v$} & \multicolumn{2}{c}{\hspace{-0.2cm}$L^{Bol} _{X,500}$}  & \hspace{-0.4cm}$kT$	 	&\hspace{-0.15cm}Selected via &\hspace{-0.5cm}Merging\\
		&		&	&  	    & \multicolumn{2}{c}{(km s$^{-1}$)}   & 	             & \multicolumn{2}{c}{(km s$^{-1}$)}    &\multicolumn{2}{c}{\hspace{-0.2cm}($10^{44}$ erg s$^{-1}$)} 	& \hspace{-0.4cm}(keV)	 		&  	       		     &\hspace{-0.4cm}system\\ [0.5ex]
\hline \vspace{-0.2cm}\\
MS1137		&\hspace{-0.3cm} Do99, Et04	& 0.785 & 23	    &  884 & 150			  & 23		     & 1022 & 111			    & 15.2  & 0.4				   &\hspace{-0.4cm}6.9\,$\pm$\,0.5  			&\hspace{-0.15cm}X-ray emission \\
RXJ1716		&\hspace{-0.3cm} Gi99, Et04	& 0.813 & 37	    &  1522& 180			  & 33		     & 1334 & 132			    & 13.9  & 1.0				   &\hspace{-0.4cm}6.8\,$\pm$\,1.0  			&\hspace{-0.15cm}X-ray emission &\hspace{+0.1cm}\checkmark\\
RXJ1821		&\hspace{-0.3cm} Gi04, Re11 	& 0.816 & 20	    & 	775& 150  			  & 19  	     & 854  & 126			    & 10.4  & 1.5				   &\hspace{-0.4cm}4.7\,$\pm$\,1.2  			&\hspace{-0.15cm}X-ray emission \\	 
MS1054		&\hspace{-0.3cm} Tr99, Et04	& 0.833 & 32	    &  1170& 160			  & 31		     & 1131 & 137			    & 28.4  & 3.0				   &\hspace{-0.55cm}10.2\,$\pm$\,1.0  			&\hspace{-0.15cm}X-ray emission &\hspace{+0.1cm}\checkmark\\
ACT0102		&\hspace{-0.3cm} Me12 		& 0.870 & 89        &  1321& 106  			  & n.a.	     & \multicolumn{2}{c}{n.a.}   	    & 136.0 & 6.8       			   &\hspace{-0.55cm}14.5\,$\pm$\,1.0   		  	&\hspace{-0.15cm}SZ effect 	  &\hspace{+0.1cm}\checkmark\\
Cl1604		&\hspace{-0.3cm} Lu04, Re11	& 0.897 & 22	    &  1226& 200			  & n.a.	     & \multicolumn{2}{c}{n.a.}		    & 2.0   & 0.4				   &\hspace{-0.4cm}2.5\,$\pm$\,1.1  			&\hspace{-0.15cm}Optical overdensity &\hspace{+0.1cm}\checkmark\\
X1229	        &\hspace{-0.3cm} Sa09 		& 0.975 & 27        &   683& 62   			  & 27  	     & 675  & 138			    & 8.8   & 1.5				   &\hspace{-0.4cm}6.4\,$\pm$\,0.7  			&\hspace{-0.15cm}X-ray emission \\
X1230	        &\hspace{-0.3cm} Fa11 		& 0.975 & 65        &   658& 277  			  & 63  	     & 807  & 109			    & 6.5   & 0.7				   &\hspace{-0.4cm}5.3\,$\pm$\,0.7  			&\hspace{-0.15cm}X-ray emission &\hspace{+0.1cm}\checkmark\\
SPT0546	        &\hspace{-0.3cm} Br10, Re11 	& 1.067 & 21        &  1181& 215  			  & 20  	     & 1041 & 167			    & 18.5  & 1.7				   &\hspace{-0.4cm}7.5\,$\pm$\,1.7  			&\hspace{-0.15cm}SZ effect	  \\
SPT2106	        &\hspace{-0.3cm} Fo11, Re11	& 1.132 & 18        &  1230& 225  			  & 17  	     & 868  & 186			    & 74.2  & 5.3				   &\hspace{-0.4cm}8.5\,$\pm$\,2.6  			&\hspace{-0.15cm}SZ effect	  &\hspace{+0.1cm}\checkmark\\
RDCS1252	&\hspace{-0.3cm} Ro04, Et04 	& 1.237 & 38        &   747& 79   			  & 38  	     & 752  & 81			    & 6.6   & 1.1				   &\hspace{-0.4cm}5.2\,$\pm$\,0.7  			&\hspace{-0.15cm}X-ray emission \\
SpARCS0035	&\hspace{-0.3cm} Wi09, Fa11 	& 1.335 & 10        &  1050& 230  			  & 9		     & 1105 & 125			    & 1.8   & 0.5				   &\hspace{-0.4cm}4.5\,$\pm$\,3.0  			&\hspace{-0.15cm}MIR overdensity&\hspace{+0.1cm}\checkmark\\
X2235		&\hspace{-0.3cm} Mu05 		& 1.396 & 30        &   802& 63   			  & n.a.	     & \multicolumn{2}{c}{n.a.}   	    & 10.0  & 0.8       			   &\hspace{-0.4cm}8.6\,$\pm$\,1.3   		  	&\hspace{-0.15cm}X-ray emission \\
ISCSJ1438	&\hspace{-0.3cm} Br11, An11	& 1.410 & 17        &   757& 223  			  & 15  	     & 782  & 170			    & 2.2   & 0.7				   &\hspace{-0.4cm}3.3\,$\pm$\,1.9  			&\hspace{-0.15cm}MIR overdensity\\
X2215a	        &\hspace{-0.3cm} Hi10, Re11	& 1.457 & 44        &   720& 110  			  & 31  	     & 750  & 100			    & 2.9   & 0.3				   &\hspace{-0.4cm}4.1\,$\pm$\,0.9  			&\hspace{-0.15cm}X-ray emission \\	[0.5ex] 
\hline                                   
\end{tabular}
\tablebib{Do99: \cite{Donahue1999}; Et04: \cite{Ettori2004a}; Gi99: \cite{Gioia1999}; Tr99: \cite{Tran1999}; Gi04: \cite{Gioia2004}; Re11: \cite{Reichert2011}; Me12: cluster \textit{``El Gordo''}, \cite{Menanteau2012}; Lu04: \cite{Lubin2004}; Sa09: \cite{Santos2009a}; Fa11: \cite{Fassbender2011a0}; Br10: \cite{Brodwin2010a}; Fo11: \cite{Foley2011}; Ro04: \cite{Rosati2004a}; Wi09: \cite{Wilson2009a}; Mu05: \cite{Mullis2005a}; Br11: \cite{Brodwin2011}; An11: \cite{Andreon2011}; Hi10: \cite{Hilton2010a}.}
\vspace{5mm}
\caption{Properties of the newly discovered or with newly published individual member redshifts, XDCP clusters selected for our study. The 5th column reports the imaging color(s) used for the photometric identification. The last column provides the identification codes of the used spectroscopic FORS2 programs.}
\centering
\label{Tab:Cluster_XDCP}     
\begin{tabular}{c c c c c c r@{\,$\pm$\,}l r@{\,$\pm$\,}l c}        
\hline\hline  
\vspace{-0.25cm}\\
Cluster ID					&\hspace{-0.2cm}RA	       &\hspace{-0.2cm}DEC	     &z       &Follow-up&\hspace{-0.2cm} N$^{clip}_{gal}$&\multicolumn{2}{c}{$\sigma^{clip} _v$} &\multicolumn{2}{c}{$L^{Bol}_{X,500}$} &\hspace{-0.2cm}Program ID     \\ 
						&\hspace{-0.2cm}J2000	       &\hspace{-0.2cm}J2000	     &	      &color&			  &\multicolumn{2}{r}{(km s$^{-1}$)}&\multicolumn{2}{c}{($10^{44}$ erg s$^{-1}$)}  &\\ [0.5ex]
\hline \vspace{-0.2cm}\\
\hspace{1ex}XDCP\,J1450.1+0904 - cl01$^{\ast}$  &\hspace{-0.2cm}14:50:09.2     &\hspace{-0.1cm}+09:04:39.1   &0.642   &R$-$z&\hspace{-0.2cm} 13 	  &414  &	   136  	 &1.05  	  &   0.20			    &\hspace{-0.2cm}085.A-0647(B)   \\
XDCP\,J1119.1+1300 - cl02			&\hspace{-0.2cm}11:19:07.7     &\hspace{-0.1cm}+13:00:23.8   &0.676   &z$-$H&\hspace{-0.2cm} 13 	  &499  &	    87  	 &0.35  	  &   0.09			    &\hspace{-0.2cm}079.A-0634(D) \\
XDCP\,J1044.7$-$0119 - cl03			&\hspace{-0.2cm}10:44:43.7     &\hspace{-0.1cm}$-$01:19:54.3 &0.755   &R$-$z&\hspace{-0.2cm} 17 	  &795  &	   188  	 &1.70  	  &   0.60			    &\hspace{-0.2cm}085.A-0647(B) \\
XDCP\,J0002.2$-$3556 - cl04			&\hspace{-0.2cm}00:02:16.1     &\hspace{-0.1cm}$-$35:56:33.8 &0.770   &z$-$H&\hspace{-0.2cm} 13 	  &1089 &      144		 &2.00  	  &   0.25			    &\hspace{-0.2cm}081.A-0332(B)   \\
XDCP\,J1243.2+1313 - cl05			&\hspace{-0.2cm}12:43:12.0     &\hspace{-0.1cm}+13:13:09.6   &0.791   &R$-$z&\hspace{-0.2cm} 25 	  &840  &	      139		 &1.70  	 &   0.50			   &\hspace{-0.2cm}084.A-0844(B)  \\
XDCP\,J0954.2+1738 - cl06			&\hspace{-0.2cm}09:54:17.1     &\hspace{-0.1cm}+17:38:05.9   &0.828   &R$-$z, z$-$H&\hspace{-0.2cm}  10   &992  &	175		 &6.70  	 &   0.75			   &\hspace{-0.2cm}084.A-0844(B)  \\
XDCP\,J0010.7$-$1127 - cl07			&\hspace{-0.2cm}00:10:42.4     &\hspace{-0.1cm}$-$11:27:46.0 &0.828   &R$-$z&\hspace{-0.2cm} 15 	  &416  &	   74		     &10.30	      &   1.60  			 &\hspace{-0.2cm}080.A-0659(A)  \\
\hspace{1ex}XDCP\,J0152.6$-$1338 - cl08$^{\ast}$&\hspace{-0.2cm}01:52:41.3     &\hspace{-0.1cm}$-$13:38:54.3 &0.829   &z$-$H&\hspace{-0.2cm} 12 	  &483  &	   98		     &2.50	      &   0.50  			 &\hspace{-0.2cm}084.A-0844(B)   \\
XDCP\,J2356.2$-$3441 - cl09			&\hspace{-0.2cm}23:56:16.5     &\hspace{-0.1cm}$-$34:41:41.8 &0.939   &z$-$H&\hspace{-0.2cm} 20 	  &624  &	   146  	     &6.50	      &   1.50  			 &\hspace{-0.2cm}081.A-0332(B)   \\
\hspace{1ex}XDCP\,J2215.9$-$1751 - cl10$^{\dag}$&\hspace{-0.2cm}22:15:56.9     &\hspace{-0.1cm}$-$17:51:40.9 &1.224   &z$-$H&\hspace{-0.2cm} 10 	  &493  &	   114  	     &0.55	      &   0.07  			 &\hspace{-0.2cm}080.A-0659(A)    \\[0.5ex]
\hline
\multicolumn{11}{c}{
  \begin{minipage}{\textwidth}%
    \vspace{0.1cm}
    \tiny$^{\ast}$Clusters cl01 and cl08 are published in the XMM Cluster Survey - Data Release 1 (XCS-DR1). Specifically, for cl01 a consistent photometric redshift of z $=$ 0.60 is provided in the XCS-DR1 web page table, accessible from \href{http://xcs-home.org/datareleases}{http://xcs-home.org/datareleases}. The X-ray and spectroscopic properties of cl08 are, instead, discussed in \cite{Mehrtens2012}. In both cases, however, the detailed lists of member redshifts were not provided before this work. $^{\dag}$The system XDCP\,J2215.9$-$1751 was previously published in \cite{Fassbender2011b} and \cite{deHoon2013}, but here we provide the value of $\sigma _v$ and perform its kinematic analysis.
  \end{minipage}}
\end{tabular}
\end{table*}
% \end{minipage}
\section{Sample selection}
\label{par:sampleSelec}
\subsection{The distant cluster sample}
\label{par:sample}
In our study, we consider a total of 25 galaxy clusters with both X-ray observations and optical spectroscopy for their member galaxies. The majority (19 out of 25) of the systems are X-ray selected with 14 of them drawn from the \XMM Distant Cluster Project sample \citep[][]{Fassbender2011b}. 
Ten of the 14 XDCP clusters are newly discovered and are labelled as ``XDCP sample'' throughout the paper. Their main properties and the redshift list of their members are provided in Table\,\ref{Tab:Cluster_XDCP} and Appendix A. The other 15 systems are already published and represent the ``literature sample'' as discussed in Sec.\,\ref{par:Liter_sample} and listed in Table\,\ref{Tab:Cluster_literature}.

All the clusters selected for our study have more than 10 spectroscopically confirmed members to guarantee a relatively reliable measurement of their $\sigma _v$ \citep{Biviano2006a}. In addition, we also include clusters with clear signs of ongoing merging in our sample because we do not expect that the presence of dynamically disturbed systems would produce a change in the estimated slope of the scaling relations but only a boost of their scatter. This is shown by \cite{Sifon2012} for a sample of SZ selected systems and also in \cite{Zhang2011} for the HIFLUGCS sample itself. The systems extracted from the literature and showing clear signs of an ongoing major merging event are marked by red squares in our plots.

\subsubsection{The literature sample}
\label{par:Liter_sample}
We drew a sample of 15 distant (z $>$ 0.6) clusters from the literature (updated to July 2012) without constraining the way in which the systems were selected. We only imposed that the X-ray properties of the systems were also well characterized and that their authors provided an estimate of $\sigma_v$, which is computed with at least ten members. In this way, we obtain a sample of X-ray, IR, optically, and SZ selected clusters whose main properties are summarized in Table\,\ref{Tab:Cluster_literature}.

For all the above systems (except ACT0102, Cl1604 and X2235), the complete set of the redshift values of their member galaxies with the associated errors are provided by the authors. This allowed us to re-compute their velocity dispersion ($\sigma^{clip}_v$) independently to be consistent with the same method adopted for the XDCP clusters (Sec.\,\ref{par:XDCPsample}) as described in Sec.\,\ref{par:kinAnalysis}. A comparison of our results and those of the literature are discussed in Sec.\,\ref{par:sigmaMeasure} and visually shown in Figs.\,\ref{fig:hist_Lit_1}, \ref{fig:hist_Lit_2}, and \ref{fig:hist_Lit_3} in Appendix B.

\subsubsection{The XDCP sample}
\label{par:XDCPsample}
In our study, we also considered a sample of ten newly discovered galaxy clusters drawn from the \textit{XMM}-Newton Distant Cluster Project \citep[][]{HxB2005, Fassbender2011b}, whose main properties are listed in Table\,\ref{Tab:Cluster_XDCP}. 
XDCP is a serendipitous X-ray survey designed to find and study distant (z $\geq$ 0.8) X-ray luminous galaxy clusters. The detection strategy involves 4 steps: (i) A cluster candidate is first detected as an extended X-ray source in archival XMM-\textit{Newton} observations. (ii) Obvious counterparts and nearby groups and clusters are identified by means of digital sky images and then discarded. (iii) Blank fields are then further studied by two-band photometric imaging, and (iv) promising high redshift candidates are finally subjected to spectroscopic redshift measurements, which also provide the final confirmation of a cluster.
This approach has been very successful, and within the XDCP project, the largest sample of X-ray selected distant clusters has been compiled to date with 22 confirmed systems at z $>$ 0.9 as previously published \citep{Fassbender2011b}. This z $>$ 0.9 sample has now increased to 23 when the newly confirmed system XDCP\,J2356.2$-$3441 (cl09) in Table\,\ref{Tab:Cluster_XDCP} at z $=$ 0.939 has been included. Their details are reported in Table\,\ref{Tab:Cluster_XDCP} and in Appendix A. The selection of the members for each cluster has been performed by applying an iterative clipping, as described in Sec.\,\ref{par:kinAnalysis}.
\begin{figure}[t]
 \includegraphics[width=7.5cm]{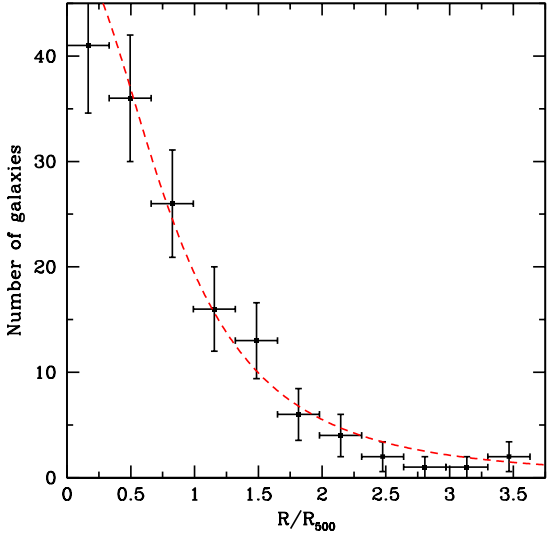}
\caption{Radial distribution of all the galaxy members of the ten considered XDCP clusters. The distances are rescaled to R$_{500}$, and the red dashed line is the best projected NFW fit. Vertical error bars represent Poisson errors. The shape of the distribution is indicative of a good radial sampling, albeit slightly thinner in the innermost and outermost regions.}
\label{fig:R200_hist}
\end{figure}
To verify that the radial distribution of the galaxies within the XDCP clusters does not show a particular bias toward the inner/outer regions, we show the \textit{stacked} radial profile for the entire sample in Fig.\,\ref{fig:R200_hist}. For a consistent comparison, the radial distances have been rescaled to R$_{500}$. The resulting distribution is then compared with a projected Navarro, Frenk, \& White \citep[NFW, ][]{Navarro1997a} profile \citep{Bartelmann1996}, which is shown as a red curve in the plot. The shape of the distribution is indicative of a good radial sampling, apart from a slightly more peaked distribution that is expected for the central bin and for R $\geq$ 2 R$_{500}$. We thus do not detect a strong bias that would indicate a severe under-sampling in the dense central regions. The mild lack of observed galaxies for R $<$ 0.5 R$_{500}$ is essentially due to the geometrical restrictions imposed by the FORS2 spectroscopic follow-up (Sec.\,\ref{par:specReduction}) that forces the placement 
of only a limited number of slits in the central, densest cluster regions (see Sec.\,\ref{par:biases}). Instead, the cluster environment at radii R $\geq$ 2 R$_{500}$ is more affected by the reduced success in the spectroscopic member confirmation because of the strong interloper contamination.

\subsection{The HIFLUGCS sample}
\label{par:HIFLUGCS}
The HIghest X-ray FLUx Galaxy Cluster Sample \citep[HIFLUGCS, ][]{Reiprich2002a} is a complete sample comprising of 64 galaxy clusters drawn from the \textit{ROSAT} All Sky Survey \citep[RASS, ][]{HxB2004a} with an X-ray flux of $f_{0.1 - 2.4keV} > 2\cdot10^{-11}$ erg s$^{-1}$cm$^{-2}$ and a galactic latitude of $|b| \geq$ 20.0 deg. The sample covers an area of two-thirds of the sky and includes objects up to z $\approx$ 0.2 with a median redshift of \textlangle z\textrangle\,=\,0.05. \XMM archive data are also available for 63 clusters, resulting in $\sim$1\,Ms clean observations. The X-ray observables of the HIFLUGCS sample have been accurately measured by combining \XMM and ROSAT data \cite{Zhang2009}.

The optical spectroscopic data used in \cite{Zhang2011} for estimating the galaxy velocity dispersions have been drawn from the literature \citep[e.g.,][]{Andernach2005} and have produced a total of 13,439 galaxies for 62 out of 64 clusters with a number of spectroscopically confirmed members that ranges from a minimum of 20 to a maximum of 972 (for the Coma cluster).
The X-ray bolometric luminosities of HIFLUGCS clusters span the range $\sim10^{42} - 10^{46}$ erg\ s$^{-1}$, which is one order of magnitude wider than the one of our distant sample, as described in the next section. In Fig.\,\ref{fig:LxBol_hist}, we compare the $L_X^{Bol}$ distribution of the two samples.

In the next sections, we present the analysis of X-ray (Sec.\,\ref{par:X_Reduction}) and optical spectroscopic data (Sec.\,\ref{par:specReduction}) for the XDCP sample.

\section{Data analysis}
\label{par:dataAnalysis}
\subsection{X-ray analysis}
\label{par:X_Reduction}
By definition of the XDCP strategy, all clusters listed in Table\,\ref{Tab:Cluster_XDCP} have been detected as extended sources in \XMM archive observations. The source extraction was carried out by means of \texttt{SAS} v6.5, applying a strict two-step flare cleaning process for the removal of high background periods. For most of the sources, the clean exposure time is $>$10 ksec for at least two of the \XMM detectors.

For the flux measurements, we applied the growth curve analysis \citep[GCA,][]{HxB2000a} method in the soft 0.5 - 2 keV band. This energy range is the ``classical'' one adopted in the literature to estimate X-ray fluxes of galaxy clusters as it minimizes the Galaxy contribution in the soft band and maximizes the sensitivity to clusters emission. \cite{Scharf2002a} demonstrated that galaxy clusters with temperatures greater than 2\,keV and redshift z\,$\leq$\,1 in such a band can be detected by Chandra and XMM-\textit{Newton} with the best signal-to-noise ratio. In the GCA method, the radial function of the cumulative source counts with background subtraction is determined and the total observed source count rate is measured from the plateau of this curve. In subsequent iterations, we use the X-ray luminosity to estimate the mass and the overdensity radius R$_{500}$ by means of the scaling relations given in \cite{Pratt2009a}. We then obtain the net source count rate from the growth curve inside an aperture 
of R$_{500}$. To determine the flux and 0.5 - 2 keV rest frame X-ray luminosity, we estimate the cluster ICM temperature using the relations in \cite{Pratt2009a} and determine the appropriate conversion factor with the \texttt{XSPEC} software. Specifically, we use a \texttt{mekal} plasma emission model with absorption by assuming a metallicity of 0.3 solar and an interstellar hydrogen column density taken from \cite{Kalberla2005}.

The bolometric luminosities $L^{Bol} _{X,500}$ have been finally obtained with \texttt{XSPEC} by extrapolating the energy distribution observed in the 0.5 - 2 keV band to 0.01 - 100 keV and by assuming an ICM metallicity of Z $=$ 0.3 Z$_{\odot}$ for all the clusters. Further details on the iterative procedure described above are provided in \cite{Suhada2012}. 

The values of $L^{Bol} _{X,500}$ with the associated errors are reported in the 7th column of Table\,\ref{Tab:Cluster_XDCP}.

\subsection{Spectroscopic reduction}
\label{par:specReduction}
All XDCP clusters have been followed-up using the VLT-FORS2 spectrograph \citep{Appenzeller1998} in the multi-object spectroscopy (MXU) configuration with an average of $\sim$\,50 1\arcsec\ width slits per mask. The observations were made by using the grism 300I+11, which provides a resolution of $R = 660$ and a wavelength coverage of $6000$\,\AA{} $\leq$ $\lambda$ $\leq 11000$\,\AA{}. The optical spectroscopic data of the newly released XDCP clusters, except for cl10, of Table\,\ref{Tab:Cluster_XDCP} have been reduced with a new dedicated pipeline: \textit{F-}VIPGI, an adapted version of VIPGI \citep{Scodeggio2005} for FORS2 data and described in detail in \cite{Nastasi2013}. The reduction processes included all the standard steps (bias subtraction, flat-fielding, and wavelength calibration), and the final redshift values were assessed by cross-correlating each 1D extracted spectrum with a library of different templates using the software package EZ \citep{Garilli2010}.

In Appendix\,A, we provide a table containing the redshifts of all observed member galaxies of the newly discovered XDCP clusters included in the kinematic analysis. Part of the spectroscopic details of XDCP\,J2215.9$-$1751 (cl10) is also available in \cite{deHoon2013}.

\subsection{Kinematic analysis}
\label{par:kinAnalysis}
As already mentioned in the previous sections for the XDCP sample, we considered only those clusters having more than ten spectroscopically confirmed members. For member selection and the following velocity dispersion computation, we adopted a two-step procedure. Namely, we applied a first member selection as described in \cite{Halliday2004}. That is, we cut a redshift window of $\pm$\,0.015 that is centered on the redshift of the brightest central galaxy (BCG) or on the median of the redshift peak of the galaxies found within 1\arcmin\, from the X-ray emission center and with a relative rest-frame velocity offset of $|\Delta v_{rest}| <$ 3000 km s$^{-1}$ if such a galaxy was not clearly identifiable. On this new galaxy subsample\footnote{We highlight that those galaxies that are discarded as members with this first cut cannot enter the analysis at later stages.} ($\vec{z}_{0.015}$), we then compute the velocity dispersion $\sigma_v$ by applying the method ``2$_{300}$'', that is defined and used by \cite{
Milvang2008A} on a sample of 21 EDisCS clusters at redshifts of 0.40 $\leq$ z $\leq$ 0.96 with 4 $\leq$ N$_{gal}$ $\leq$ 50. This method consists of a refinement of the member selection via an iterative 3$\sigma$ clipping on $\vec{z}_{0.015}$ and was found by those authors to be the only method always able to provide the most (visually judged) correct and robust results after various tests. The process starts with a first guess on z$_{cl}$ and $\sigma_v$ where the former is given by the biweight location estimator \citep{Beers1990a} of $\vec{z}_{0.015}$, and the latter is assumed $\sigma^{guess}_v =$ 300 km s$^{-1}$. Therefore, the procedure is iterated by selecting those galaxies with $-$3$\sigma^{guess}_v <$\ v$_{rest}$\ $< +$3$\sigma^{guess}_v$ and then recomputed using $\sigma_v ^{guess}$ for this new subsample. After that, the above clipping criterion is applied again by using the last estimate of $\sigma_v ^{guess}$ and allowing the previously clipped galaxies to re-enter in the analysis. The values of 
the velocity dispersion are computed via the biweight scale estimator if $N^{clip}_{gal} >$ 10 or the gapper scale estimator otherwise. In addition, we also applied the correction due to the uncertainties associated with the redshift measurements for each computed $\sigma^{guess}_v$, as prescribed by \cite{Danese1980a}. We found that convergence is usually reached within four steps and the results of our analysis are summarized in Table\,\ref{Tab:Cluster_XDCP}, as shown in the Appendix A. The quoted error on each $\sigma _v$ has been estimated as the rms of 1,000 bootstrapped realizations created from the final clipped redshift set, which is also subjected to the same iterative-clipping process described above.

We also remind the reader here that \cite{Beers1990a} extensively demonstrated the robustness and efficiency of the biweight estimators, which are able to provide robust results independently of the assumed nature of the population and the presence of outliers. This is provided that the studied sample had a size of N $>$ 10. This result has mainly driven our choice of restricting the analysis to those clusters with a final N$^{clip}_{gal} > $ 10.
\begin{figure}[t]
 \includegraphics[width=7.5cm]{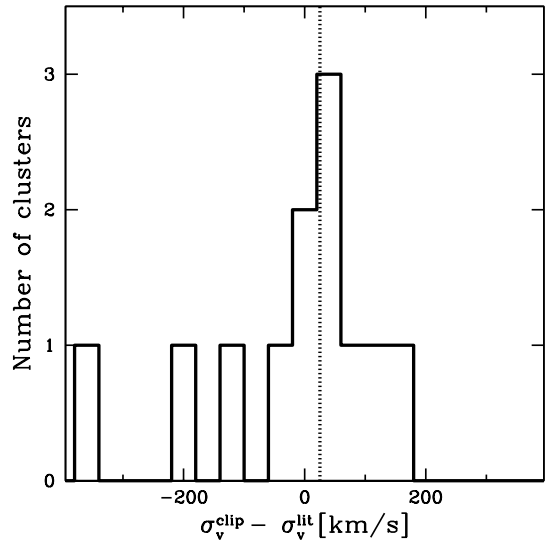}
\caption{The absolute differences between the $\sigma_v$ estimated with our two-step iterative clipping procedure, as described in Sec.\,\ref{par:kinAnalysis}, and the ones provided by the literature. The distribution has a median offset (marked by the dotted line) of $\Delta\sigma_v$= 25\,$\pm$\,59 km s$^{-1}$, which is consistent with zero, and a rms offset of \textlangle($\Delta \sigma_v$)$^2$\textrangle$^{1/2}$\,=\,141 km s$^{-1}$. The latter value corresponds to a relative difference of \textlangle($\Delta \sigma_v/\sigma_v$)$^2$\textrangle$^{1/2}$\,=\,0.15.}
\label{fig:diffSigma}
\end{figure}
\subsubsection{Velocity bias considerations}
\label{par:biases}
In XDCP, the spectroscopic targets are preferentially selected among the most luminous galaxies with colors consistent with the observed red-sequence. As explained in \cite{Fassbender2011b}, this strategy is able to maximize the number of real cluster members finally recovered from the spectroscopic follow-up. Because of the geometrical restrictions imposed on the slit positions of the FORS2 masks, the number of slits that can be used to sample the innermost regions of a cluster is very small (a maximum of 6 within the central 30\arcsec for typical 6\arcsec\ length slits), and hence, the densest regions of the systems tend to be slightly \textit{under-}sampled, as shown in Fig.\,\ref{fig:R200_hist}. This effect can introduce a bias in our galaxy sample with the most central members preferentially being the most luminous and red. A $\sigma _v$ estimate entirely based on such a sample could be significantly biased toward lower values because of the dynamical friction effect, which only acts effectively on the 
most massive galaxies \citep{Biviano2006a, Saro2013}. However, an inspection of our data shows that we have a mixture of bright ($H_{Vega} \lesssim 18$) and faint ($H_{Vega} \sim 20$) galaxies also in the center. As shown in \cite{Saro2013}, a sample with small (N $\leq$ 30) numbers of redshifts and ``randomly selected'' galaxies significantly reduces the bias on $\sigma _v$ compared to a sample with the same number of galaxies but ranked by luminosity.
\begin{figure*}[t]
\includegraphics[width=9.5cm]{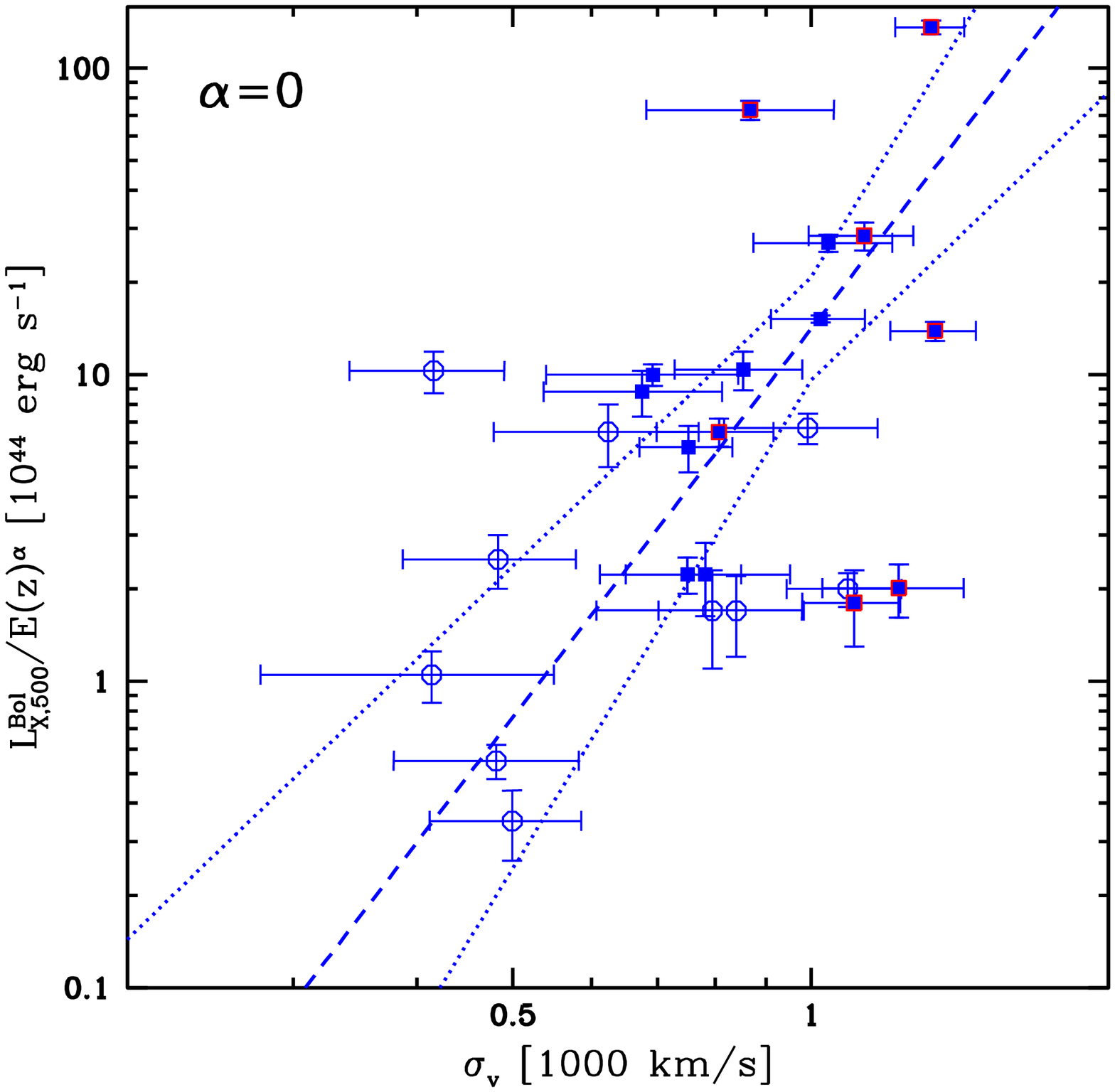}
 \includegraphics[width=9.5cm]{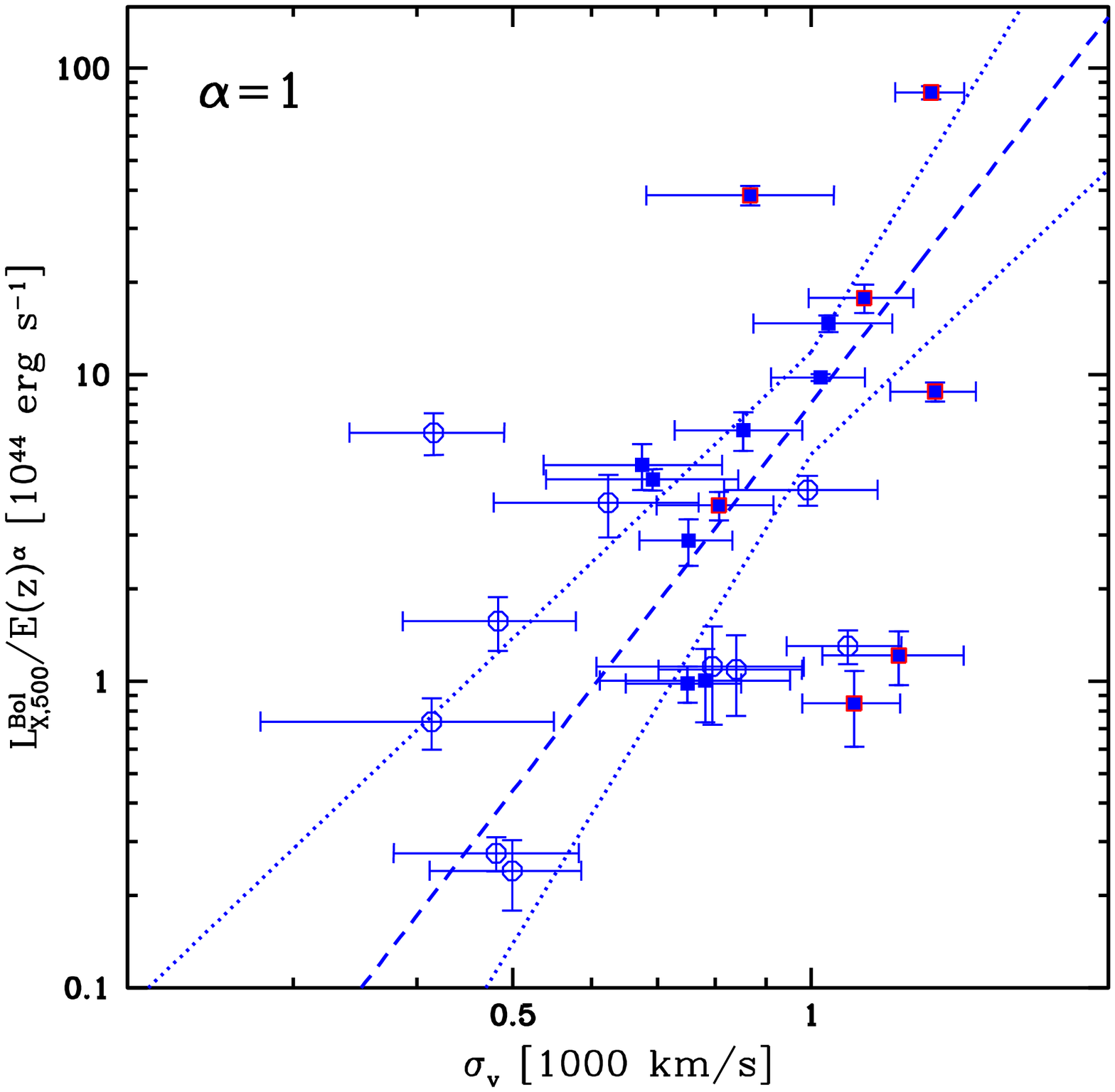}
\caption{$L_{X,500}^{Bol} - \sigma _v$ relation of our distant sample for no ($\alpha$\,=\,0, \textit{left}) and self-similar ($\alpha$\,=\,1, \textit{right}) evolution. The filled squares mark the ``literature sample'', whereas the empty circles the ``XDCP sample''. The best fit is marked by the dashed lines and its 1$\sigma$ uncertainties by the dotted ones. Red squares refer to the merging systems of Table\,\ref{Tab:Cluster_literature}.}
\label{fig:Lx_sigma_distOnly}
\end{figure*}
In addition, we also considered literature clusters clearly experiencing major mergers events in our study. These systems are indicated in Table\ref{Tab:Cluster_literature}, and their positions in Figs.\,\ref{fig:Lx_sigma_distOnly}, \ref{fig:Lx_sigma_HIFLUGCS_samedL}, \ref{fig:Lx_Tx}, and \ref{fig:sigma_Tx} are marked with red squares. As can be appreciated in the above plots, these clusters do not appear to introduce any systematic bias toward steeper or shallower slopes, but they generally follow the same distribution of the other distant, dynamically relaxed systems around the global fitted relations. This behaviour is similar to the one reported in \cite{Zhang2011} for the HIFLUGCS clusters, where the authors did not detect any statistical difference between the slopes found for the disturbed, non-disturbed, and cool core clusters. Therefore, we decided to include systems in a merging phase in our study as they are not expected to significantly affect the fitted slopes of the relations but, possibly, 
only increase their scatter.

\subsubsection{Test on accuracy of velocity dispersion measurements}
\label{par:sigmaMeasure}
\begin{figure*}[hb]
\includegraphics[width=17cm]{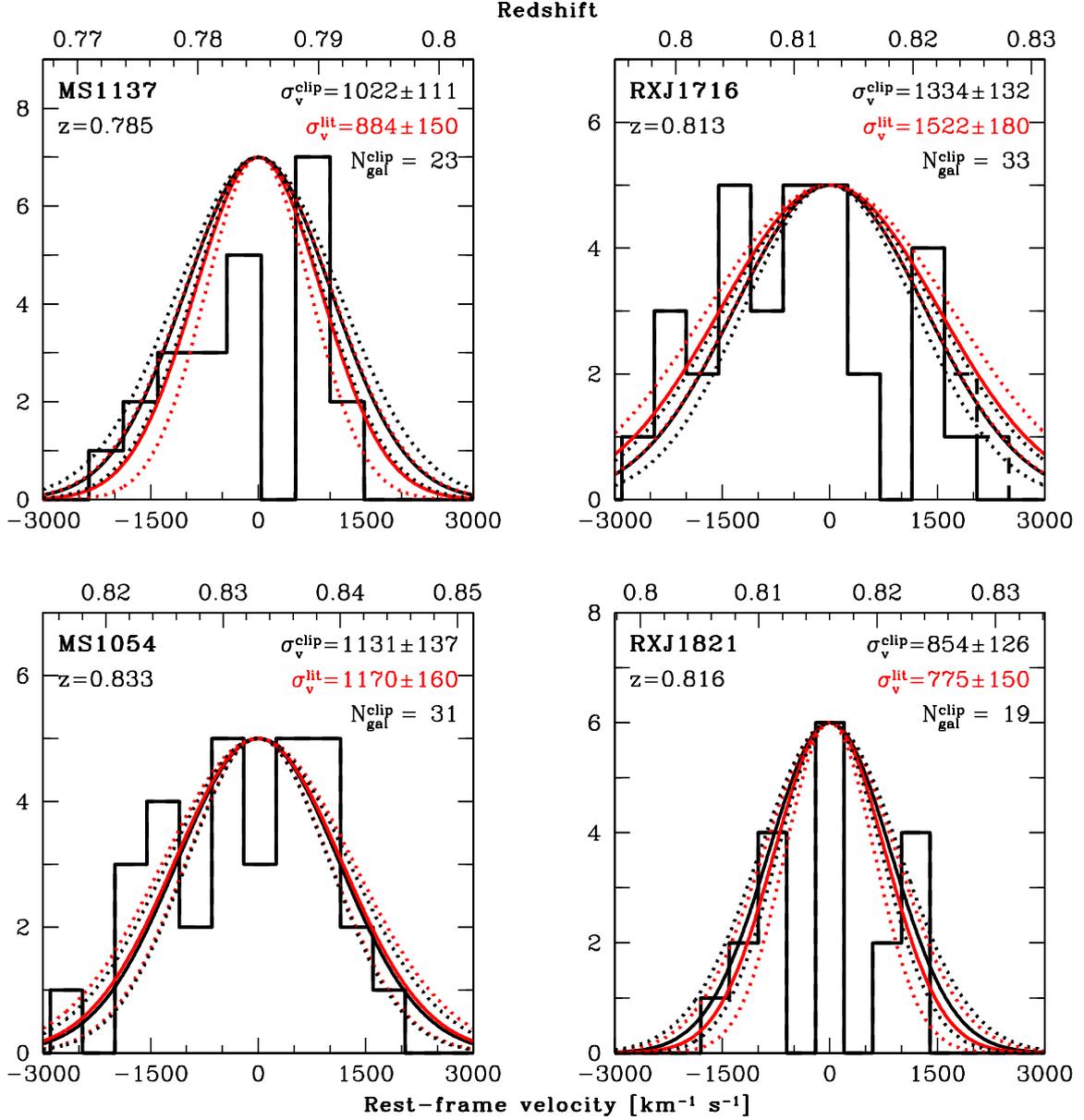}
\caption{Rest-frame velocity histograms of the ``literature sample'' clusters with a public redshift set. In addition to the same details presented in Appendix A, we also report here the public value of $\sigma^{lit}_v$ (and its uncertainty) for each cluster as listed in Table\,\ref{Tab:Cluster_literature}, and we show them with red Gaussian curves. Although our analysis sometimes produced slightly different values of z$_{cl}$ with respect to the ones stated by the authors, such differences are always limited to $c \cdot \Delta z_{cl}/(1+z_{cl})  \lesssim 100$ km s$^{-1}$ and hence the two Gaussian curves of each cluster show the same central values. Our estimates of $\sigma_v$ agree with $\sigma^{lit}_v$ within the errors albeit sometimes the distributions are far from being Gaussian, especially for those systems experiencing major merging events (see e.g. SpARCS0035).}
\label{fig:hist_Lit_1}
\end{figure*}

\begin{figure*}[hb]
 \includegraphics[width=18cm]{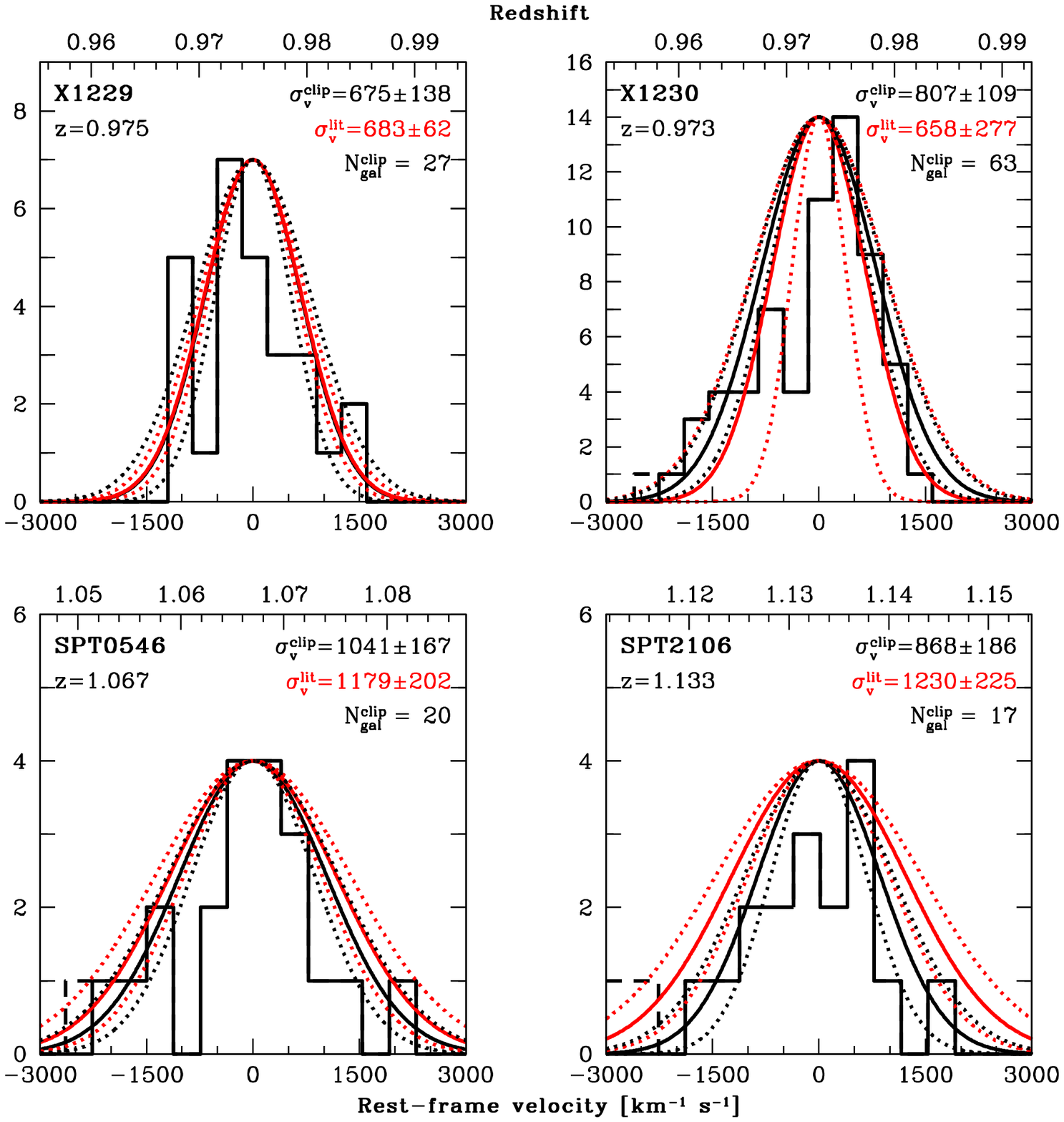}
\caption{Similar to Fig.\,\ref{fig:hist_Lit_1}.}
\label{fig:hist_Lit_2}
\end{figure*}

We applied the method described in Sec.\,\ref{par:kinAnalysis} to the redshift sets of the ``literature sample''. In Figs.\,\ref{fig:hist_Lit_1} and \ref{fig:hist_Lit_2}, we show the comparison of the $\sigma_v$ values computed with our method (black Gaussian) with the ones provided by the different authors (red Gaussian). Although we see some discrepancies in the numbers of the finally selected members, the different estimates agree within the errors, and as shown in Fig.\,\ref{fig:diffSigma}, the systematic median offset is consistent with zero with relative rms differences of $\pm$\,15\%.
\begin{figure*}[t]
\includegraphics[width=9.5cm]{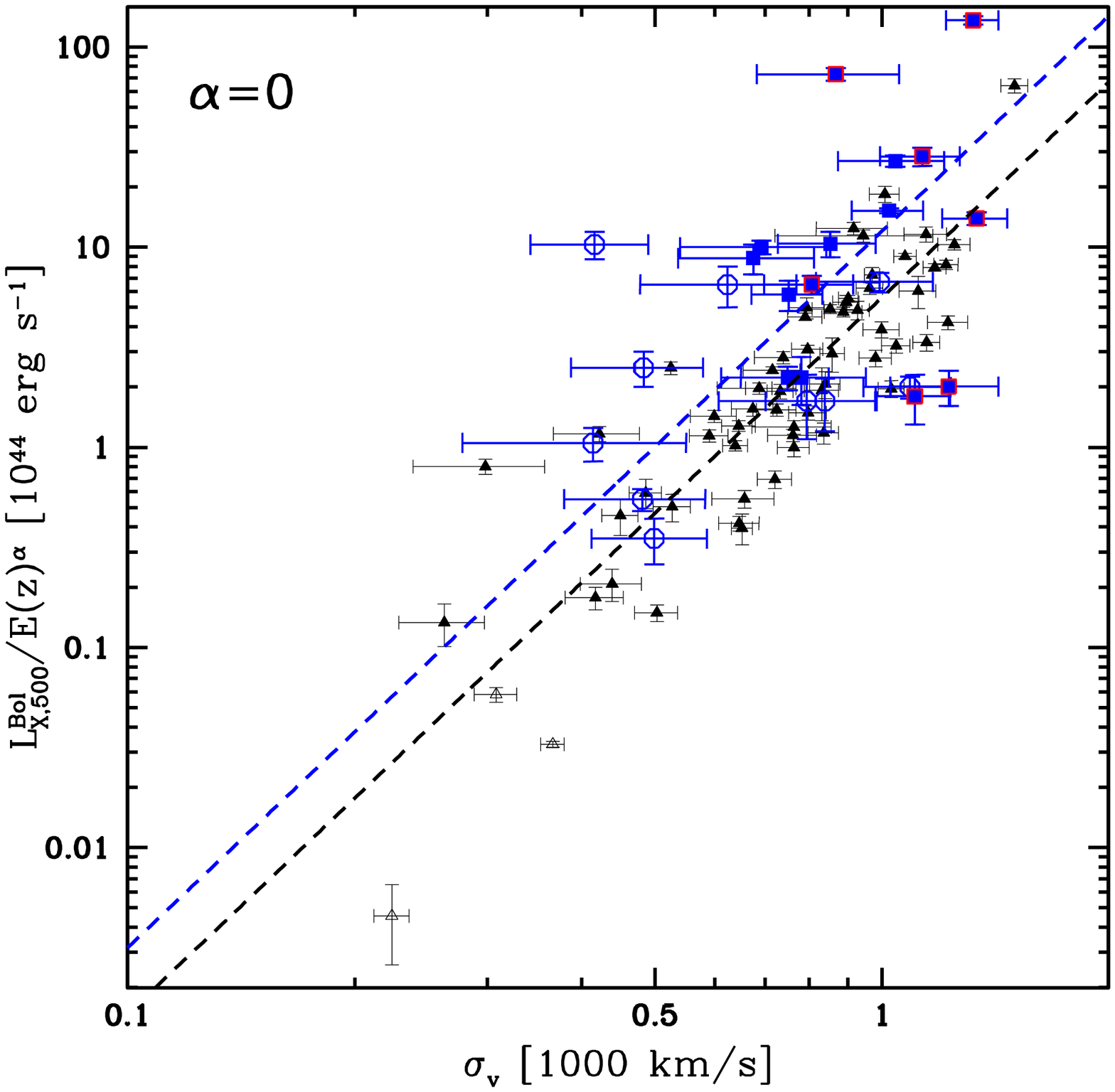}
\includegraphics[width=9.5cm]{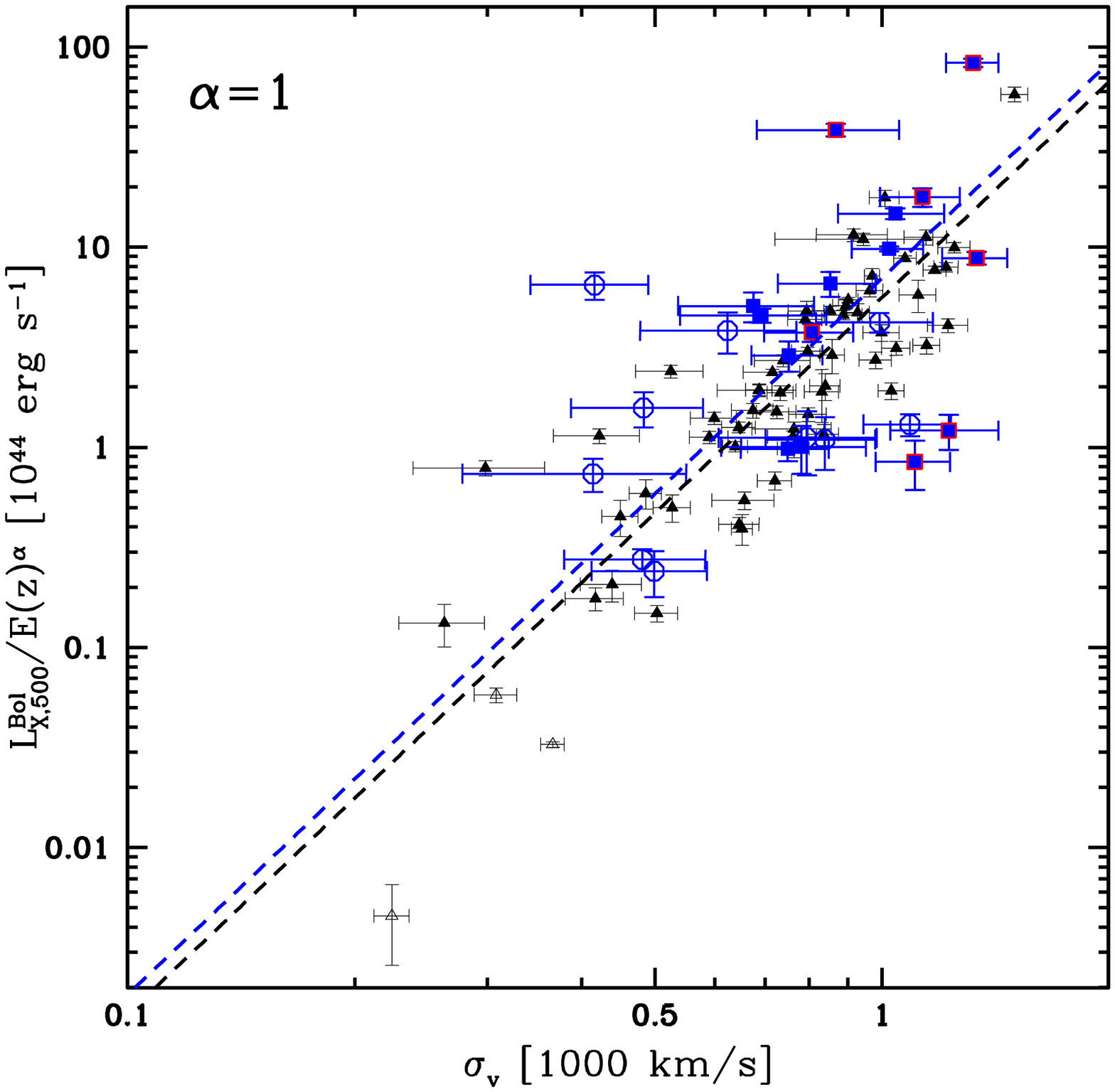}
\caption{$L_{X,500}^{Bol} - \sigma _v$ with the same convention of Fig.\,\ref{fig:Lx_sigma_distOnly}. The black triangles represent the HIFLUGCS sample with filled symbols referring to those objects in the common luminosity range $L_{X,500}^{Bol} > 10^{43}$ erg s$^{-1}$ that yeld a best fit relation with a slope of B $=$ 3.58 (black, dashed line). The best fit of the distant sample assuming the above slope is represented by the blue, dashed line, and the 1$\sigma$ error is not shown for clarity.} 
\label{fig:Lx_sigma_HIFLUGCS_samedL}
\end{figure*}
\section{Results}
\label{par:results}
In this section, we discuss the results of the fitting procedure of the $L_X - \sigma_v$, $L_X - T_X$, and $\sigma_v - T_x$ relations for our sample of 25 distant galaxy clusters, where $T_X$ is the X-ray determined temperature of the ICM. We always use the BCES regression fitting method as it correctly accounts for heteroscedastic errors (i.e. varying randomly and independently from point to point) on both variables \citep{Akritas1996}. Specifically, we adopt the BCES bisector method for all the scaling relations except for $L_X - T_X$, where we use the BCES orthogonal method to consistently compare our results with the ones found by \cite{Pratt2009a}. The X-ray bolometric luminosity and velocity dispersion values are normalised to $10^{44}$ erg s$^{-1}$ and 1000 km s$^{-1}$, respectively, whereas the relations involving $T_X$ assume this quantity normalised to 5 keV. Therefore, the fitted relations are in the form
\begin{equation}
 \label{eq:lum_sigma_fit}
\log\left(\frac{L_{X,500}^{Bol}}{E(z)^{\alpha}\cdot10^{44}\,\mbox{erg/s}}\right) = B\cdot \log\left(\frac{\sigma _v}{1000\,\mbox{km/s}}\right) + A
\end{equation}
\begin{equation}
\label{eq:lum_T_fit}
\log\left(\frac{L_{X,500}^{Bol}}{E(z)^{\alpha}\cdot10^{44}\,\mbox{erg/s}}\right) = B\cdot \log\left(\frac{T_X}{5\,\mbox{keV}}\right) + A
\end{equation}
\begin{equation}
\label{eq:sigma_T_fit}
\log\left(\frac{\sigma _v}{1000\,\mbox{km/s}}\right) = B\cdot \log\left(\frac{T_X}{5\,\mbox{keV}}\right) + A\mbox{,}
\end{equation}
where $\alpha$ parametrizes the evolutionary behaviour of the relation.
In Sec.\,\ref{par:fit_distOnly}, we show the best fit obtained for the 25 distant clusters. We provide two fits, one for a value of the exponent of the evolution parameter $\alpha$\,=\,0 (no evolution of the relation) and one for $\alpha$\,=\,1 (self-similar evolution). The choice of these values comes from the observed relation between $L_X$ and $T_X$. $T_X$ is analogous to $\sigma^2$, and we expect both parameters to behave similarly in relation to $L_X$. 

According to the studies of \cite{Reichert2011}, there is no significant evolution of the $L_X - T_X$ relation with redshift, but uncertainties are so large that a significant positive evolution cannot be ruled out. We therefore bracket the range of ignorance by the two different values of $\alpha$.

\begin{figure}[h]
\includegraphics[width=7.5cm]{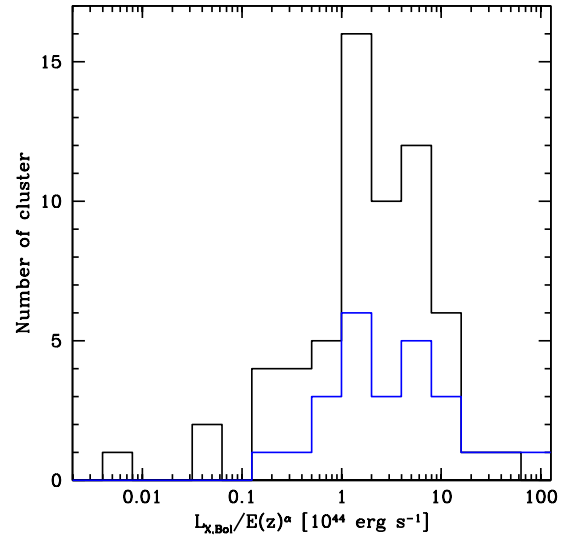}
\caption{Histogram of the $L^{Bol} _{X,500}$ for both HIFLUGCS (black) and our sample of 25 distant clusters (blue). The common luminosity range considered here is $L_{X,500}^{Bol} > 10^{43}$ erg s$^{-1}$.}
\label{fig:LxBol_hist}
\end{figure}

In Sec.\,\ref{par:fit_distOnly}, we compare our sample with that of HIFLUGCS at z $\sim$ 0.05 \citep{Zhang2011} by fitting the high-z scaling relation with the same slope found for those local clusters residing in the same $L_X$ range. As shown in Fig.\,\ref{fig:LxBol_hist}, the range of $L_{X,500}^{Bol}$ spanned by the distant and nearby sample is quite different with the latter having objects that are two orders of magnitude less luminous than the faintest cluster at z $>$ 0.6. Since comparing objects with too different luminosities may induce a bias in the studied relations, we took into account the HIFLUGCS slope computed in the common luminosity range $L_{X,500}^{Bol} > 10^{43}$ erg s$^{-1}$. We also considered cases with $\alpha$\,=\,0 and $\alpha$\,=\,1, although the difference is negligible for the local sample.

Finally, we study how the X-ray temperature of the ICM correlates with $L_X$ and $\sigma _v$, respectively, in the ``literature sample'' clusters in Sec.\,\ref{par:fit_L_T} and \ref{par:fit_sigma_T}. We limited the analysis to the literature systems because the X-ray data of the ``XDCP clusters'' do not allow for a reliable measurement of their $T_X$.

A summary of all the measurements and the fitting methods adopted for the different cases is provided in Table\,\ref{Tab:fit_Parameters}.
\subsection{The $L_X$ - $\sigma_v$ relation}
\subsubsection{Distant sample alone}
\label{par:fit_distOnly}
We applied the BCES bisector analysis to the $L^{Bol}_{X,500}$ and $\sigma _v$ measured for our distant sample of 25 clusters. As mentioned in Sec.\,\ref{par:results}, we consider two cases assuming no evolution ($\alpha$\,=\,0) and self-similar evolution ($\alpha$\,=\,1) for the scaling relations. The results are shown in Fig.\,\ref{fig:Lx_sigma_distOnly}.

As summarized in Table\,\ref{Tab:fit_Parameters}, we find a slope of B $\simeq$ 4.2 for both the self-similar and no evolution models, although with a large ($\sim$26\%) associated uncertainty, that is fully consistent with the findings of \cite{Zhang2011} and \cite{Ortiz-Gil2004} for clusters in the local universe.

\subsubsection{Distant sample with HIFLUGCS slope}
\label{par:fit_HIFLUGCS_common_Lx} 
To investigate the differences in normalisation between the $L_X - \sigma_v$ relation for distant and local clusters, we repeated the analysis done in Sec.\,\ref{par:fit_distOnly} but by freezing the slope to the one holding for the HIFLUGCS clusters. However, the HIFLUGCS sample reaches an X-ray luminosity limit that is two orders of magnitude less luminous than the faintest cluster in our distant sample, as shown in the histogram in Fig.\,\ref{fig:LxBol_hist}. To alleviate the possible bias introduced by considering objects so different in luminosity, we considered the slope computed for only those HIFLUGCS objects residing in the common luminosity range $L_{X,500}^{Bol} > 10^{43}$ erg s$^{-1}$. Although this selection, we excluded only three HIFLUGCS clusters. Our fit on this HIFLUGCS subsample produced a flatter slope (B $=$ 3.58) with respect to the entire sample but still agrees with the original result within the uncertainty limits. The results are shown in Fig.\,\ref{fig:Lx_sigma_HIFLUGCS_samedL}.
\begin{figure}[t]
 \includegraphics[width=8cm]{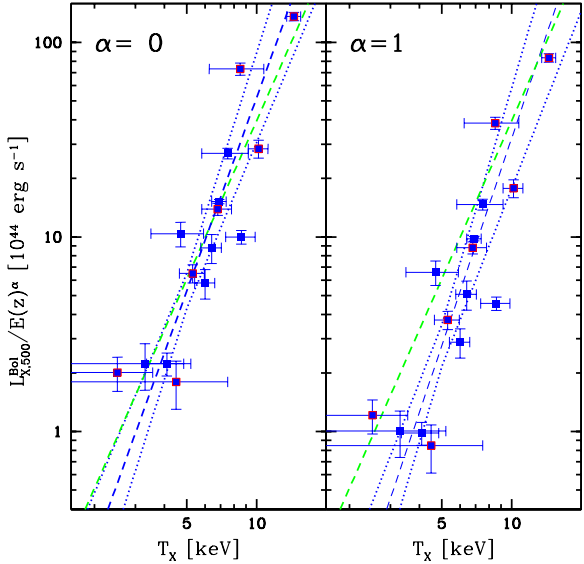}
\caption{The $L_X - T_X$ relation found for the distant sample of literature clusters. In green, the empirical relation of \cite{Pratt2009a} is shown. The symbols used are similar in meaning as that as of Fig.\,\ref{fig:Lx_sigma_distOnly}.} 
\label{fig:Lx_Tx}
\end{figure}

\begin{figure}[h]
 \includegraphics[width=8cm]{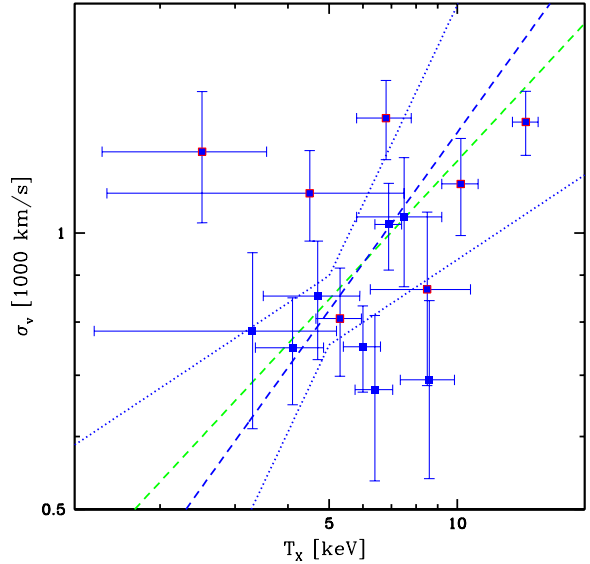}
\caption{The $\sigma _v - T_X$ relation found for 15 clusters of the literature sample. In green, the theoretical expectation of $\sigma_v \propto T_X^{0.5}$ is shown. The symbols used here are defined the same ways as in Fig.\,\ref{fig:Lx_sigma_distOnly}.}
\label{fig:sigma_Tx}
\end{figure} 

The normalisations in the two cases differ from the local sample by $\Delta A/A$\,=\,0.44\,$\pm$\,0.22 and $\Delta A/A$\,=\,0.13\,$\pm$\,0.22 (i.e., consistent with zero) for $\alpha$\,=\,0 and $\alpha$\,=\,1, respectively. 
\subsection{The $L_X - T_X$ relation}
\label{par:fit_L_T}
We also investigated the relation between $L_{X,500}^{Bol}$ and the X-ray temperature ($T_X$) of the ICM only for the sample of the literature clusters. For these objects, the values of $T_X$ were measured directly by the authors and are listed in Table\,\ref{Tab:Cluster_literature}. In this case, we used the BCES orthogonal method for our analysis. We then compared our results with the trends observed in the local Universe by considering the study of \cite{Pratt2009a} on a sample of 31 nearby (z\,$<$\,0.2) clusters of the REXCESS survey \citep{HXB2007a}. We highlight that \cite{Zhang2011} did not carry out such a study on the HIFLUGCS sample because of the inhomogeneous range of projected distances they used to measure the cluster temperature.

Our results are shown in Fig.\,\ref{fig:Lx_Tx} and summarized in Table\,\ref{Tab:fit_Parameters}. In the same plot, we also show in green the relation found by \cite{Pratt2009a} at z $\sim$ 0:
\begin{equation}
\label{eq:L_T_pratt}
h(z)^{-1} L_X = (6.07 \pm 0.58) (T_X/5\mbox{keV})^{2.70 \pm 0.24} 10^{44} \mbox{erg}\ \mbox{s}^{-1}\mbox{.}
\end{equation}

We find quite steep slopes but with large uncertainties ($\sim$\,14\%), mainly due to the small number of points. However, the best agreement between the distant and local relation is recovered for $\alpha$\,=\,0  as shown in Fig.\,\ref{fig:Lx_Tx}, which favours a scenario where no significant evolution with redshift is expected in the $L_X - T_X$ relation.

\subsection{The $\sigma_v - T_X$ relation}
\label{par:fit_sigma_T}
Finally, we investigated the $\sigma_v - T_X$ relation for those (12) literature clusters for which we could compute $\sigma_v$. These two quantities are important as both probe the depth of the cluster potential well estimated by using baryons as tracers. Since the gas particles of the ICM and the cluster galaxies feel the same potential under the assumption that they both have the same specific kinetic energy, it is expected that
 \begin{equation}
\hspace{3.3cm}
\label{eq:sigma_T_equilibrium}
\beta = \frac{\sigma_v^2 \mu m_p}{k_B T_X} \simeq 1\mbox{,}
\label{eq:beta}
\end{equation}
\begin{table*}[ht]
\caption{Summary of the fitting parameters for the $L_X - \sigma_v$, $L_X - T_X$, and $\sigma_v - T_X$ relations, assuming the form: log($Y$)\,$=$\,B$\cdot$log($X$)+A.}
\label{Tab:fit_Parameters}     
\centering
\begin{tabular}{c c c c c c c l l l}       
\hline
\vspace{-0.3cm}            \\
\multicolumn{10}{c}{$L_X - \sigma_v$} 																			\\
\vspace{-0.3cm}            \\
\hline\hline  
\vspace{-0.25cm}            \\
B	     & Err(B)	&     A      &   Err(A)      & 	Sample			& \# of clusters 	& Fitting Method	& $\alpha$ &   $L^{Bol} _{X,500}$ range & Figure	  						\\
\vspace{-0.25cm}            \\
\hline
\vspace{-0.25cm}            \\
4.210	     &  1.080   &  1.150     &    0.168      &  Distant         	&25			& BCES bisector		&  0   	   &   $> 10^{43}$ erg s$^{-1}$ & \ref{fig:Lx_sigma_distOnly} $-$ \textit{left}       \\
4.200	     &  1.100   &  0.907     &    0.167      &  Distant         	&25			& BCES bisector		&  1  	   &   $> 10^{43}$ erg s$^{-1}$ & \ref{fig:Lx_sigma_distOnly} $-$ \textit{right}       \\
4.010	     &  0.334   &  0.782     &    0.048	     &  HIFLUGCS	        &62			& BCES bisector		&  1  	   &   All		     	& \ref{fig:Lx_sigma_HIFLUGCS_samedL}       \\
3.580 	     &  0.357   &  0.749     &    0.048	     &  HIFLUGCS	        &59			& BCES bisector		&  1   	   &   $> 10^{43}$ erg s$^{-1}$ & \ref{fig:Lx_sigma_HIFLUGCS_samedL}       \\
3.580	     &  0.974   &  1.080     &    0.158      &  Distant - fixed slope   &25			& BCES bisector		&  0  	   &   $> 10^{43}$ erg s$^{-1}$ & \ref{fig:Lx_sigma_HIFLUGCS_samedL} $-$ \textit{left}     \\
3.580	     &  0.964   &  0.845     &    0.156      &  Distant - fixed slope   &25			& BCES bisector		&  1  	   &   $> 10^{43}$ erg s$^{-1}$ & \ref{fig:Lx_sigma_HIFLUGCS_samedL} $-$ \textit{right}     \\
\vspace{-0.25cm}            \\
\hline 
\vspace{-0.3cm}            \\
\multicolumn{10}{c}{$L_X - T_X$} \\
\vspace{-0.3cm}            \\
\hline\hline  
\vspace{-0.25cm}            \\
3.330	     &  0.466   &  0.723     &    0.090      &  Literature         	&15			& BCES orthogonal	&  0  	   &   $> 10^{43}$ erg s$^{-1}$ & \ref{fig:Lx_Tx} $-$ \textit{left}       \\
3.540	     &  0.502   &  0.441     &    0.107      &  Literature         	&15			& BCES orthogonal	&  1  	   &   $> 10^{43}$ erg s$^{-1}$ & \ref{fig:Lx_Tx} $-$ \textit{right}       \\
\vspace{-0.25cm}            \\
\hline
\vspace{-0.3cm}            \\
\multicolumn{10}{c}{$\sigma_v - T_X$} \\
\vspace{-0.3cm}            \\
\hline\hline  
\vspace{-0.25cm}            \\
0.643	     &  0.335   & -0.084     &   0.038	     &  Literature         	&15			& BCES bisector		&  $-$     &   $> 10^{43}$ erg s$^{-1}$ & \ref{fig:sigma_Tx}       \\
\vspace{-0.25cm}            \\
\hline   
\end{tabular}
\end{table*}
\noindent where $m_p$ is the proton mass, $k_B$ the Boltzmann constant, and $\mu$ is the mean molecular weight. If the above condition holds, one would expect that $\sigma_v \propto T_X^{0.5}$.
We tested this assumption by comparing the computed best fit of the data with the self-similar relation $\sigma_v \propto T_X^{0.5}$. The results are shown in Fig.\,\ref{fig:sigma_Tx} and reported in Table\,\ref{Tab:fit_Parameters}.

The slope we find (B\,=\,0.643\,$\pm$\,0.335) is slightly steeper than the self-similar expectations but still consistent with B\,=\,0.5. However, an almost identical result but with a much higher significance was reported by \cite{Xue2000} for a sample of 274 low-z clusters drawn from the literature. Despite the associated uncertainty, our findings may indeed indicate a systematic deviation from the self-similarity that deserves to be investigated further.
Additionally, we computed the ratios $\beta$ of the kinetic specific energy in galaxies and gas for the literature sample, by assuming $\mu$\,=\,0.59 in \ref{eq:beta}. We find this quantity spans the range 0.3\,$<$\,$\beta$\,$<$\,1.7 (with the only exception of the supercluster Cl1604 for which $\beta$\,=\,3.7) with a median value of $\beta$\,=\,0.85\,$\pm$\,0.28. These values agree with the ones typically reported in the literature \citep[see e.g., ][]{Wu1998}, and albeit with a large scatter with the theoretical expectation $\beta$\,=\,1.

All the the parameters of the relations discussed in Sec.\,\ref{par:results} are summarized in Table\,\ref{Tab:fit_Parameters}.

\section{Discussion}
\label{par:discussion}
In the presented work, we used a set of 25 galaxy clusters at redshift 0.64 $\leq$ z $\leq$ 1.46 to investigate how the X-ray properties ($L_X$, $T_X$) of the ICM correlate among each other and with the galaxy velocity dispersion ($\sigma_v$) in the distant Universe. To detect possible evolutionary effects on the above relations, we compared our results with the ones observed in a set of 64 clusters at \textlangle z\textrangle\,=\,0.05 (the HIFLUGCS sample), which are representative of the conditions in the local Universe. Our findings on the \textit{slopes} of the relations show a $L_X - \sigma_v$ trend consistent with the local observations (B $\sim$\,4) although with a large uncertainty. However, this slope is shallower than the expectations from the typical $L_X - T_X$ trend. The slope typically observed for such a relation is around 2.5\,-\,3.0 for clusters up to z\,$\sim$\,1.3\footnote{See \cite{HxB2012} for an updated compilation of literature values for a set of clusters at 0.1\,$\leq$\,z\,$\leq$\,1.
3.}. If the self-similar assumption $\sigma_v \propto T_X^{0.5}$ holds, the above $L_X \propto T_X^{2.5\mbox{\,-\,}3}$ would translate into $L_X \propto \sigma_v^{5\mbox{\,-\,}6}$, which is much steeper than our findings. However, if the equipartition constrain is alleviated and $\sigma_v \propto T_X^{0.64}$ is assumed, then a dependence of $L_X \propto \sigma_v^{4\mbox{\,-\,}4.7}$ would be justified. A slope of B\,$=$\,0.64 for the $\sigma_v - T_X$ relation is exactly what we find in the presented work, although the associated uncertainty makes the equipartition solution still acceptable. However, an identical result at a significance that is ten times higher was obtained by \cite{Xue2000} from their study on a sample of 274 low-z clusters drawn from the literature. The slope we find for the $\sigma_v - T_X$ relation may hence indeed reveal the presence of non-gravitational effects, which is responsible for the deviation of the ICM from the isothermal equilibrium with the underlying cluster potential \citep{
Xue2000, Rumbaugh2013}. Studies on real and simulated data typically ascribe to gas cooling and central sources of heating, like AGN and supernovae (SNe), the observed deviations of the ICM scaling relations from self-similar expectations \citep[e.g., ][]{Voit2005b, McCarthy2011}. These assumptions are also consistent with our results on the $L_X - T_X$ relation, where we found a consistent slope within 1$\sigma$ with the value commonly reported in the literature, but deviating more than 3$\sigma$ from the self-similar expectations. This discrepancy may be justified by assuming an additional source of energy, heating the ICM more effectively in low-mass clusters \citep{McCarthy2010, Stott2012}.

Concerning the study of \textit{normalisation}, we measure an offset of $\Delta A/A$\,$\sim$\,0 and $\sim$\,44\% ($\alpha$\,=\,1 and $\alpha$\,=\,0, respectively) between the distant and HIFLUGCS $L_X - \sigma_v$ relation. This finding was obtained with clusters in the same luminosity range $L_X\,>\,10^{43}$\,erg\,s$^{-1}$ and seems to favour a scenario where self-similar evolution of the scaling relations is indeed relevant. However, part of the offset found in the $\alpha$\,=\,0 case may be due to $bias$ effects introduced by the flux limited nature of the cluster selection. As shown by \cite{Reichert2011} at redshift z $\geq$ 0.7, the correction factor on the luminosities could amount to a maximum of $\sim$\,25\%. An opposite result is obtained for the $L_X - T_X$ relation. In this case, the best match with the z\,$\sim$\,0 trend is found for $\alpha$\,=\,0 and points toward an absence of evolution. The latter finding is consistent with the results reported in other recent works on distant galaxy clusters 
\citep{Reichert2011, Hilton2012, Rumbaugh2013}, where zero or negative evolution with redshift was found for the $L_X - T_X$ scaling relation. By comparing the observed trends with the ones predicted by different sets of simulations, the authors inferred that the majority of the energy injected into the ICM had occurred at high redshift and that models where the ICM is heated by AGN and SNe only at late times can be ruled out. Hence, the observed redshift evolution of the $L_X - T_X$ relation suggests a scenario where \textit{preheating} mechanisms increased the energy of the gas already at z\,$>$\,3 before its accretion onto the cluster. An important consequence of such a scenario is that the gas mass fraction for a given cluster mass is expected to \textit{decrease} toward higher redshift, resulting into galaxy clusters having the same $T_X$ with lower luminosities. This effect would produce a lower number of observable distant galaxy clusters with respect to the self-similar expectations and, therefore, 
may heavily affect the future X-ray and SZE surveys \citep{Reichert2011, HxB2012}.

In conclusion, we currently cannot make any definitive statement on which of the two possible scenarios (self-similar evolution/no-evolution) is the best supported by the observations, given the typical uncertainties of our results. 

\section{Summary}
\label{par:summary}
In this paper, we provided the kinematic and X-ray properties of a sample of ten newly discovered, X-ray selected galaxy clusters drawn from the XMM-\textit{Newton} Distant Cluster Project survey. These new systems reside in the redshift range 0.65 $\leq$ z $\leq$ 1.23 with X-ray luminosities 0.35 $\leq L_{X,500}^{Bol}/(10^{44}\mbox{erg}\ \mbox{s}^{-1}) \leq$ 10.3. They complement the XDCP sample reported by \cite{Fassbender2011b}, which is thus expanded to 31 spectroscopically confirmed clusters in the redshift range 0.6 $<$ z $<$ 1.6.

We analysed the correlations between the cluster X-ray properties and the galaxy velocity dispersion of the ten new XDCP systems with a sample of 15 distant clusters drawn from the literature. We also compared the $L_X - \sigma_v$ results with the trend typically observed in the local Universe, taking the findings of \cite{Zhang2011} for the HIFLUGCS sample at \textlangle z\textrangle\,=\,0.05 as reference.

In summary, we found that
\begin{itemize}
\item[$\bullet$] the \textit{slope} of the $L_X - \sigma_v$ relation appears consistent with the trend observed in the local Universe (B\,$\sim$\,4), when assuming either no or self-similar evolution ($\alpha$\,=\,0 and $\alpha$\,=\,1, respectively). Fixing the slope, we find an offset in the \textit{normalisation} between the distant and the HIFLUGCS sample clusters in the same luminosity range of $\Delta A/A$\,$\sim$\,0 and $\sim$\,44\% assuming $\alpha$\,=\,1 and $\alpha$\,=\,0, respectively. This finding seems to favour a scenario where self-similar evolution of the scaling relations is indeed relevant. However, we currently cannot make any definitive statement on which of the two possible scenarios (self-similar evolution/no-evolution) is the best supported by the observations, given the typical uncertainties of our results.
\item[$\bullet$] The $L_X - T_X$ relation appears consistent (within the uncertainties) with the ones typically reported in the literature ($L_X \propto T_X^{2.5\mbox{-}3}$) and are 3$\sigma$ from the expectations ($L_X \propto T_X^2$) of an ICM purely heated by gravitational processes. This would favour a scenario where additional sources of energy like AGN and SNe, which heat the ICM more effectively in low-mass systems, must be considered. 

A direct comparison of our data with nearby clusters suggests a better match for $\alpha$\,=\,0, hence pointing toward an absence of redshift evolution. This finding is consistent with the ones recently reported by many authors and suggests the presence of some \textit{preheating} mechanisms, which are able to increase the energy of the gas already at z\,$>$\,3, before its accretion onto the cluster.

\item[$\bullet$] The $\sigma_v - T_X$ relation appears slightly steeper than the self-similar expectations and closely resembles the findings of \cite{Xue2000}, which are obtained from a sample of 274 clusters at \textlangle z\textrangle\,=\,0.03. This result also may be an indication of a deviation from an isothermal equilibrium between the galaxies and the intracluster gas particles due to non-gravitational sources of heating.
\end{itemize}

The results reported here demonstrate that the galaxy velocity dispersion can be established as a useful mass proxy for distant clusters on a similar level as X-ray luminosity.
Much larger smaples are, however, required to obtain a reliable calibration of the studied relations. These data will be delivered by the increasing efforts in deeper X-ray, SZ, and optical/infrared surveys.

\begin{acknowledgements}
We thank the referee for the helpful and thoughtful comments that helped to improve the clarity of this paper.
The \XMM project is an ESA Science Mission with instruments and contributions directly funded by ESA Member States and the USA (NASA).
The XMM-Newton project is supported by the Bundesministerium f\"ur Wirtschaft und Technologie/Deutsches Zentrum f\"ur Luft- und Raumfahrt (BMWI/DLR, FKZ 50 OX 0001), the Max-Planck Society and the Heidenhain-Stiftung. 
This research has made use of the NASA/IPAC Extragalactic Database (NED) which is operated by the Jet Propulsion Laboratory, California Institute of Technology, under contract with the National Aeronautics and Space Administration. 
This work was supported by the Munich Excellence Cluster ``Structure and Evolution of the Universe'' (www.universe-cluster.de), by the DFG under grants Schw536/24-1, Schw536/24-2, BO 702/16, SPP 1177 and through the TR33. GWP acknowledges ANR grant ANR-11-BD56-015. We acknowledge the excellent support provided by Calar Alto and ESO-VLT staff in carrying out the service observations.
\end{acknowledgements}
%%%%%%%%%%%%%%%%%%%%%%%%%%%%%%%%%%%%%%%%%%%%%%%%%%%%%%%%%%%%%%%%%%%%%%%%%%%%%%%%%%%%%%%%%%%%%%%%%%%%%%%%%
\newpage
\onecolumn
\section*{Appendix A}
\label{par:append_A}
In this Appendix, we provide the details of the ``XDCP sample'' clusters listed in Table\,\ref{Tab:Cluster_XDCP}, seven of which are newly published systems. For each system, we show its z- or H-band image with the X-ray contours overlaid in blue, and the spectroscopic members marked by red regions. We also show the rest-frame velocity histograms of the cluster members and provide their detailed redshift list in table form.

%%%%%% cl01
\begin{figure*}[ht]
\centering
\includegraphics[height=9cm, clip=true]{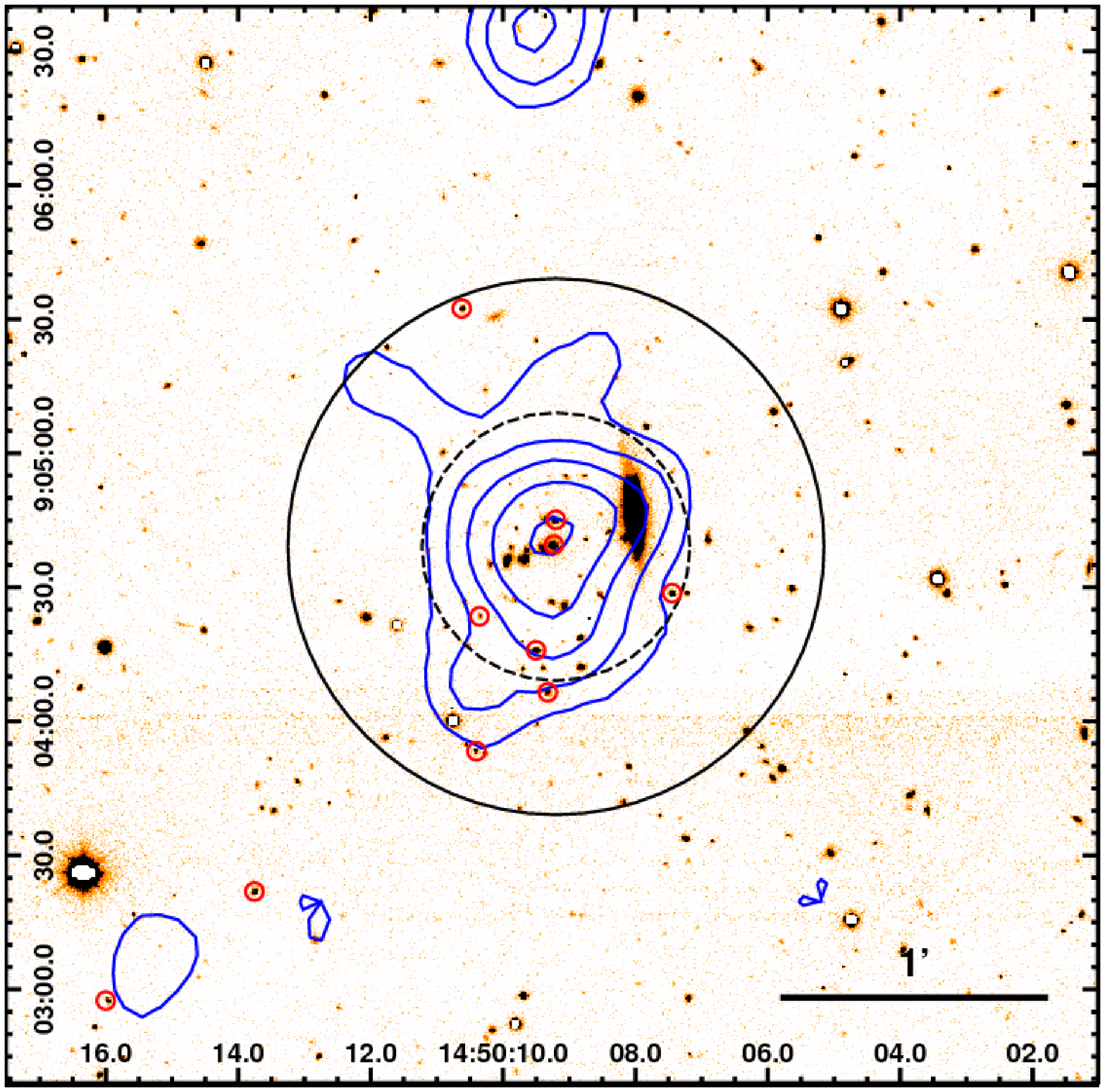}
\includegraphics[height=9cm, clip=true]{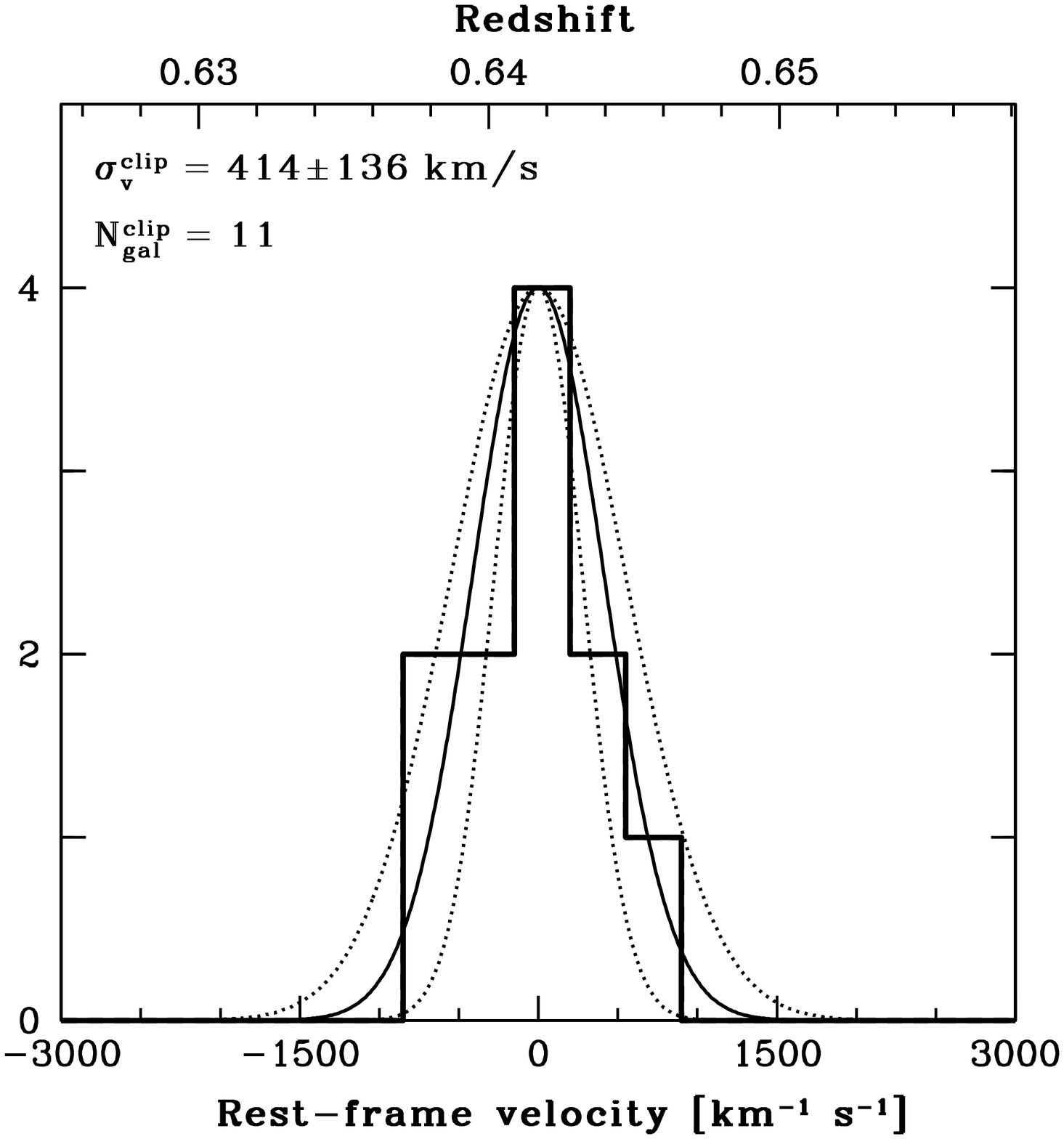}
\caption{\textit{Left:} 4\arcmin$\times$\,4\arcmin\ z-Band image of the cluster XDCP\,J1450.1+0904 - cl01 at z = 0.6417. X-ray contours are overlaid in blue, whereas red circles mark the member galaxies used in our kinematic analysis. The solid (dashed) black circle indicates a radius of 1\arcmin (0.5\arcmin) centered on the X-ray emission. \textit{Right:} Rest-frame velocity histogram of the cluster galaxies. The peculiar velocity (in km/s, bottom side) and the redshift values (top side) are reported in the panel with the estimated $\sigma_v$, its uncertainty, and the number of the clipped members $N^{clip}_{gal}$ contributing to the $\sigma_v$ estimate. The solid curve is the best Gaussian fit of the distribution with variance equal to $\sigma_v$, whereas the dotted ones represent its $\pm$\,1$\sigma$ uncertainty.}
\label{fig:117yray}
\end{figure*}
% \vspace{-1cm}
\begin{table*}[b]
\centering
\caption{Spectroscopic details of the galaxies of the cluster cl01. The ``QF'' column reports the quality flag associated to each spectroscopic redshift with QF $=$ 2, 3, and 4 corresponding to a confidence level for the estimated z value of $>$75\%, $>$90\%, and 100\%, respectively, whereas QF $=$ 1 is for tentative estimates. A check mark symbol in the last column indicates that the object has been rejected as cluster member by the $\sigma$-clipping procedure described in the text. All objects with QF $=$ 1 are excluded from the analysis.}
\begin{tabular}{c c c c c c c c c c}
\hline \hline  
\vspace{-0.3cm}            \\
R$_{500}$  & RA	 	&     DEC        &   z    & z$_{err}$ & QF 	&   \multicolumn{3}{c}{Distance from X-ray centroid} & Clipped	\\
(kpc)      & (J2000)	&     (J2000)    &        & 	      &    	&     (\arcsec)	& 	(kpc)	&	(r/R$_{500}$)&	out	\\
\vspace{-0.3cm}            \\
\hline
\vspace{-0.25cm}            \\
	\multicolumn{10}{c}{\textbf{XDCP\,J1450.1+0904 - cl01}}\\
	\hline \vspace{-0.3cm}            \\
	\textbf{644}& \textbf{14:50:09.2} & \textbf{+09:04:39.1}   & \textbf{0.6417} &	   &		&		&		 &			 \\
	\hline \vspace{-0.3cm}            \\
	     & 14:50:09.3 & +09:04:39.2   & 0.6419 & 0.0002   &  3 	&  	 1	&  	    6 &  0.009\\
	     & 14:50:09.2 & +09:04:45.1   & 0.6425 & 0.0002   &  3 	&  	 6	&  	    42 &  0.065\\
	     & 14:50:10.4 & +09:04:23.5   & 0.6418 & 0.0002   &  2 	&  	 23	&  	    160 &  0.248\\
	     & 14:50:09.5 & +09:04:15.9   & 0.6430 & 0.0002   &  3 	&  	 24	&  	    163 &  0.253\\
	     & 14:50:07.4 & +09:04:28.7   & 0.6429 & 0.0002   &  3 	&  	 28	&  	    192 &  0.298\\
	     & 14:50:09.3 & +09:04:06.5   & 0.6377 & 0.0002   &  3 	&  	 33	&  	    225 &  0.349\\
	     & 14:50:10.4 & +09:03:53.3   & 0.6462 & 0.0003   &  2 	&  	 49	&  	    339 &  0.526\\
	     & 14:50:10.6 & +09:05:32.4   & 0.6379 & 0.0002   &  3 	&  	 57	&  	    394 &  0.612\\
	     & 14:50:13.8 & +09:03:21.9   & 0.6407 & 0.0002   &  3 	&  	 103	&  	    707 &  1.098\\
	     & 14:50:16.0 & +09:02:57.5   & 0.6420 & 0.0002   &  2 	&  	 143	&  	    984 &  1.528\\
	     & 14:50:04.9 & +09:06:55.0   & 0.6405 & 0.0002   &  3 	&  	 150	&  	    1032 &  1.602\\
\vspace{-0.3cm}            \\
\hline\hline  
\end{tabular}
\end{table*}

%%%%%% cl02
\begin{figure*}[ht]
\centering
\includegraphics[height=9cm, clip=true]{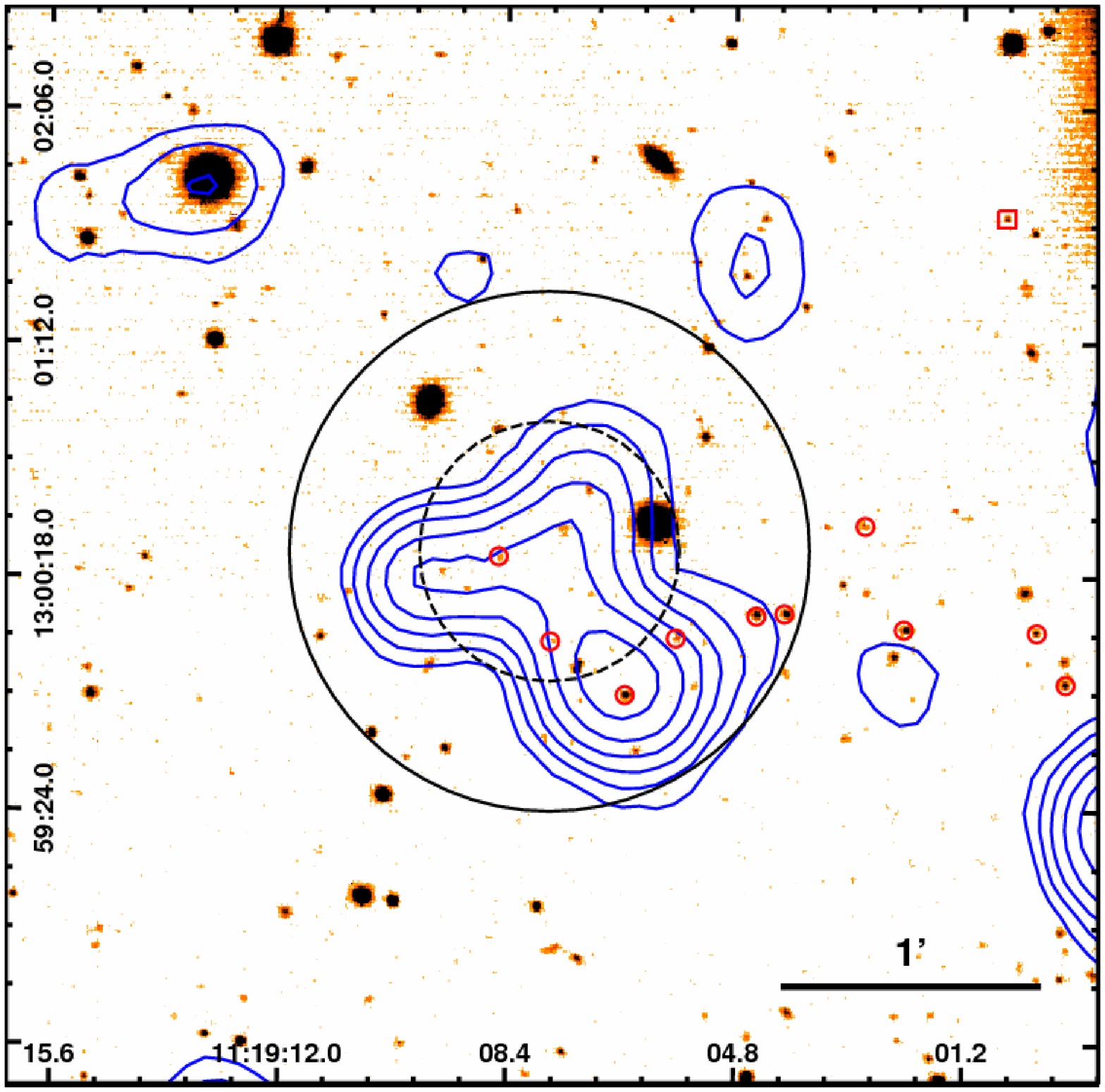}
\includegraphics[height=9cm, clip=true]{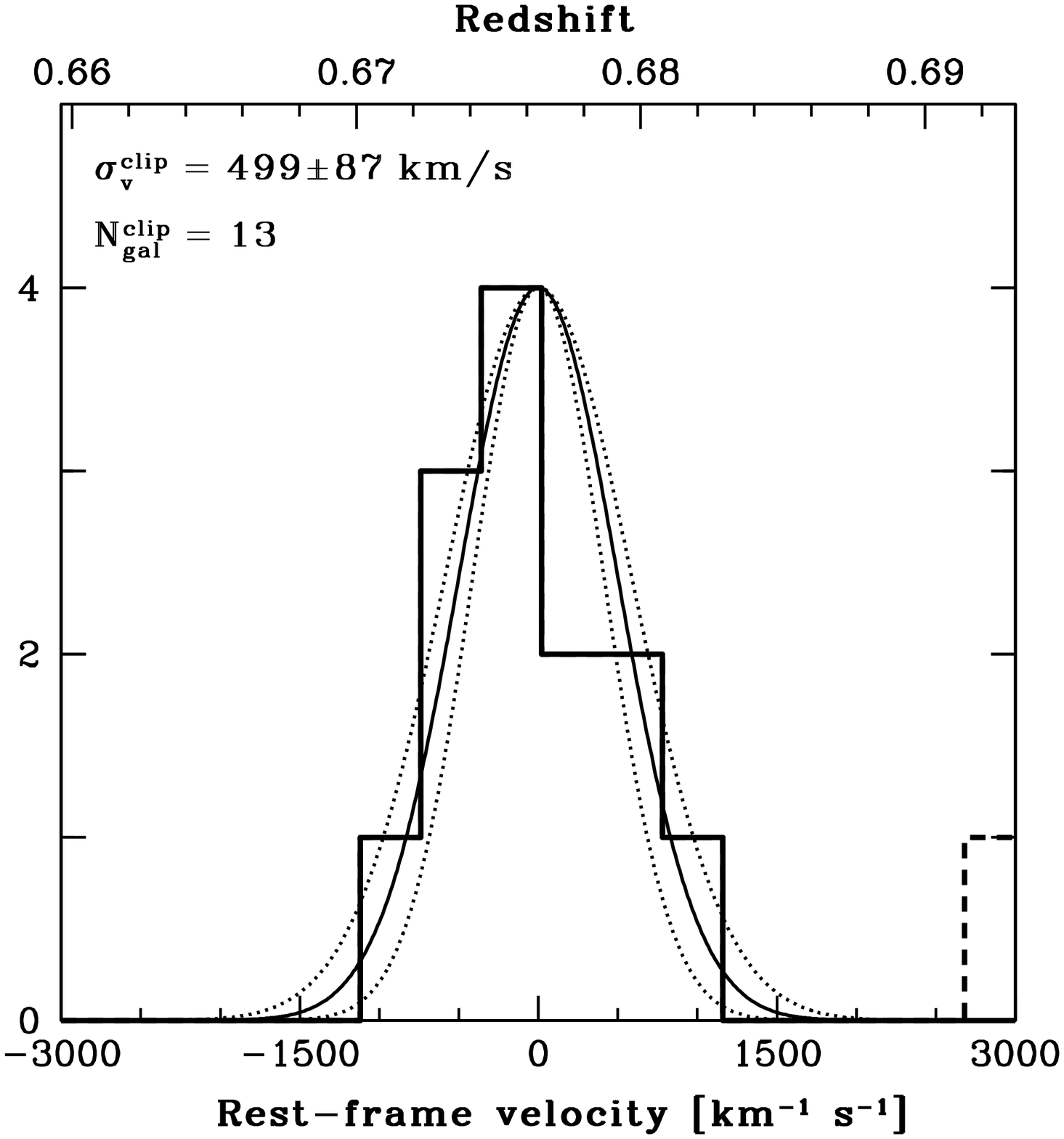}
\caption{\textit{Left:} A 4\arcmin$\times$4\arcmin\ wide z-Band image of cluster XDCP\,J1119.1+1300 - cl02 at z = 0.6764. Symbols and colors have the same meaning as in Fig.\,\ref{fig:117yray}. Red squares mark those galaxies excluded from the analysis because their measured redshift is uncertain or because they are rejected by the $\sigma$-clipping procedure described in the text. \textit{Right:} Rest-frame velocity histogram of the cluster galaxies. Here the dashed histograms refer to those galaxies with peculiar velocities $|v| < 3000$ km s$^{-1}$ from $z_{cl}$ but which are discarded as members by the iterative 3$\sigma$ clipping process as described in Sec.\,\ref{par:sigmaMeasure}.}
\label{fig:134com}
\end{figure*}
% \vspace{-1cm}
\begin{table*}[b]
\centering
\caption{Spectroscopic details of the galaxies of the cluster cl02.}
\begin{tabular}{c c c c c c c c c c}
\hline \hline  
\vspace{-0.3cm}            \\
R$_{500}$  & RA	 	&     DEC        &   z    & z$_{err}$ & QF 	&   \multicolumn{3}{c}{Distance from X-ray centroid} & Clipped	\\
(kpc)      & (J2000)	&     (J2000)    &        & 	      &    	&     (\arcsec)	& 	(kpc)	&	(r/R$_{500}$)&	out	\\
\vspace{-0.3cm}            \\
\hline
\vspace{-0.25cm}            \\
	\multicolumn{10}{c}{\textbf{XDCP\,J1119.1+1300 - cl02}}\\
	\hline \vspace{-0.3cm}            \\
	\textbf{488}	   & \textbf{11:19:07.7} & \textbf{+13:00:23.8}   & \textbf{0.6764} &	   &		&		&		 &			 \\
	\hline \vspace{-0.3cm}            \\
	     & 11:19:08.5 & +13:00:22.6   & 0.6725 & 0.0002   &  3 	&  	 11	&  	    80 &  0.164\\
	     & 11:19:07.7 & +13:00:03.0   & 0.6721 & 0.0002   &  2 	&  	 21	&  	    146 &  0.299\\
	     & 11:19:05.7 & +13:00:03.8   & 0.6764 & 0.0002   &  3 	&  	 36	&  	    251 &  0.514\\
	     & 11:19:06.5 & +12:59:50.7   & 0.6773 & 0.0002   &  3 	&  	 38	&  	    265 &  0.543\\
	     & 11:19:04.4 & +13:00:08.0   & 0.6762 & 0.0002   &  3 	&  	 50	&  	    351 &  0.719\\
	     & 11:19:04.0 & +13:00:09.3   & 0.6764 & 0.0002   &  3 	&  	 57	&  	    399 &  0.818\\
	     & 11:19:02.7 & +13:00:29.0   & 0.6732 & 0.0002   &  4 	&  	 73	&  	    513 &  1.051\\
	     & 11:19:02.1 & +13:00:05.9   & 0.6790 & 0.0002   &  3 	&  	 84	&  	    594 &  1.217\\
	     & 11:19:00.0 & +13:00:05.2   & 0.6742 & 0.0002   &  3 	&  	 114	&  	    800 &  1.639\\
	     & 11:18:59.5 & +12:59:53.3   & 0.6784 & 0.0002   &  3 	&  	 123	&  	    867 &  1.777\\
	     & 11:19:00.5 & +13:01:40.9   & 0.6921 & 0.0003   &  2 	&  	 131	&  	    922 &  1.889 & \checkmark \\ 
	     & 11:18:58.1 & +13:00:01.3   & 0.6810 & 0.0003   &  3 	&  	 142	&  	    998 &  2.045\\
	     & 11:18:55.5 & +12:59:28.7   & 0.6803 & 0.0006   &  2 	&  	 187	&  	    1315 &  2.695\\
	     & 11:18:52.7 & +13:00:51.5   & 0.6763 & 0.0002   &  3 	&  	 221	&  	    1557 &  3.191\\
\vspace{-0.3cm}            \\
\hline\hline  
\end{tabular}
\end{table*}

% %%%%%% cl03
\begin{figure*}[ht]
\centering
\includegraphics[height=9cm, clip=true]{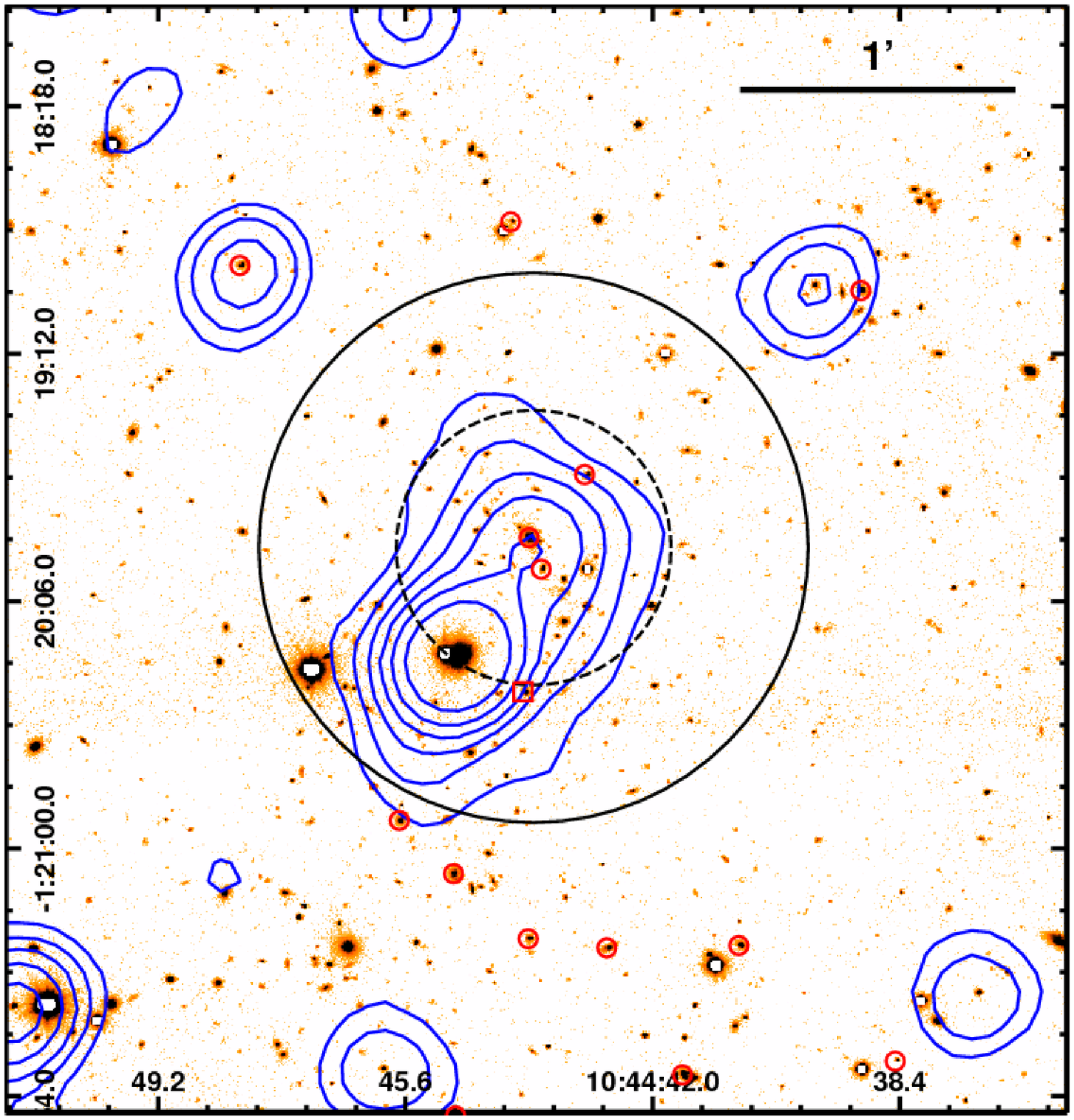}
\includegraphics[height=9cm, clip=true]{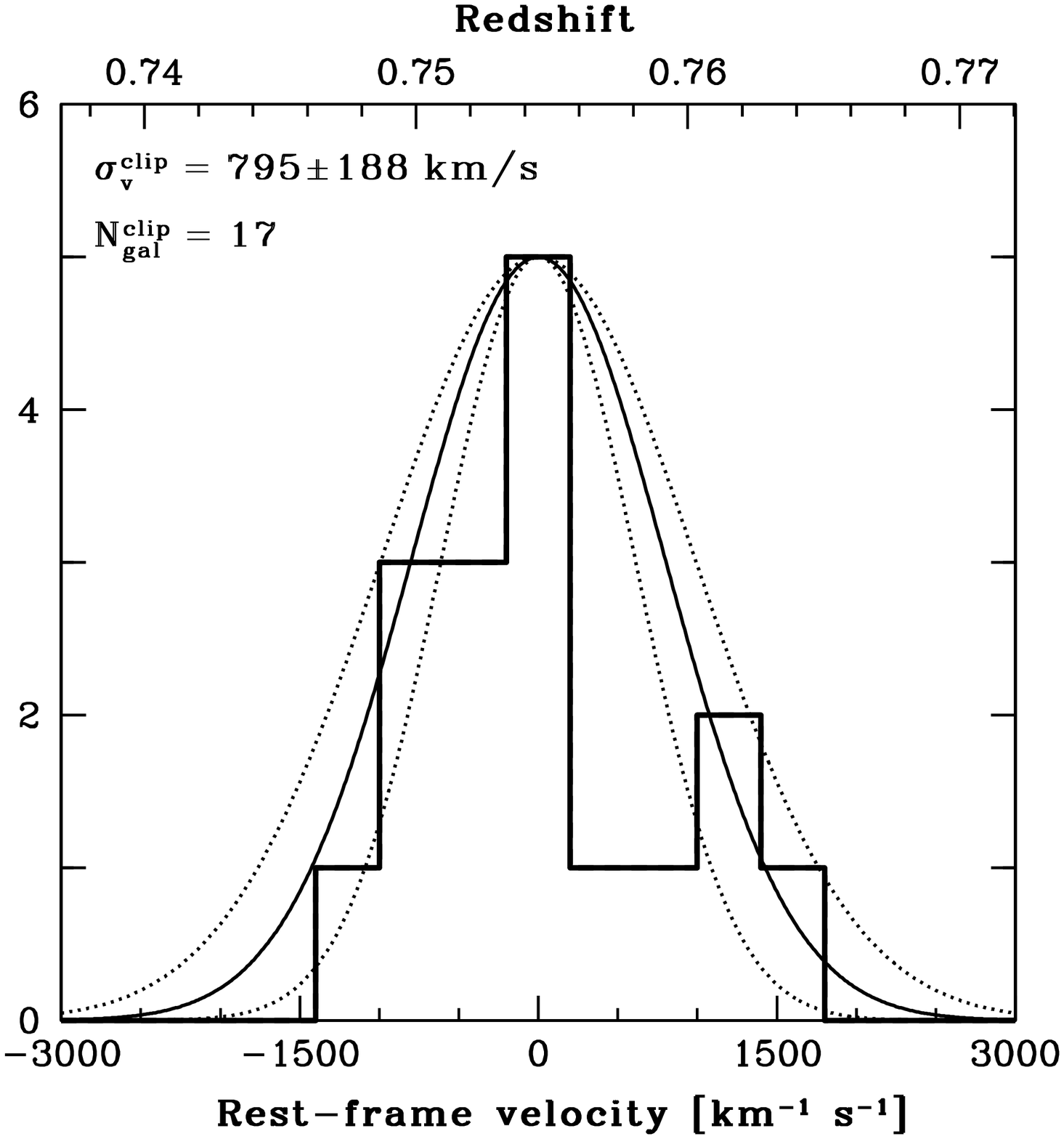}
\caption{\textit{Left:} A 4\arcmin$\times$4\arcmin\ wide z-Band image of cluster XDCP\,J1044.7$-$0119 - cl03 at z = 0.7545. Symbols and colors have the same meaning as in Fig.\,\ref{fig:134com}. \textit{Right:} Rest-frame velocity histogram of the cluster galaxies.}
\end{figure*}
% \vspace{-1cm}
\begin{table*}[b]
\centering
\caption{Spectroscopic details of the galaxies of the cluster cl03.}
\begin{tabular}{c c c c c c c c c c}
\hline \hline  
\vspace{-0.3cm}            \\
R$_{500}$  & RA	 	&     DEC        &   z    & z$_{err}$ & QF 	&   \multicolumn{3}{c}{Distance from X-ray centroid} & Clipped	\\
(kpc)      & (J2000)	&     (J2000)    &        & 	      &    	&     (\arcsec)	& 	(kpc)	&	(r/R$_{500}$)&	out	\\
\vspace{-0.3cm}            \\
\hline
\vspace{-0.25cm}            \\
	\multicolumn{10}{c}{\textbf{XDCP\,J1044.7$-$0119 - cl03}}    \\
	\hline \vspace{-0.3cm}            \\
	\textbf{669}	   & \textbf{10:44:43.7} & \textbf{$-$01:19:54.3}   & \textbf{0.7545} &	   &		&		&		 &			 \\
	\hline \vspace{-0.3cm}            \\
	     & 10:44:43.9 & $-$01:19:53.1   & 0.7548 & 0.0002	&  3	  &	   3	  &	      20 &  0.030\\
	     & 10:44:43.7 & $-$01:19:59.0   & 0.7508 & 0.0002	&  2	  &	   6	  &	      42 &  0.063\\
	     & 10:44:43.0 & $-$01:19:39.7   & 0.7490 & 0.0002	&  3	  &	   18	  &	      130 &  0.194\\
	     & 10:44:43.9 & $-$01:20:27.1   & 0.7686 & 0.0002	&  1	  &	   33	  &	      242 &  0.362 & \checkmark \\ 
	     & 10:44:45.7 & $-$01:20:55.2   & 0.7542 & 0.0002	&  3	  &	   68	  &	      502 &  0.750\\
	     & 10:44:44.1 & $-$01:18:44.2   & 0.7621 & 0.0002	&  3	  &	   70	  &	      517 &  0.773\\
	     & 10:44:45.0 & $-$01:21:06.8   & 0.7513 & 0.0002	&  3	  &	   75	  &	      551 &  0.824\\
	     & 10:44:43.8 & $-$01:21:20.0   & 0.7545 & 0.0003	&  3	  &	   87	  &	      637 &  0.952\\
	     & 10:44:48.0 & $-$01:18:53.7   & 0.7593 & 0.0002	&  3	  &	   89	  &	      654 &  0.978\\
	     & 10:44:39.0 & $-$01:18:59.2   & 0.7618 & 0.0002	&  3	  &	   89	  &	      656 &  0.981\\
	     & 10:44:42.7 & $-$01:21:23.0   & 0.7546 & 0.0003	&  3	  &	   90	  &	      662 &  0.990\\
	     & 10:44:40.8 & $-$01:21:22.5   & 0.7509 & 0.0002	&  3	  &	   99	  &	      725 &  1.084\\
	     & 10:44:41.6 & $-$01:21:50.4   & 0.7569 & 0.0002	&  3	  &	   120    &	      885 &  1.323\\
	     & 10:44:44.9 & $-$01:21:59.3   & 0.7511 & 0.0003	&  2	  &	   126    &	      928 &  1.387\\
	     & 10:44:38.5 & $-$01:21:47.4   & 0.7529 & 0.0002	&  3	  &	   138    &	      1012 &  1.513\\
	     & 10:44:50.8 & $-$01:17:48.1   & 0.7477 & 0.0002	&  3	  &	   165    &	      1215 &  1.816\\
	     & 10:44:46.3 & $-$01:17:11.0   & 0.7552 & 0.0003	&  2	  &	   168    &	      1235 &  1.846\\
	     & 10:44:45.9 & $-$01:22:44.5   & 0.7556 & 0.0006	&  1	  &	   173    &	      1275 &  1.906 & \checkmark \\ 
	     & 10:44:54.6 & $-$01:17:24.8   & 0.7636 & 0.0004	&  2	  &	   221    &	      1625 &  2.429\\
\vspace{-0.3cm}            \\
\hline\hline  
\end{tabular}
\end{table*}

%%%%%% cl04
\begin{figure*}[ht]
\centering
\includegraphics[height=9cm, clip=true]{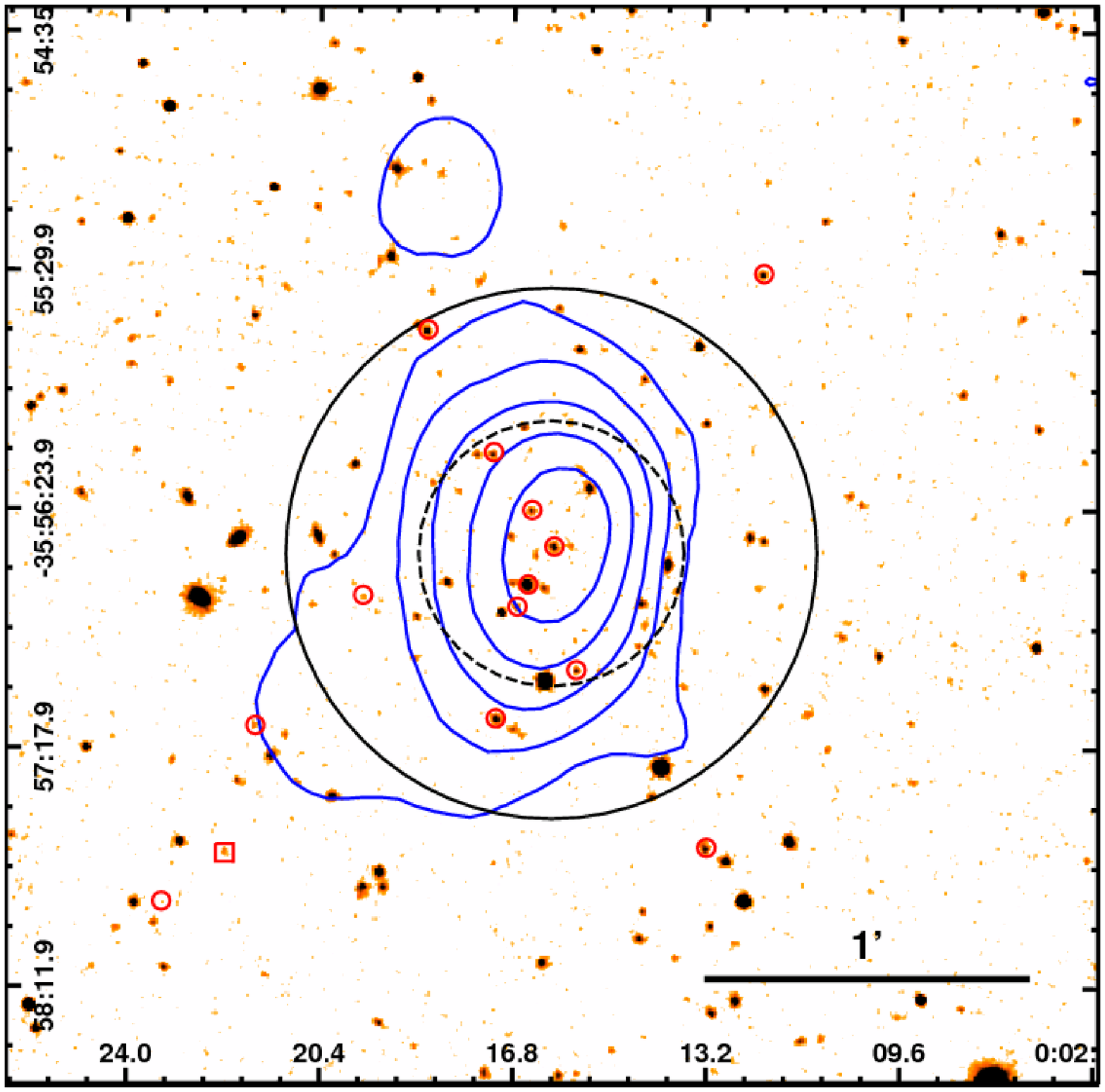}
\includegraphics[height=9cm, clip=true]{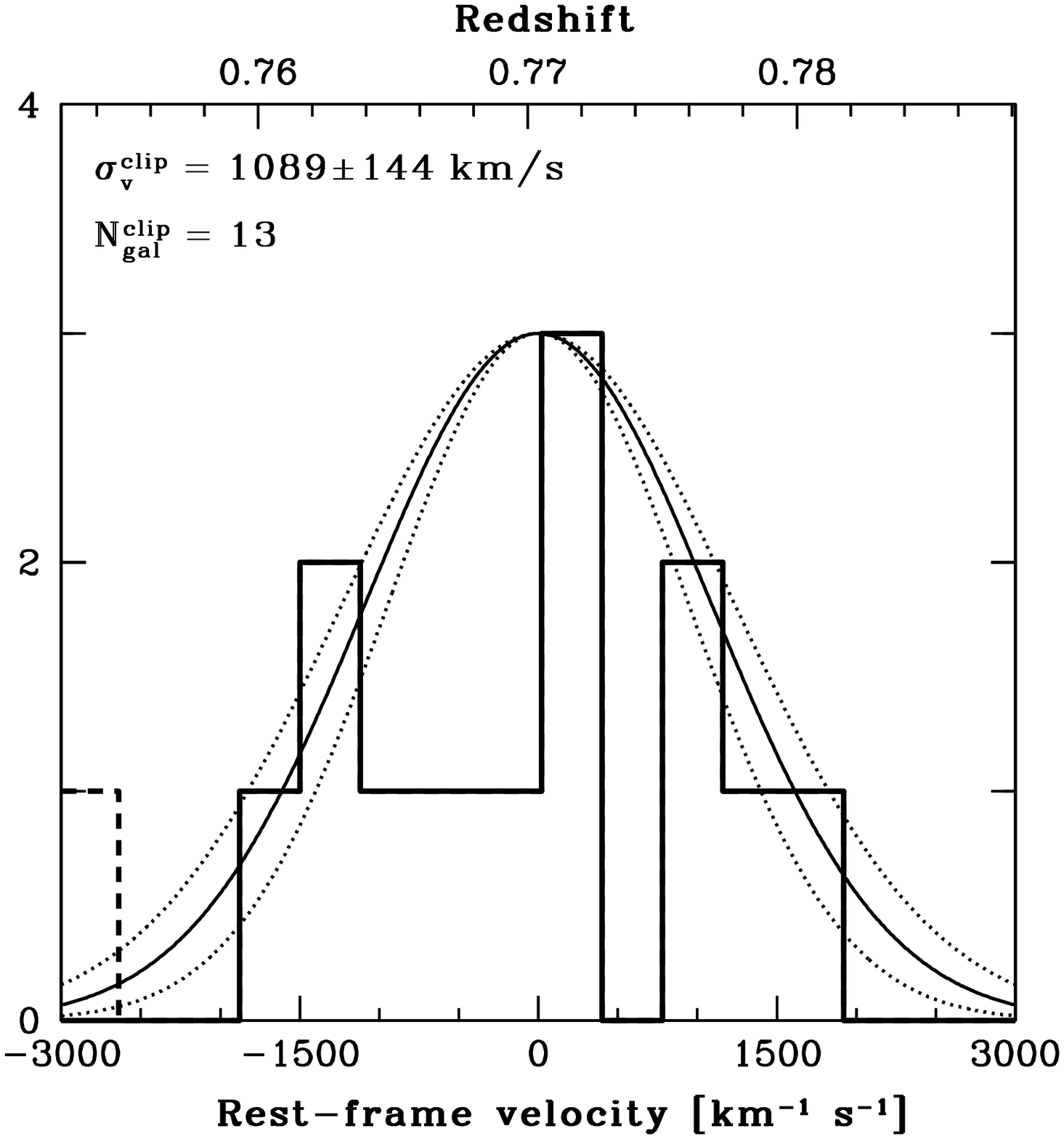}
\caption{\textit{Left:} A 4\arcmin$\times$4\arcmin\ wide H-Band image of cluster XDCP\,J0002.2$-$3556 - cl04 at z = 0.7704. Symbols and colors have the same meaning as in Fig.\,\ref{fig:134com}. \textit{Right:} Rest-frame velocity histogram of the cluster galaxies.}
\end{figure*}
% \vspace{-1cm}
\begin{table*}[b]
\centering
\caption{Spectroscopic details of the galaxies of the cluster cl04.}
\begin{tabular}{c c c c c c c c c c}
\hline \hline  
\vspace{-0.3cm}            \\
R$_{500}$  & RA	 	&     DEC        &   z    & z$_{err}$ & QF 	&   \multicolumn{3}{c}{Distance from X-ray centroid} & Clipped	\\
(kpc)      & (J2000)	&     (J2000)    &        & 	      &    	&     (\arcsec)	& 	(kpc)	&	(r/R$_{500}$)&	out	\\
\vspace{-0.3cm}            \\
\hline
\vspace{-0.25cm}            \\
	\multicolumn{10}{c}{\textbf{XDCP\,J0002.2$-$3556 - cl04}}    \\
	\hline \vspace{-0.3cm}            \\
	\textbf{674}	   & \textbf{00:02:16.1} & \textbf{-35:56:33.8}   & \textbf{0.7704} &	   &		&		&		 &			 \\
	\hline \vspace{-0.3cm}            \\
	     & 00:02:16.1 & $-$35:56:31.9   & 0.7757 & 0.0002	&  3	  &	   2	  &	      14 &  0.021\\
	     & 00:02:16.5 & $-$35:56:40.5   & 0.7629 & 0.0002	&  4	  &	   9	  &	      64 &  0.095\\
	     & 00:02:16.4 & $-$35:56:23.3   & 0.7716 & 0.0002	&  3	  &	   11	  &	      84 &  0.125\\
	     & 00:02:16.7 & $-$35:56:45.5   & 0.7786 & 0.0002	&  3	  &	   14	  &	      104 &  0.154\\
	     & 00:02:17.2 & $-$35:56:10.6   & 0.7767 & 0.0002	&  3	  &	   27	  &	      198 &  0.294\\
	     & 00:02:15.6 & $-$35:56:59.8   & 0.7812 & 0.0002	&  3	  &	   27	  &	      198 &  0.294\\
	     & 00:02:17.1 & $-$35:57:11.1   & 0.7631 & 0.0002	&  4	  &	   39	  &	      291 &  0.432\\
	     & 00:02:19.6 & $-$35:56:43.3   & 0.7662 & 0.0003	&  2	  &	   44	  &	      322 &  0.478\\
	     & 00:02:18.4 & $-$35:55:42.9   & 0.7686 & 0.0002	&  3	  &	   58	  &	      432 &  0.641\\
	     & 00:02:13.2 & $-$35:57:40.2   & 0.7660 & 0.0002	&  3	  &	   75	  &	      558 &  0.828\\
	     & 00:02:21.6 & $-$35:57:12.8   & 0.7610 & 0.0002	&  2	  &	   77	  &	      573 &  0.850\\
	     & 00:02:12.1 & $-$35:55:29.8   & 0.7726 & 0.0002	&  3	  &	   80	  &	      593 &  0.880\\
	     & 00:02:22.2 & $-$35:57:41.6   & 0.7535 & 0.0002   &  3 	&  	 101	&  	    745 &  1.105 & \checkmark \\ 
	     & 00:02:23.4 & $-$35:57:52.9   & 0.7710 & 0.0002   &  4 	&  	 118	&  	    877 &  1.301\\
\vspace{-0.3cm}            \\
\hline\hline  
\end{tabular}
\end{table*}

%%%%%% cl05
\begin{figure*}[ht]
\centering
\includegraphics[height=9cm, clip=true]{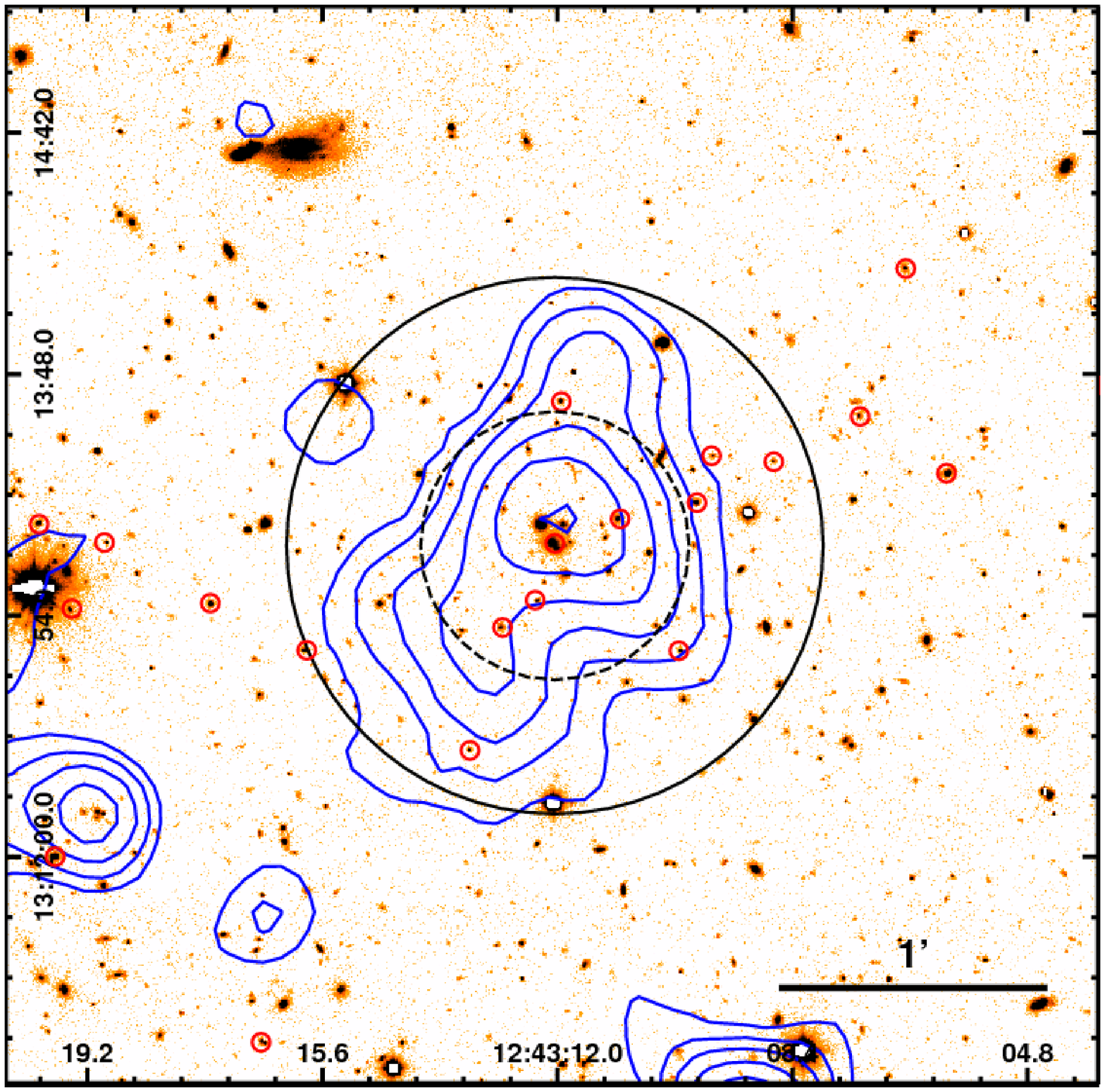}
\includegraphics[height=9cm, clip=true]{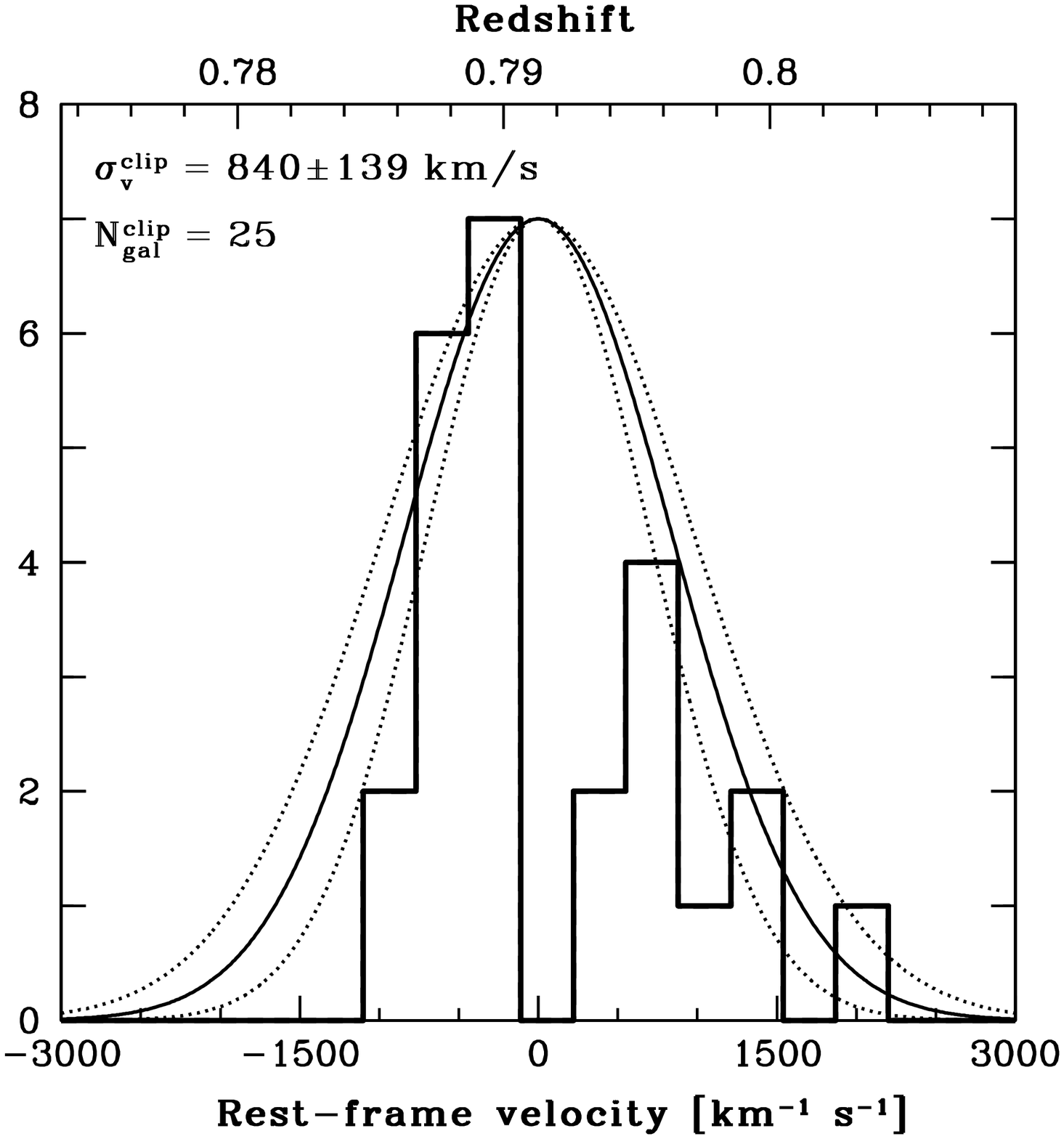}
\caption{\textit{Left:} A 4\arcmin$\times$4\arcmin\ wide z-Band image of cluster XDCP\,J1243.2+1313 - cl05 at z = 0.7913. Symbols and colors have the same meaning as in Fig.\,\ref{fig:134com}. \textit{Right:} Rest-frame velocity histogram of the cluster galaxies.}
\end{figure*}
% \vspace{-1cm}
\begin{table*}[b]
\centering
\caption{Spectroscopic details of the galaxies of the cluster cl05.}
\begin{tabular}{c c c c c c c c c c}
\hline \hline  
\vspace{-0.3cm}            \\
R$_{500}$  & RA	 	&     DEC        &   z    & z$_{err}$ & QF 	&   \multicolumn{3}{c}{Distance from X-ray centroid} & Clipped	\\
(kpc)      & (J2000)	&     (J2000)    &        & 	      &    	&     (\arcsec)	& 	(kpc)	&	(r/R$_{500}$)&	out	\\
\vspace{-0.3cm}            \\
\hline
\vspace{-0.25cm}            \\
	\multicolumn{10}{c}{\textbf{XDCP\,J1243.2+1313 - cl05}} \\
	\hline \vspace{-0.3cm}            \\
	\textbf{651}	   & \textbf{12:43:12.0} & \textbf{+13:13:09.6}   & \textbf{0.7913} &	   &		&		&		 &			 \\
	\hline \vspace{-0.3cm}            \\
	     & 12:43:12.0 & +13:13:10.1   & 0.7890 & 0.0002   &  4 	&  	 1	&  	    6 &  0.009\\
	     & 12:43:12.3 & +13:12:57.4   & 0.7883 & 0.0002   &  2 	&  	 13	&  	    96 &  0.147\\
	     & 12:43:11.0 & +13:13:15.6   & 0.7881 & 0.0002   &  3 	&  	 15	&  	    114 &  0.175\\
	     & 12:43:12.8 & +13:12:51.3   & 0.7951 & 0.0002   &  2 	&  	 22	&  	    164 &  0.252\\
	     & 12:43:11.9 & +13:13:41.8   & 0.7986 & 0.0002   &  3 	&  	 32	&  	    241 &  0.370\\
	     & 12:43:09.8 & +13:13:19.2   & 0.7883 & 0.0002   &  3 	&  	 33	&  	    246 &  0.378\\
	     & 12:43:10.1 & +13:12:46.1   & 0.7869 & 0.0002   &  3 	&  	 37	&  	    274 &  0.421\\
	     & 12:43:09.6 & +13:13:29.7   & 0.7965 & 0.0002   &  2 	&  	 41	&  	    304 &  0.467\\
	     & 12:43:13.3 & +13:12:23.8   & 0.7855 & 0.0002   &  3 	&  	 50	&  	    372 &  0.571\\
	     & 12:43:08.6 & +13:13:28.7   & 0.7944 & 0.0002   &  2 	&  	 53	&  	    393 &  0.604\\
	     & 12:43:15.8 & +13:12:46.2   & 0.7891 & 0.0002   &  4 	&  	 61	&  	    454 &  0.697\\
	     & 12:43:07.3 & +13:13:38.5   & 0.7973 & 0.0002   &  2 	&  	 74	&  	    553 &  0.849\\
	     & 12:43:17.3 & +13:12:56.4   & 0.7896 & 0.0002   &  4 	&  	 78	&  	    585 &  0.899\\
	     & 12:43:06.0 & +13:13:25.8   & 0.7881 & 0.0002   &  4 	&  	 89	&  	    667 &  1.025\\
	     & 12:43:06.6 & +13:14:11.6   & 0.7905 & 0.0002   &  2 	&  	 100	&  	    747 &  1.147\\
	     & 12:43:18.9 & +13:13:10.3   & 0.7906 & 0.0002   &  2 	&  	 100	&  	    750 &  1.152\\
	     & 12:43:19.4 & +13:12:55.5   & 0.7948 & 0.0002   &  3 	&  	 109	&  	    819 &  1.258\\
	     & 12:43:19.9 & +13:13:14.4   & 0.7868 & 0.0002   &  3 	&  	 116	&  	    864 &  1.327\\
	     & 12:43:16.5 & +13:11:18.5   & 0.8033 & 0.0002   &  3 	&  	 129	&  	    963 &  1.479\\
	     & 12:43:03.5 & +13:13:45.6   & 0.7934 & 0.0002   &  1 	&  	 129	&  	    965 &  1.482 & \checkmark \\ 
	     & 12:43:19.7 & +13:11:59.0   & 0.7897 & 0.0002   &  4 	&  	 132	&  	    985 &  1.513\\
	     & 12:43:17.7 & +13:10:59.0   & 0.8000 & 0.0002   &  2 	&  	 155	&  	    1159 &  1.780\\
	     & 12:43:23.4 & +13:11:58.4   & 0.7858 & 0.0002   &  3 	&  	 181	&  	    1354 &  2.080\\
	     & 12:43:22.6 & +13:11:34.7   & 0.7899 & 0.0002   &  3 	&  	 182	&  	    1358 &  2.086\\
	     & 12:43:22.7 & +13:11:31.0   & 0.7962 & 0.0002   &  3 	&  	 185	&  	    1381 &  2.121\\
	     & 12:43:25.7 & +13:11:45.2   & 0.7937 & 0.0002   &  2 	&  	 217	&  	    1621 &  2.490\\
\vspace{-0.3cm}            \\
\hline\hline  
\end{tabular}
\end{table*}

%%%%%% cl06
\begin{figure*}[ht]
\centering
\includegraphics[height=9cm, clip=true]{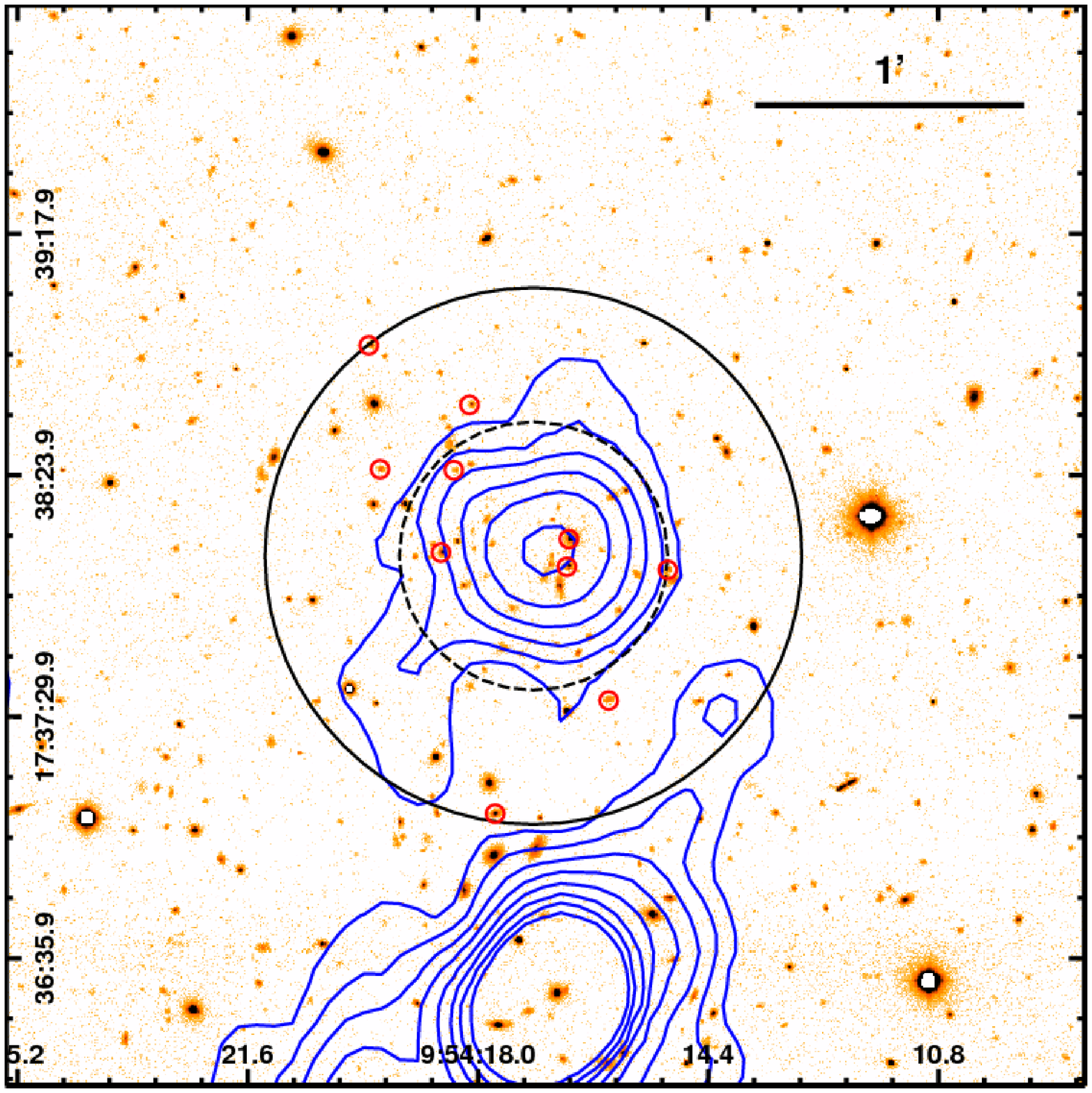}
\includegraphics[height=9cm, clip=true]{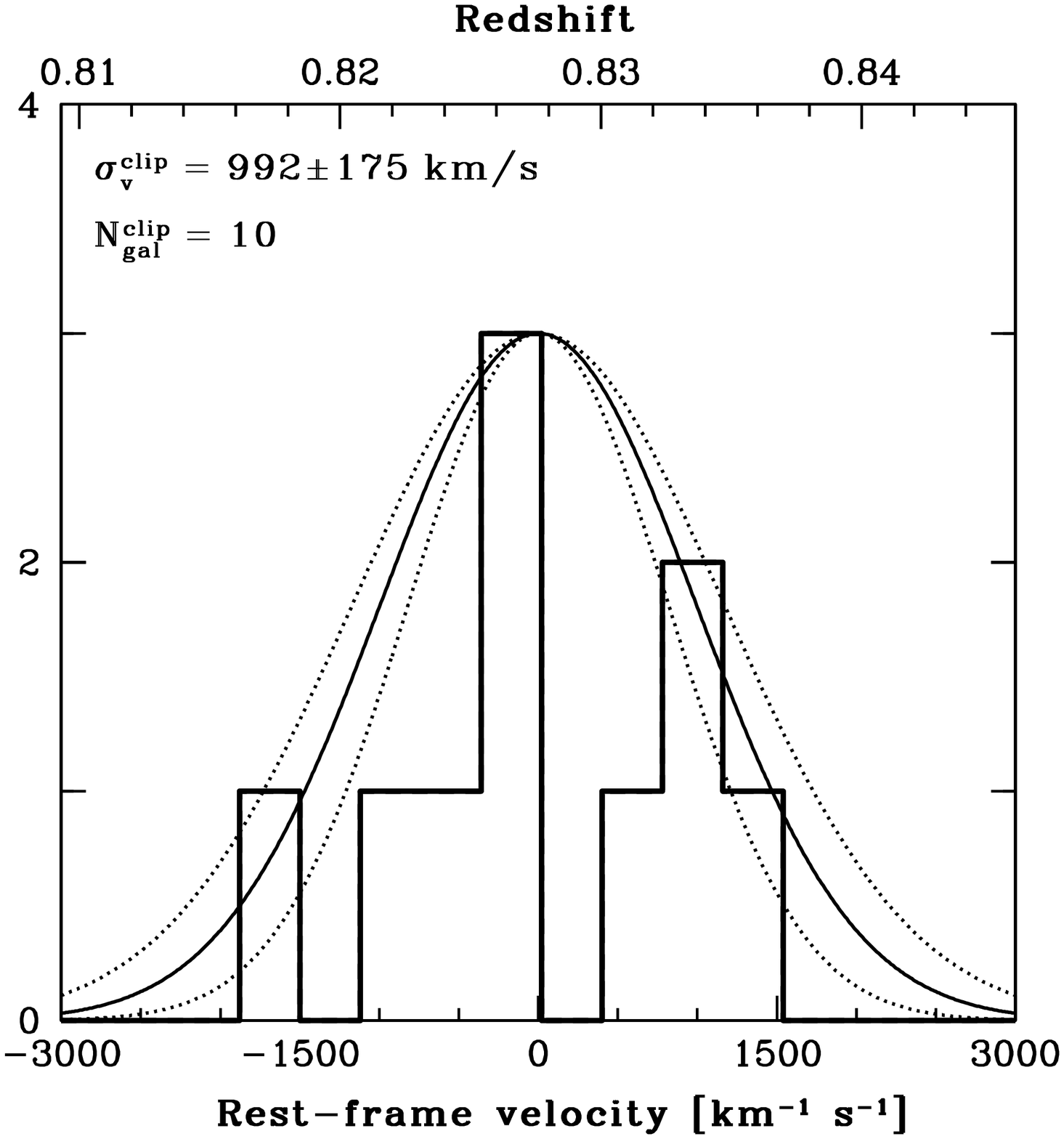}
\caption{\textit{Left:} A 4\arcmin$\times$4\arcmin\ wide z-Band image of cluster XDCP\,J0954.2+1738 - cl06 at z = 0.8276. Symbols and colors have the same meaning as in Fig.\,\ref{fig:134com}. \textit{Right:} Rest-frame velocity histogram of the cluster galaxies.}
\end{figure*}
% \vspace{-1cm}
\begin{table*}[b]
\centering
\caption{Spectroscopic details of the galaxies of the cluster cl06.}
\begin{tabular}{c c c c c c c c c c}
\hline \hline  
\vspace{-0.3cm}            \\
R$_{500}$  & RA	 	&     DEC        &   z    & z$_{err}$ & QF 	&   \multicolumn{3}{c}{Distance from X-ray centroid} & Clipped	\\
(kpc)      & (J2000)	&     (J2000)    &        & 	      &    	&     (\arcsec)	& 	(kpc)	&	(r/R$_{500}$)&	out	\\
\vspace{-0.3cm}            \\
\hline
\vspace{-0.25cm}            \\
	\multicolumn{10}{c}{\textbf{XDCP\,J0954.2+1738 - cl06}}      \\
	\hline \vspace{-0.3cm}            \\
	\textbf{808}	   & \textbf{09:54:17.1} & \textbf{+17:38:05.9}   & \textbf{0.8276} &	   &		&		&		 &			 \\
	\hline \vspace{-0.3cm}            \\
	     & 09:54:16.6 & +17:38:03.5   & 0.8219 & 0.0002   &  3 	&  	 8	&  	    60 &  0.074\\
	     & 09:54:16.5 & +17:38:09.7   & 0.8258 & 0.0002   &  4 	&  	 9	&  	    68 &  0.084\\
	     & 09:54:18.5 & +17:38:06.0   & 0.8253 & 0.0002   &  3 	&  	 21	&  	    156 &  0.193\\
	     & 09:54:18.3 & +17:38:25.1   & 0.8269 & 0.0002   &  2 	&  	 26	&  	    199 &  0.246\\
	     & 09:54:15.0 & +17:38:02.8   & 0.8335 & 0.0002   &  2 	&  	 30	&  	    229 &  0.283\\
	     & 09:54:15.9 & +17:37:33.6   & 0.8265 & 0.0002   &  4 	&  	 36	&  	    275 &  0.340\\
	     & 09:54:18.1 & +17:38:39.7   & 0.8324 & 0.0002   &  3 	&  	 37	&  	    278 &  0.344\\
	     & 09:54:19.5 & +17:38:25.3   & 0.8360 & 0.0002   &  3 	&  	 40	&  	    301 &  0.373\\
	     & 09:54:17.7 & +17:37:08.3   & 0.8166 & 0.0002   &  3 	&  	 58	&  	    442 &  0.547\\
	     & 09:54:19.7 & +17:38:52.0   & 0.8310 & 0.0002   &  3 	&  	 60	&  	    453 &  0.561\\
	     & 09:54:15.9 & +17:35:18.0   & 0.8233 & 0.0002   &  1 	&  	 168	&  	    1274 &  1.577 & \checkmark \\ 
\vspace{-0.3cm}            \\
\hline\hline  
\end{tabular}
\end{table*}

%%%%%% cl07
\begin{figure*}[ht]
\centering
\includegraphics[height=9cm, clip=true]{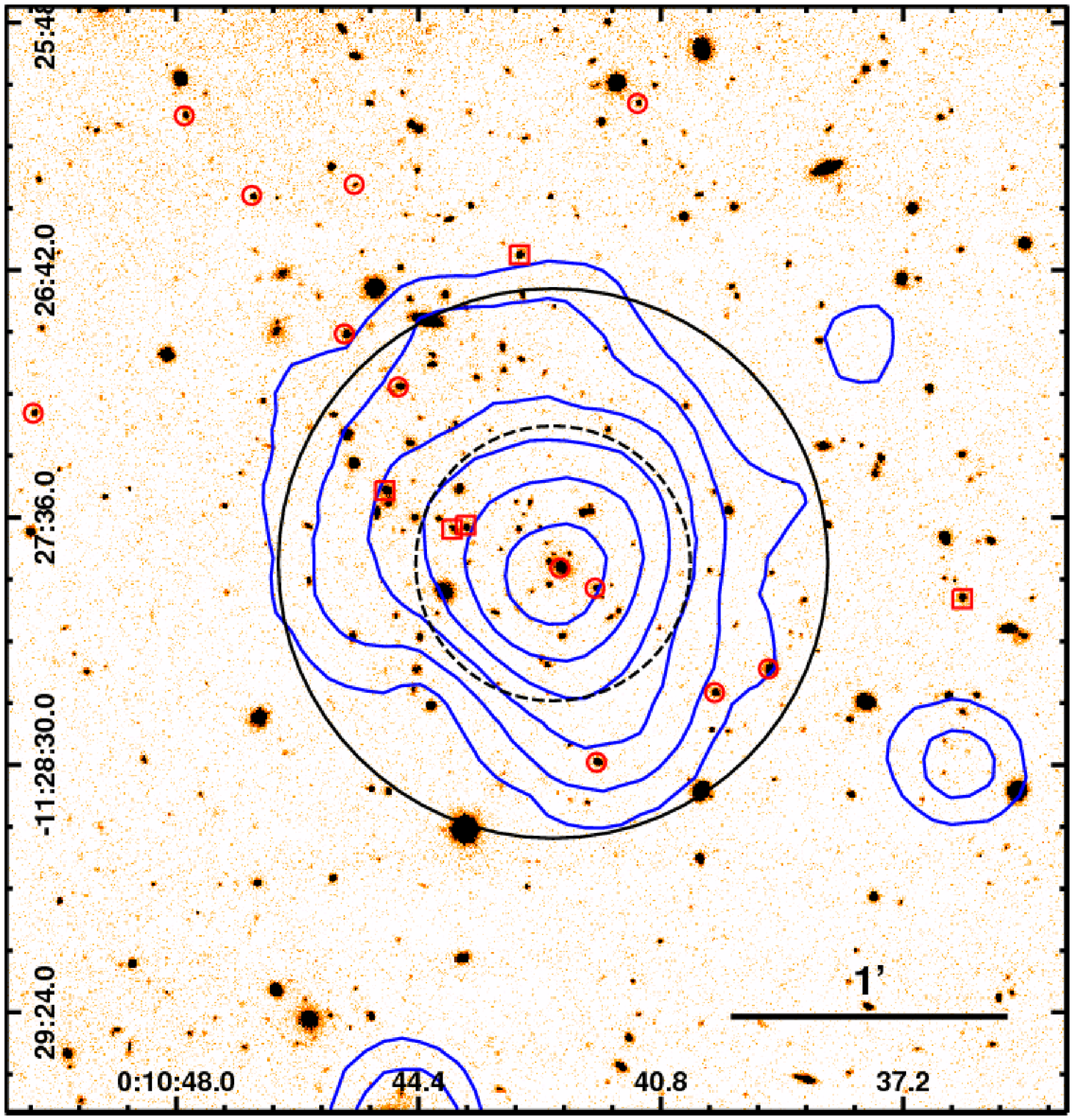}
\includegraphics[height=9cm, clip=true]{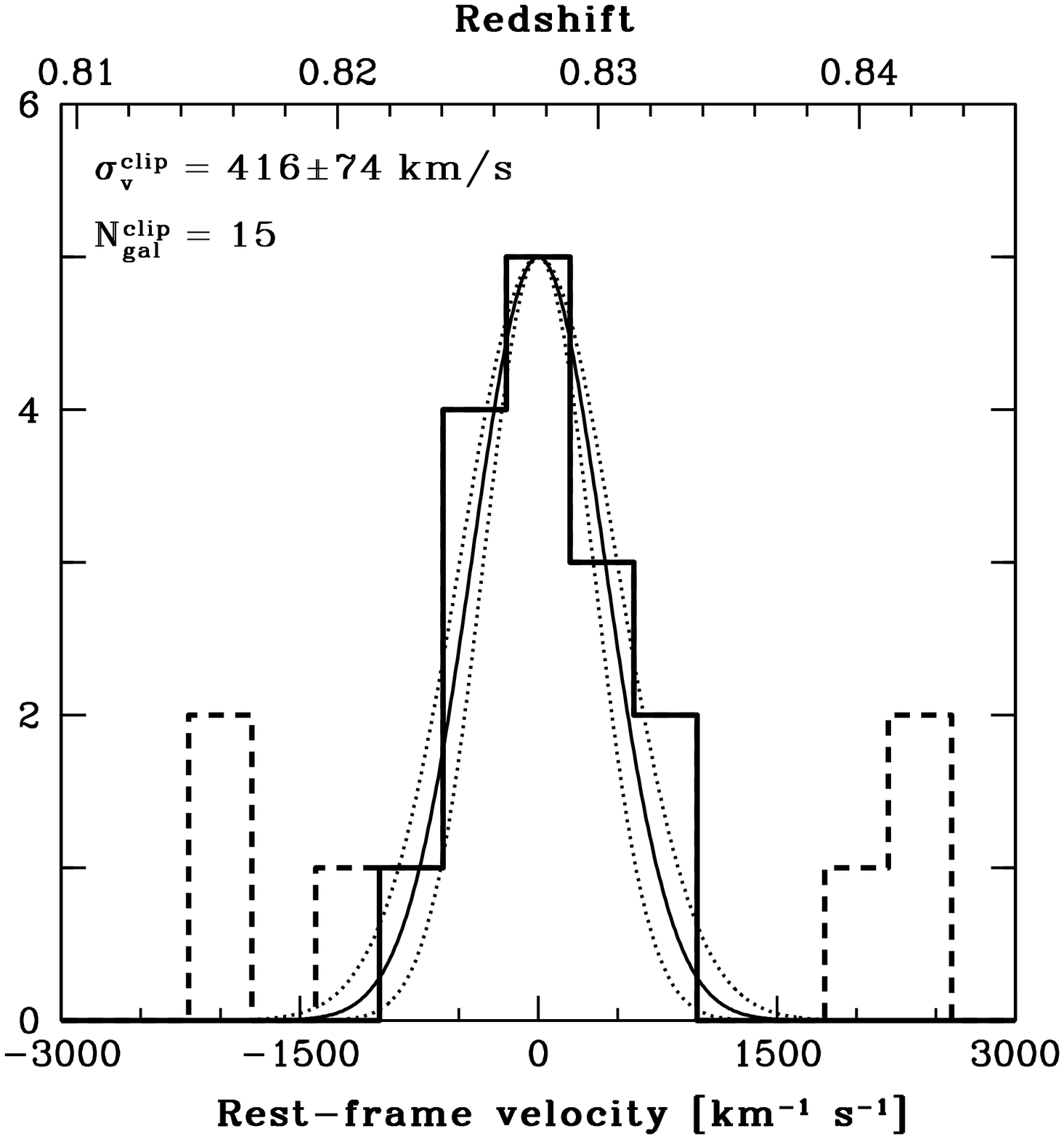}
\caption{\textit{Left:} A 4\arcmin$\times$4\arcmin\ wide z-Band image of cluster XDCP\,J0010.7$-$1127 - cl07 at z = 0.8277. Symbols and colors have the same meaning as in Fig.\,\ref{fig:134com}. \textit{Right:} Rest-frame velocity histogram of the cluster galaxies.}
\end{figure*}
% \vspace{-1cm}
\begin{table*}[b]
\centering
\caption{Spectroscopic details of the galaxies of the cluster cl07.}
\begin{tabular}{c c c c c c c c c c}
\hline \hline  
\vspace{-0.3cm}            \\
R$_{500}$  & RA	 	&     DEC        &   z    & z$_{err}$ & QF 	&   \multicolumn{3}{c}{Distance from X-ray centroid} & Clipped	\\
(kpc)      & (J2000)	&     (J2000)    &        & 	      &    	&     (\arcsec)	& 	(kpc)	&	(r/R$_{500}$)&	out	\\
\vspace{-0.3cm}            \\
\hline
\vspace{-0.25cm}            \\
	\multicolumn{10}{c}{\textbf{XDCP\,J0010.7$-$1127 - cl07}}      \\
	\hline \vspace{-0.3cm}            \\
	\textbf{860}	   & \textbf{00:10:42.4} & \textbf{$-$11:27:46.0}   & \textbf{0.8277} &	   &		&		&		 &			 \\
	\hline \vspace{-0.3cm}            \\
	     & 00:10:42.3 & $-$11:27:46.0   & 0.8273 & 0.0002	&  4	  &	   2	  &	      14 &  0.016\\
	     & 00:10:41.8 & $-$11:27:51.4   & 0.8226 & 0.0002	&  3	  &	   11	  &	      81 &  0.094\\
	     & 00:10:43.7 & $-$11:27:37.0   & 0.8412 & 0.0005	&  2	  &	   21	  &	      157 &  0.183 & \checkmark \\ 
	     & 00:10:43.9 & $-$11:27:38.5   & 0.8152 & 0.0002	&  4	  &	   23	  &	      176 &  0.205 & \checkmark \\ 
	     & 00:10:44.9 & $-$11:27:30.1   & 0.8434 & 0.0002	&  4	  &	   40	  &	      302 &  0.351 & \checkmark \\ 
	     & 00:10:41.8 & $-$11:28:29.4   & 0.8263 & 0.0002	&  3	  &	   44	  &	      337 &  0.392\\
	     & 00:10:40.0 & $-$11:28:14.2   & 0.8283 & 0.0002	&  3	  &	   45	  &	      342 &  0.398\\
	     & 00:10:44.7 & $-$11:27:07.5   & 0.8283 & 0.0002	&  4	  &	   51	  &	      388 &  0.451\\
	     & 00:10:39.2 & $-$11:28:09.0   & 0.8265 & 0.0002	&  4	  &	   52	  &	      398 &  0.463\\
	     & 00:10:45.5 & $-$11:26:55.9   & 0.8314 & 0.0002	&  4	  &	   68	  &	      513 &  0.597\\
	     & 00:10:42.9 & $-$11:26:38.6   & 0.8151 & 0.0002	&  4	  &	   68	  &	      515 &  0.599 & \checkmark \\ 
	     & 00:10:36.3 & $-$11:27:53.4   & 0.8397 & 0.0002	&  4	  &	   90	  &	      680 &  0.791 & \checkmark \\ 
	     & 00:10:45.4 & $-$11:26:23.3   & 0.8294 & 0.0003	&  3	  &	   93	  &	      709 &  0.824\\
	     & 00:10:41.1 & $-$11:26:05.6   & 0.8303 & 0.0002	&  3	  &	   102    &	      775 &  0.901\\
	     & 00:10:46.9 & $-$11:26:25.0   & 0.8303 & 0.0002	&  3	  &	   103    &	      786 &  0.914\\
	     & 00:10:50.1 & $-$11:27:13.2   & 0.8249 & 0.0002	&  3	  &	   118    &	      897 &  1.043\\
	     & 00:10:47.9 & $-$11:26:08.3   & 0.8260 & 0.0002	&  3	  &	   127    &	      961 &  1.117\\
	     & 00:10:33.3 & $-$11:27:44.7   & 0.8276 & 0.0002	&  3	  &	   134    &	      1020 &  1.186\\
	     & 00:10:31.1 & $-$11:28:18.4   & 0.8195 & 0.0002	&  3	  &	   170    &	      1289 &  1.499 & \checkmark \\ 
	     & 00:10:30.4 & $-$11:28:44.2   & 0.8314 & 0.0002	&  3	  &	   186    &	      1415 &  1.645\\
	     & 00:10:28.2 & $-$11:27:43.7   & 0.8247 & 0.0002	&  3	  &	   208    &	      1581 &  1.838\\
\vspace{-0.3cm}            \\
\hline\hline  
\end{tabular}
\end{table*}

%%%%%% cl08
\newpage
\begin{figure*}[th]
\centering
\includegraphics[height=9cm, clip=true]{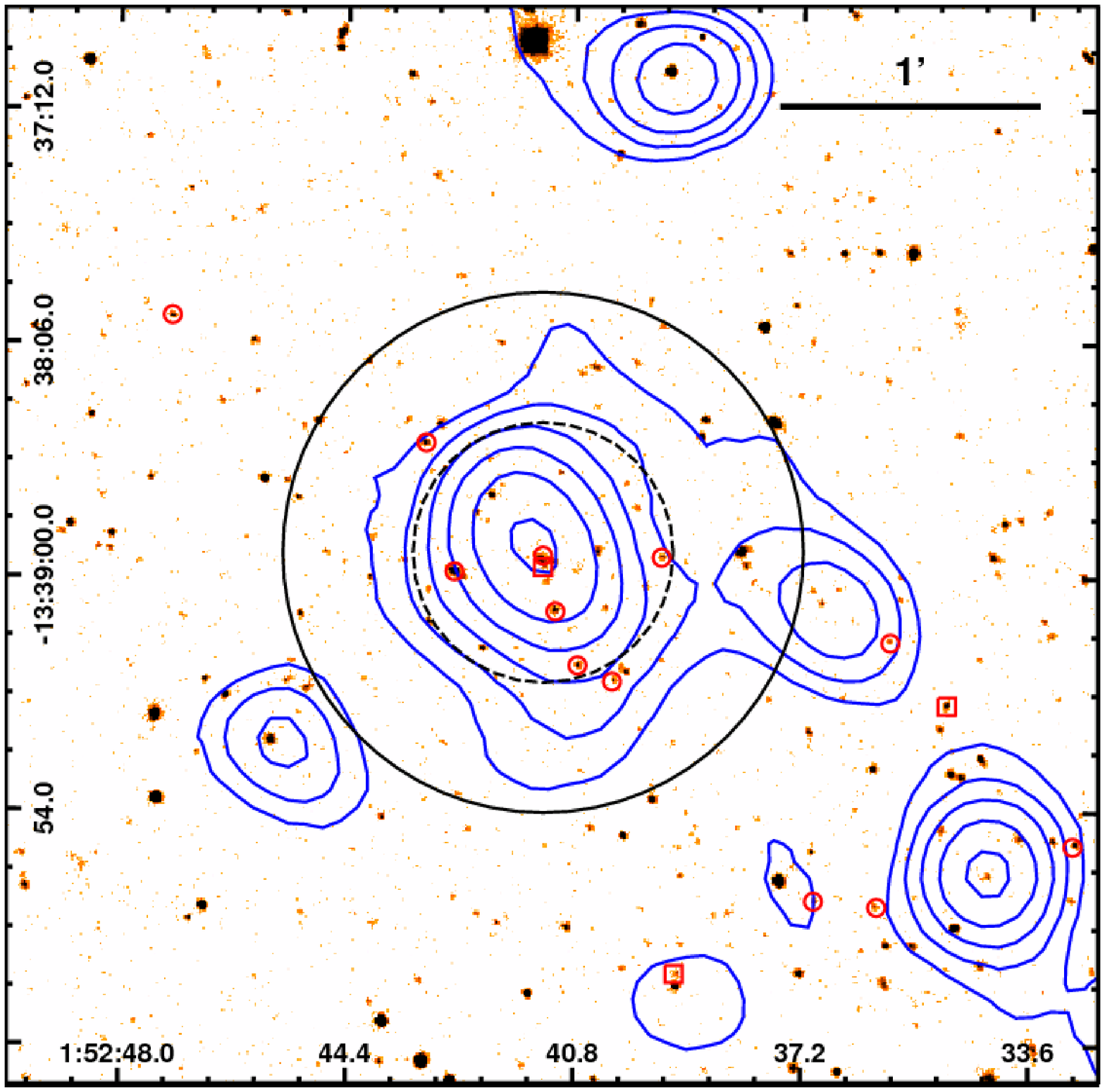}
\includegraphics[height=9cm, clip=true]{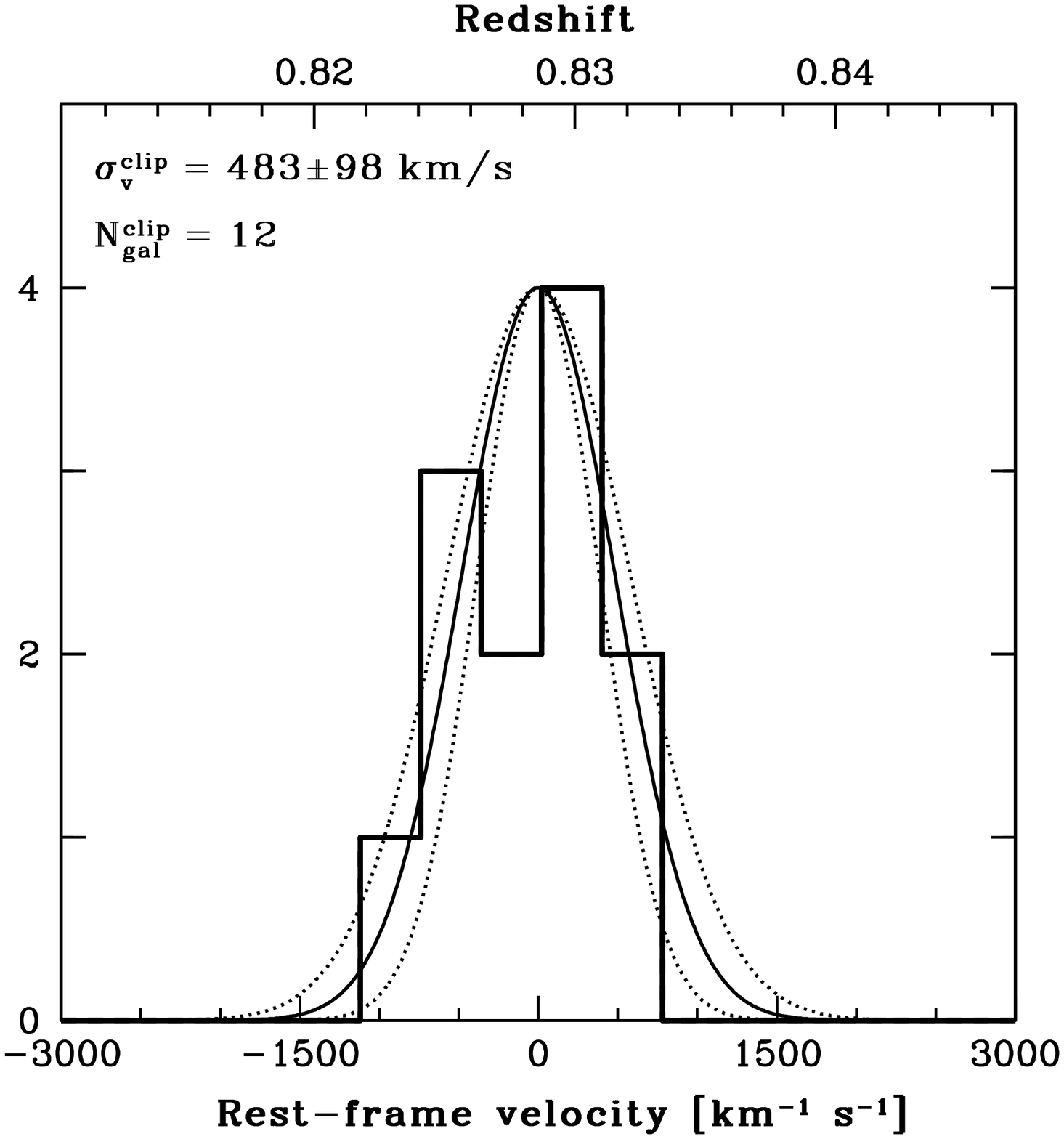}
\caption{\textit{Left:} A 4\arcmin$\times$4\arcmin\ wide z-Band image of cluster XDCP\,J0152.6$-$1338 - cl08 at z = 0.8289. Symbols and colors have the same meaning as in Fig.\,\ref{fig:134com}. \textit{Right:} Rest-frame velocity histogram of the cluster galaxies.}
\end{figure*}
% \vspace{-1cm}
\begin{table*}[b]
\centering
\caption{Spectroscopic details of the galaxies of the cluster cl08.}
\begin{tabular}{c c c c c c c c c c}
\hline \hline  
\vspace{-0.3cm}            \\
R$_{500}$  & RA	 	&     DEC        &   z    & z$_{err}$ & QF 	&   \multicolumn{3}{c}{Distance from X-ray centroid} & Clipped	\\
(kpc)      & (J2000)	&     (J2000)    &        & 	      &    	&     (\arcsec)	& 	(kpc)	&	(r/R$_{500}$)&	out	\\
\vspace{-0.3cm}            \\
\hline
\vspace{-0.25cm}            \\
	\multicolumn{10}{c}{\textbf{XDCP\,J0152.6$-$1338 - cl08}}     \\
	\hline \vspace{-0.3cm}  	  \\
	\textbf{695}	   & \textbf{01:52:41.3} & \textbf{$-$13:38:54.3}   & \textbf{0.8289} &	   &		&		&		 &			 \\
	\hline \vspace{-0.3cm}            \\
	     & 01:52:41.3 & $-$13:38:55.0   & 0.8307 & 0.0004	&  2	  &	   2	  &	      14 &  0.020\\
	     & 01:52:41.3 & $-$13:38:57.7   & 0.8265 & 0.0012	&  1	  &	   3	  &	      26 &  0.037 & \checkmark \\ 
	     & 01:52:41.1 & $-$13:39:07.9   & 0.8303 & 0.0003	&  3	  &	   14	  &	      105 &  0.151\\
	     & 01:52:42.7 & $-$13:38:59.1   & 0.8261 & 0.0002	&  4	  &	   21	  &	      160 &  0.230\\
	     & 01:52:39.4 & $-$13:38:55.7   & 0.8233 & 0.0002	&  3	  &	   27	  &	      209 &  0.301\\
	     & 01:52:40.7 & $-$13:39:21.0   & 0.8261 & 0.0003	&  2	  &	   28	  &	      212 &  0.305\\
	     & 01:52:40.2 & $-$13:39:24.0   & 0.8307 & 0.0005	&  3	  &	   34	  &	      257 &  0.370\\
	     & 01:52:43.2 & $-$13:38:29.4   & 0.8268 & 0.0002	&  3	  &	   37	  &	      281 &  0.404\\
	     & 01:52:35.8 & $-$13:39:14.9   & 0.8326 & 0.0002	&  2	  &	   82	  &	      626 &  0.901\\
	     & 01:52:34.9 & $-$13:39:29.4   & 0.8318 & 0.0002	&  1	  &	   100    &	      757 &  1.089 & \checkmark \\ 
	     & 01:52:47.2 & $-$13:38:00.1   & 0.8256 & 0.0002	&  3	  &	   102    &	      771 &  1.109\\
	     & 01:52:39.2 & $-$13:40:31.4   & 0.8222 & 0.0005	&  1	  &	   102    &	      775 &  1.115 & \checkmark \\ 
	     & 01:52:37.0 & $-$13:40:14.9   & 0.8324 & 0.0002	&  2	  &	   102    &	      777 &  1.118\\
	     & 01:52:36.0 & $-$13:40:15.9   & 0.8295 & 0.0002	&  3	  &	   112    &	      851 &  1.224\\
	     & 01:52:32.9 & $-$13:40:01.8   & 0.8279 & 0.0002	&  3	  &	   140    &	      1067 &  1.535\\
\vspace{-0.3cm}            \\
\hline\hline  
\end{tabular}
\end{table*}

%%%%%% cl09
% \clearpage
\begin{figure*}[ht]
\centering
\includegraphics[height=9cm, clip=true]{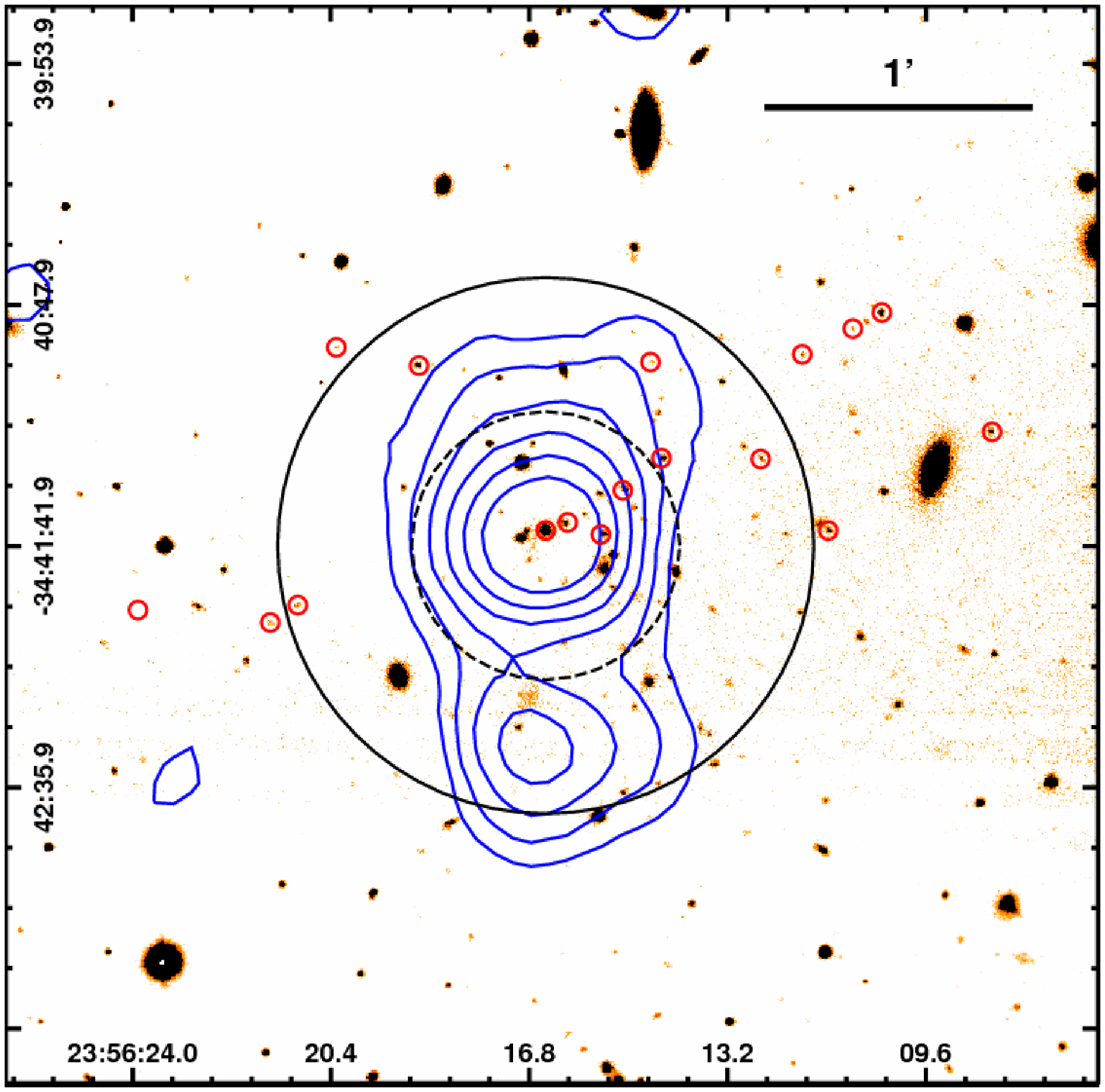}
\includegraphics[height=9cm, clip=true]{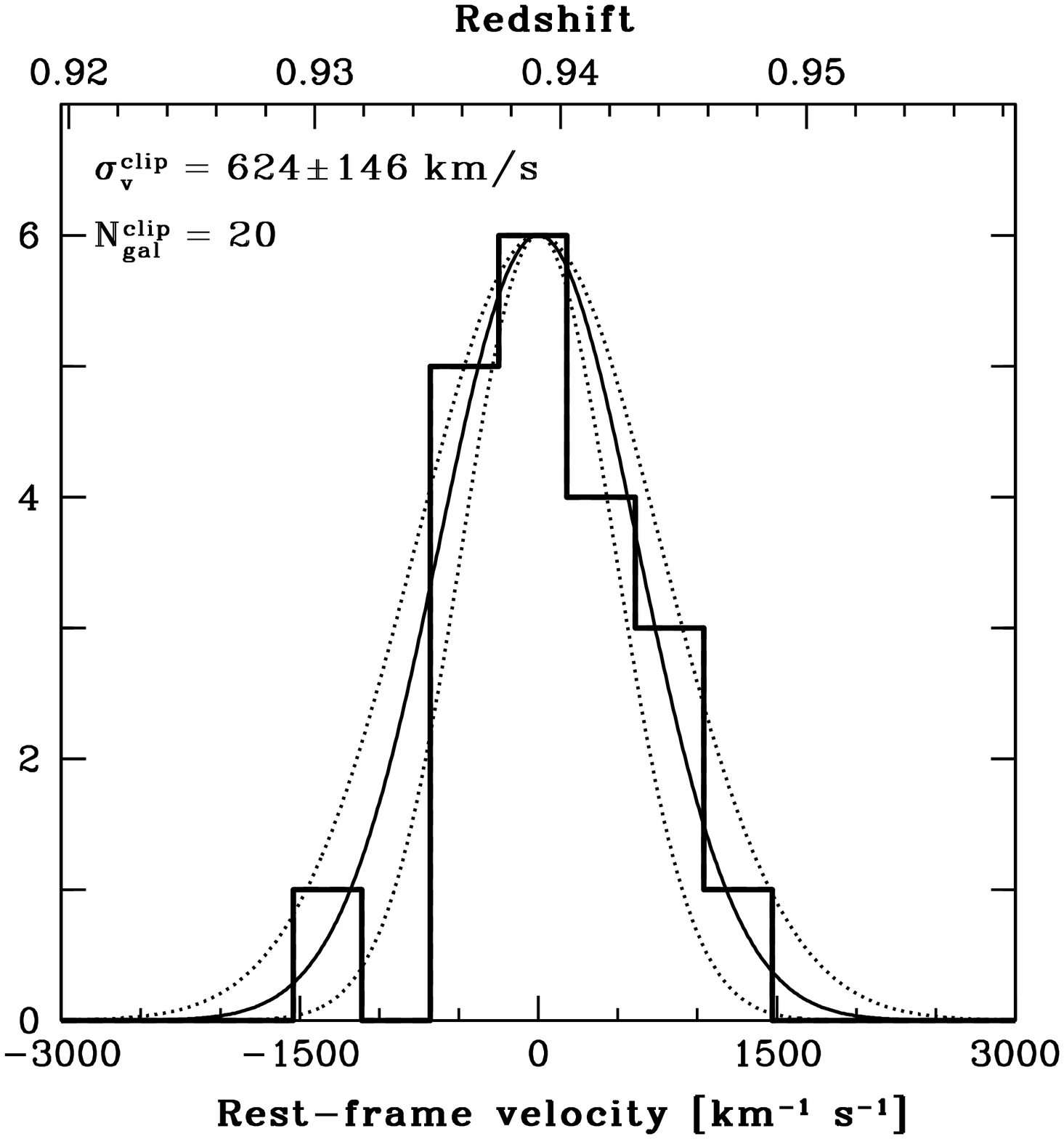}
\caption{\textit{Left:} A 4\arcmin$\times$4\arcmin\ wide z-Band image of cluster XDCP\,J2356.2$-$3441 - cl09 at z = 0.9391. Symbols and colors have the same meaning as in Fig.\,\ref{fig:134com}. \textit{Right:} Rest-frame velocity histogram of the cluster galaxies.}
\end{figure*}
% \vspace{-1cm}
\begin{table*}[b]
\centering
\caption{Spectroscopic details of the galaxies of the cluster cl09.}
\begin{tabular}{c c c c c c c c c c}
\hline \hline  
\vspace{-0.3cm}            \\
R$_{500}$  & RA	 	&     DEC        &   z    & z$_{err}$ & QF 	&   \multicolumn{3}{c}{Distance from X-ray centroid} & Clipped	\\
(kpc)      & (J2000)	&     (J2000)    &        & 	      &    	&     (\arcsec)	& 	(kpc)	&	(r/R$_{500}$)&	out	\\
\vspace{-0.3cm}            \\
\hline
\vspace{-0.25cm}            \\
	\multicolumn{10}{c}{\textbf{XDCP\,J2356.2$-$3441 - cl09}} \\
	\hline \vspace{-0.3cm}            \\
	\textbf{777}	   & \textbf{23:56:16.5} & \textbf{$-$34:41:41.8}   & \textbf{0.9391} &	   &		&		&		 &			 \\
	\hline \vspace{-0.3cm}            \\
	     & 23:56:16.5 & $-$34:41:38.4   & 0.9407 & 0.0002	&  3	  &	   3	  &	      27 &  0.035\\
	     & 23:56:16.1 & $-$34:41:36.6   & 0.9452 & 0.0002	&  3	  &	   7	  &	      54 &  0.069\\
	     & 23:56:15.5 & $-$34:41:39.3   & 0.9305 & 0.0003	&  3	  &	   13	  &	      103 &  0.133\\
	     & 23:56:15.1 & $-$34:41:29.4   & 0.9427 & 0.0002	&  3	  &	   22	  &	      170 &  0.219\\
	     & 23:56:14.4 & $-$34:41:22.2   & 0.9433 & 0.0003	&  4	  &	   33	  &	      260 &  0.335\\
	     & 23:56:14.6 & $-$34:41:00.7   & 0.9453 & 0.0002	&  3	  &	   47	  &	      374 &  0.481\\
	     & 23:56:18.8 & $-$34:41:01.5   & 0.9376 & 0.0003	&  3	  &	   49	  &	      390 &  0.502\\
	     & 23:56:12.6 & $-$34:41:22.4   & 0.9462 & 0.0002	&  4	  &	   52	  &	      411 &  0.529\\
	     & 23:56:21.0 & $-$34:41:55.1   & 0.9374 & 0.0002	&  3	  &	   57	  &	      448 &  0.577\\
	     & 23:56:11.4 & $-$34:41:38.4   & 0.9421 & 0.0002	&  4	  &	   64	  &	      501 &  0.645\\
	     & 23:56:21.5 & $-$34:41:59.0   & 0.9375 & 0.0005	&  3	  &	   64	  &	      503 &  0.647\\
	     & 23:56:20.3 & $-$34:40:57.4   & 0.9365 & 0.0002	&  3	  &	   65	  &	      509 &  0.655\\
	     & 23:56:11.9 & $-$34:40:59.3   & 0.9357 & 0.0002	&  3	  &	   71	  &	      562 &  0.723\\
	     & 23:56:10.9 & $-$34:40:53.2   & 0.9389 & 0.0002	&  2	  &	   84	  &	      663 &  0.853\\
	     & 23:56:10.4 & $-$34:40:49.6   & 0.9381 & 0.0002	&  4	  &	   91	  &	      721 &  0.928\\
	     & 23:56:23.9 & $-$34:41:56.2   & 0.9384 & 0.0002	&  3	  &	   92	  &	      724 &  0.932\\
	     & 23:56:08.4 & $-$34:41:16.3   & 0.9414 & 0.0002	&  4	  &	   103    &	      810 &  1.042\\
	     & 23:56:27.3 & $-$34:41:48.6   & 0.9376 & 0.0002	&  3	  &	   133    &	      1050 &  1.351\\
	     & 23:56:03.7 & $-$34:42:10.9   & 0.9356 & 0.0002	&  2	  &	   160    &	      1261 &  1.623\\
	     & 23:56:01.5 & $-$34:41:19.4   & 0.9374 & 0.0002	&  3	  &	   186    &	      1469 &  1.891\\
\vspace{-0.3cm}            \\
\hline\hline  
\end{tabular}
\end{table*}

% %%%%% cl10
\newpage
\begin{figure*}[ht]
\centering
\includegraphics[height=9cm, clip=true]{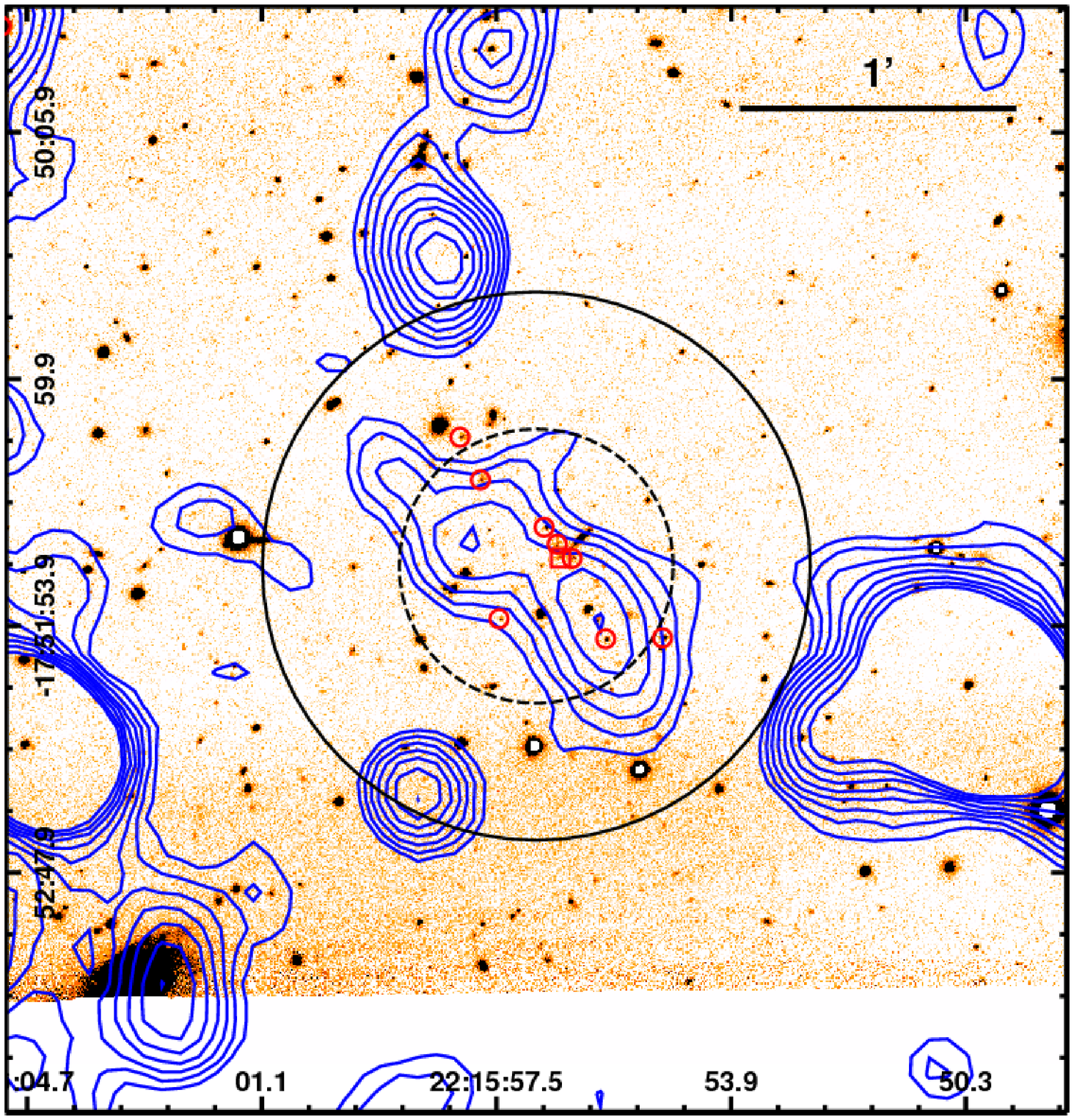}
\includegraphics[height=9cm, clip=true]{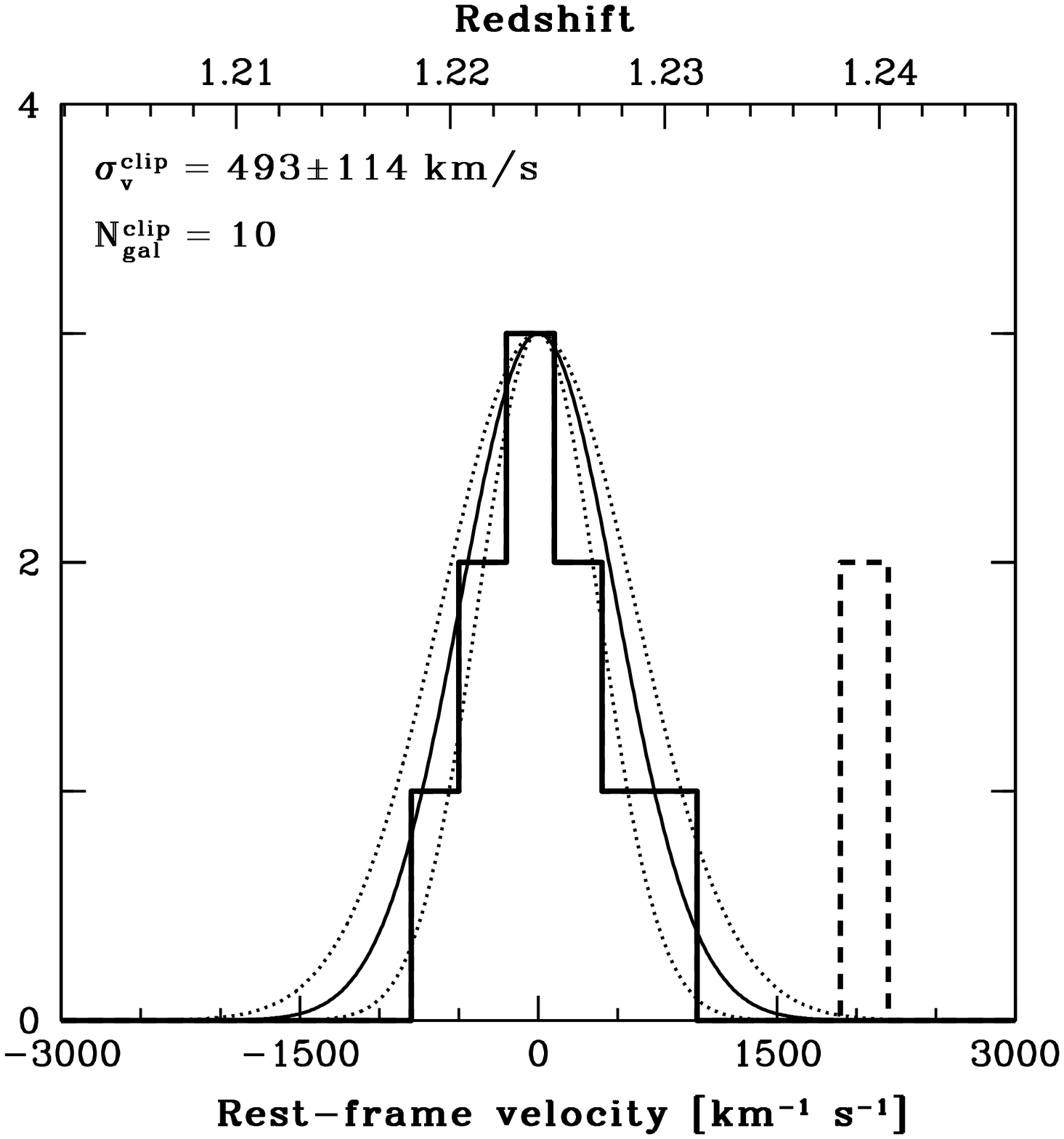}
\caption{\textit{Left:} A 4\arcmin$\times$4\arcmin\ wide z-Band image of cluster XDCP\,J2215.9$-$1751 - cl10 at z = 1.2249. Symbols and colors have the same meaning as in Fig.\,\ref{fig:134com}. \textit{Right:} Rest-frame velocity histogram of the cluster galaxies.}
\end{figure*}
% \vspace{-1cm}
\begin{table*}[b]
\centering
\caption{Spectroscopic details of the galaxies of the cluster cl10.}
\begin{tabular}{c c c c c c c c c c}
\hline \hline  
\vspace{-0.3cm}            \\
R$_{500}$  & RA	 	&     DEC        &   z    & z$_{err}$ & QF 	&   \multicolumn{3}{c}{Distance from X-ray centroid} & Clipped	\\
(kpc)      & (J2000)	&     (J2000)    &        & 	      &    	&     (\arcsec)	& 	(kpc)	&	(r/R$_{500}$)&	out	\\
\vspace{-0.3cm}            \\
\hline
\vspace{-0.25cm}            \\
	\multicolumn{10}{c}{\textbf{XDCP\,J2215.9$-$1751 - cl10}} \\
	\hline \vspace{-0.3cm}  	  \\
	\textbf{384}	   & \textbf{22:15:56.9} & \textbf{$-$17:51:40.9}   & \textbf{1.2249} &    &		&		&		 &			 \\
	\hline \vspace{-0.3cm}  	  \\
	     & 22:15:56.5 & $-$17:51:39.1   & 1.2388 & 0.0001	&  3	  &	   6	  &	      46 &  0.120 & \checkmark \\ 
	     & 22:15:56.6 & $-$17:51:36.1   & 1.2190 & 0.0006	&  3	  &	   6	  &	      53 &  0.138\\
	     & 22:15:56.4 & $-$17:51:39.2   & 1.2212 & 0.0002	&  3	  &	   8	  &	      64 &  0.167\\
	     & 22:15:56.8 & $-$17:51:32.4   & 1.2256 & 0.0002	&  3	  &	   9	  &	      72 &  0.188\\
	     & 22:15:57.4 & $-$17:51:52.2   & 1.2240 & 0.0003	&  3	  &	   14	  &	      114 &  0.297\\
	     & 22:15:55.8 & $-$17:51:56.0   & 1.2232 & 0.0002	&  3	  &	   22	  &	      184 &  0.479\\
	     & 22:15:57.8 & $-$17:51:21.7   & 1.2224 & 0.0004	&  3	  &	   23	  &	      190 &  0.495\\
	     & 22:15:54.9 & $-$17:51:56.6   & 1.2302 & 0.0007	&  3	  &	   32	  &	      266 &  0.693\\
	     & 22:15:58.1 & $-$17:51:12.7   & 1.2292 & 0.0002	&  3	  &	   33	  &	      273 &  0.711\\
	     & 22:15:46.4 & $-$17:52:23.0   & 1.2267 & 0.0001	&  3	  &	   156    &	      1300 &  3.385\\
	     & 22:16:05.1 & $-$17:49:42.3   & 1.2231 & 0.0000	&  3	  &	   167	  &	     1388 &  3.615\\
	     & 22:15:47.1 & $-$17:49:07.9   & 1.2391 & 0.0001	&  3	  &	   207	  &	     1723 &  4.487 & \checkmark \\ 
\vspace{-0.3cm}            \\
\hline\hline  
\end{tabular}
\end{table*}

\newpage
\clearpage
\section*{Appendix B}
In this Appendix, we show the rest-frame velocity histograms of the 12 ``literature sample'' clusters listed in Table\,\ref{Tab:Cluster_literature} with a public redshift set. As in Appendix\,A, we also plot the best Gaussian fit whose variance is the $\sigma^{clip}_v$ computed with the method described in Sec.\,\ref{par:sigmaMeasure}. We also report the $\sigma_v^{lit}$ values and plot the corresponding Gaussian fit in red.\vspace{1cm}
\vspace{-0.5cm}

\begin{figure*}[t]
 \includegraphics[width=18cm]{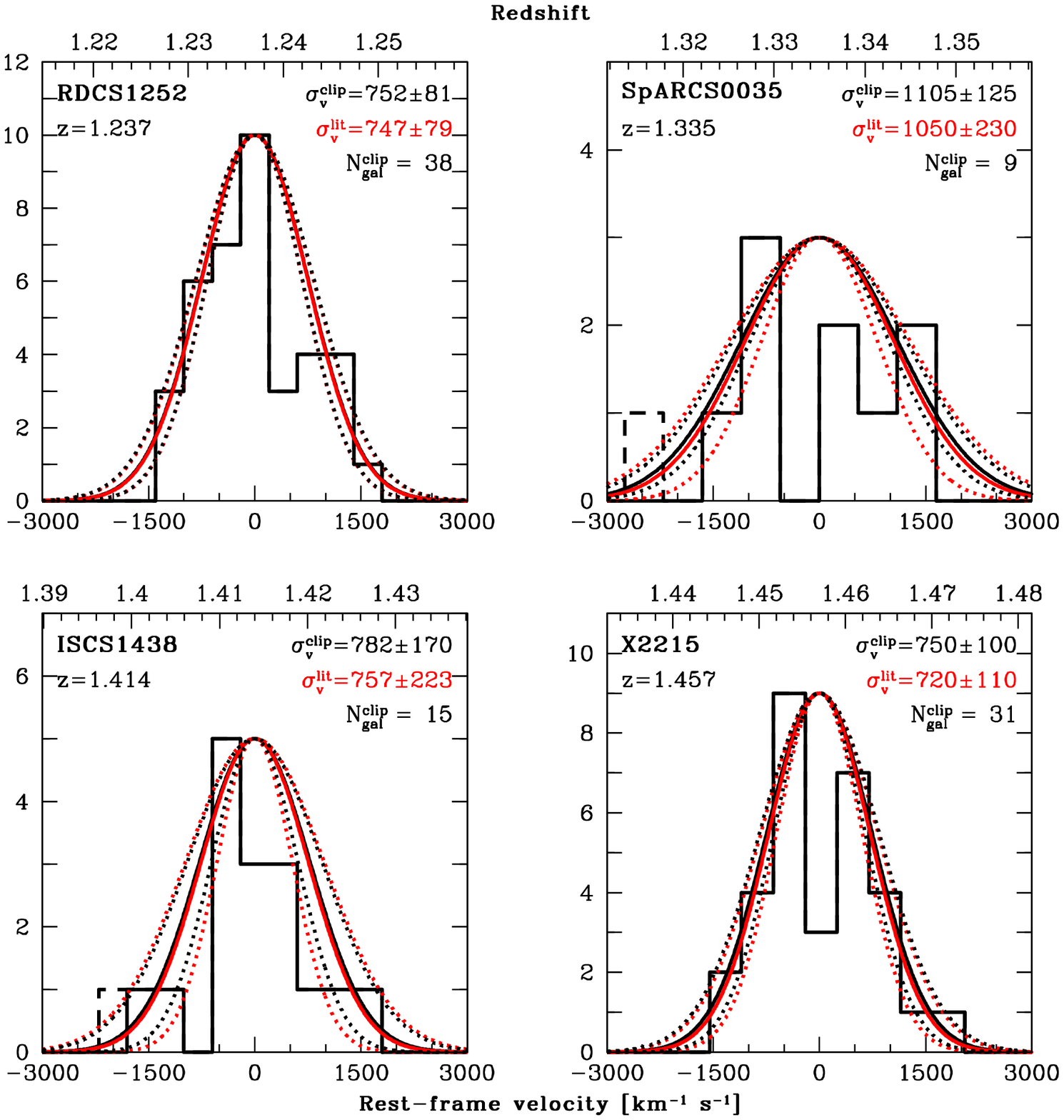}
\caption{Similar to Fig.\,\ref{fig:hist_Lit_1}.}
\label{fig:hist_Lit_3}
\end{figure*}
\newpage
\twocolumn

 \bibliographystyle{aa}
\bibliography{aa2013_22321.bib}

\end{document}